\documentclass[11pt,a4paper]{article}

\pdfoutput=1
\usepackage{jcappub}

\usepackage{amssymb}
\usepackage{amsmath}
\usepackage{bm}
\usepackage{latexsym}
\usepackage{epsfig}
\usepackage{psfrag}
\usepackage{amsmath}
\usepackage[normalem]{ulem}
\usepackage{textcomp}
\usepackage{color}
\usepackage{pstricks}
\usepackage[utf8]{inputenc}
\usepackage{calligra} 
\DeclareMathAlphabet{\mathcalligra}{T1}{calligra}{m}{n}
\DeclareFontShape{T1}{calligra}{m}{n}{<->s*[2.2]callig15}{}

\usepackage{graphicx}
\usepackage{amsmath}
\usepackage{amsfonts}
\usepackage{placeins}
\usepackage{mathtools}
\usepackage{todonotes}
\usepackage{amssymb}
\usepackage{bm}
\usepackage{csquotes}
\usepackage{epstopdf}
\usepackage{float}
\usepackage{subfigure}
\usepackage{capt-of}
\usepackage{amsmath}

\usepackage{hyperref}
\usepackage{gensymb}
\usepackage{graphicx}
\usepackage{amssymb}
\usepackage{amsmath}
\usepackage{bm}
\usepackage{latexsym}
\usepackage{epsfig}
\usepackage{psfrag}
\usepackage{color}
\usepackage[utf8]{inputenc}
\usepackage[english]{babel}
\usepackage{comment}
\usepackage{physics}

\usepackage{cancel}
\usepackage{orcidlink}

\makeatletter
\gdef\@fpheader{}
\makeatother

\def\nn{\nonumber} 
\def\pa{\partial}
\def\f{\frac}

\def\l{\left}
\def\r{\right}
\def\d{{\mathrm{d}}}
\def\Mpl{M_{_{\mathrm{Pl}}}}

\def\pt{\mathcal{P}_{_{\mathrm{T}}}}
\def\ptad{\mathcal{P}_{_{\mathrm{T}}}^\mathrm{ad}}
\def\ptr{\mathcal{P}_{_{\mathrm{T}}}^\mathrm{reg}}

\def\HI{H_{_{\mathrm I}}}
\def\He{H_{\mathrm e}}

\def\Tre{T_{\mathrm{rh}}}

\def\e{\mathrm{e}}
\def\etr{\eta_{\mathrm{tr}}}

\def\ei{\eta_{\mathrm{i}}}
\def\ee{\eta_{\mathrm{e}}}
\def\er{\eta_{\mathrm{r}}}

\def\ema{\eta_{\mathrm{m}}}

\def\eeq{\eta_{\mathrm{eq}}}

\def\ae{a_{\mathrm{e}}}
\def\ar{a_{\mathrm{r}}}
\def\am{a_{\mathrm{m}}}

\def\ke{k_{\mathrm{e}}}
\def\kre{k_{\mathrm{re}}}
\def\keq{k_{\mathrm{eq}}}

\def\Ai{\mathrm{Ai}}
\def\Bi{\mathrm{Bi}}
\def\vk{\bm k}

\def\xeq{x_{\mathrm{eq}}}

\def\th{\tilde h}

\def\bl#1\el{\begin{align}#1\end{align}}

\allowdisplaybreaks

\usepackage{hyperref}

\makeatletter
\gdef\@fpheader{}
\g@addto@macro\bfseries{\boldmath}
\makeatother

\newcommand{\deflen}[2]{%
    \expandafter\newlength\csname #1\endcsname
    \expandafter\setlength\csname #1\endcsname{#2}%
}
\subheader{}

\title{Primary gravitational waves at high frequencies~I:~Origin of 
suppression in the power spectrum}
\author[a]{Alipriyo Hoory\,\orcidlink{0009-0006-3486-2460},}
\emailAdd{alipriyo@physics.iitm.ac.in}
\author[b]{J\'{e}r\^{o}me Martin\,\orcidlink{0000-0002-6861-2092},}
\emailAdd{jmartin@iap.fr}
\author[a]{Arnab Paul\,\orcidlink{0000-0003-3498-6755},}
\emailAdd{arnab.paul@physics.iitm.ac.in}
\author[a]{and L.~Sriramkumar\,\orcidlink{0000-0003-1168-990X}\,}
\emailAdd{sriram@physics.iitm.ac.in}
\affiliation[a]{Centre for Strings, Gravitation and Cosmology,
Department of Physics, Indian Institute of Technology Madras, 
Chennai~600036, India}
\affiliation[b]{Institut d'Astrophysique de Paris, 98 bis boulevard 
Arago, F-75014 Paris, France}
\date{today}
\begin{document}
\abstract{The primary gravitational waves (PGWs) are generated in the early 
universe from the quantum vacuum during inflation.
In the standard scenario of slow roll inflation, the power spectrum (PS) of 
PGWs over large scales, which leave the Hubble radius during inflation, is 
nearly scale-invariant.
However, over very small scales, which never leave the Hubble radius, the 
PS of PGWs behaves as~$k^2$, where $k$ denotes the wave number.
In this work, we examine the PS of PGWs at such high wave numbers or frequencies 
when the PGWs are evolved post-inflation, through the epochs of radiation and
matter domination.
Firstly, we argue that the PS has to be regularized in order to truncate the 
unphysical~$k^2$ rise at high frequencies.
Assuming instantaneous transitions from inflation to the epochs of radiation 
and matter domination, we carry out the method of adiabatic regularization to
arrive at the PS of PGWs over a wide range of frequencies.
We show that the process of regularization truncates the~$k^2$ rise and that 
the regularized PS of PGWs oscillates with a fixed amplitude about a vanishing 
mean value over small scales or, equivalently, at high frequencies.
Secondly, we smooth the transition from inflation to radiation domination 
(to be precise, we smooth the `effective potential' governing the equation 
of motion of PGWs) and examine the impact of the smoothing on the regularized 
PS of PGWs.
With the help of a linear smoothing function, we explicitly show that the 
smoother transition leads to a power-law suppression in the amplitude of 
the oscillations (about the zero mean value) of the regularized PS of PGWs 
over small scales that never leave the Hubble radius during inflation.
Our analysis indicates that, when transitions are involved, regularization
as well as smooth transitions seem essential to ensure that the correlation
functions of the PGWs in real space are well behaved. 
We conclude with a discussion on the different directions in which the 
results we have obtained need to be extended.}
\maketitle


\section{Introduction}

Inflation provides an attractive mechanism to explain the origin of the
primordial perturbations (see, for example, the reviews~\cite{Mukhanov:1990me,
Martin:2003bt,Martin:2004um,Bassett:2005xm,Sriramkumar:2009kg,Baumann:2008bn,
Baumann:2009ds,Sriramkumar:2012mik,Linde:2014nna,Martin:2015dha}).
The scalar perturbations generated during inflation are primarily responsible
for the anisotropies in the cosmic microwave background~(CMB)~\cite{Planck:2015sxf,
Planck:2018jri} and the large-scale structure~\cite{eBOSS:2020yzd,eBOSS:2021pff,
DESI:2024mwx,DESI:2025zgx}.
During inflation, in addition to the scalar perturbations, the tensor 
perturbations or gravitational waves (GWs) are also generated from 
the quantum vacuum (for the initial discussions, see, for example, 
Refs.~\cite{Grishchuk:1974ny,Starobinsky:1979ty}; for recent reviews 
on the generation of primary and secondary GWs, see, for instance, 
Refs.~\cite{Guzzetti:2016mkm,Caprini:2018mtu,Domenech:2021ztg,
Roshan:2024qnv,Giovannini:2024vei}).
Since the GWs interact weakly, the primary GWs (PGWs), after being generated 
from the quantum vacuum during inflation, propagate freely thereafter.
Hence, they carry imprints of the different phases of the universe, such as the 
epochs of reheating, radiation and matter domination as well as any additional 
epochs that may have arisen in between (in this context, see, for instance, 
Refs.~\cite{Bernal:2019lpc,Bernal:2020ywq,Haque:2021dha}).
Therefore, observations of the power spectrum (PS) of the PGWs can, in 
principle, help us reconstruct the history of the early universe.

There are two important pieces of information related to inflation that we 
hope to glean from the direct and indirect observations of PGWs.
They are the energy scale at which inflation occurred and the duration of 
inflation.
We mentioned above that the scalar perturbations are primarily responsible 
for the anisotropies in the CMB.
The tensor perturbations too leave specific signatures on the CMB (for a 
detailed discussion on the topic, see the review~\cite{Kamionkowski:2015yta}).
In particular, they are solely responsible (at least on large scales or,
equivalently, at the lower multipoles) for the so-called B-mode polarization
of the CMB\footnote{On smaller scales or higher multipoles, gravitational 
lensing of the scalar perturbations can also induce the B-mode polarization
in the CMB~\cite{Planck:2015aco,Namikawa:2021gyh}.}.
The observation of the B-mode polarization in the CMB on large scales can 
immediately point to the energy scale of inflation. 
However, no observations of the B-mode polarization have yet been made over
the lower multipoles of the CMB.
As a result, there exists only an upper bound on the contributions of the 
tensor perturbations or primordial GWs to the anisotropies in the CMB (see
Refs.~\cite{Planck:2015sxf,Planck:2018jri}; for the prospects of detection
of the B-mode in the near future, see Ref.~\cite{Paoletti:2022anb}).
The latest CMB observations constrain the dimensionless tensor-to-scalar 
ratio~$r$ to be $r<0.036$~\cite{BICEP:2021xfz}.
This roughly translates to the upper bound on the Hubble scale during
inflation, say, $\HI$, to be $\HI/\Mpl\simeq 10^{-5}$, where $\Mpl$
denotes the Planck mass.

The constraints from the CMB on the primordial GWs are indirect.
We should mention that the frequency, say, $f$, of the primordial GWs
is related to the wave number~$k$ through the relation
\begin{equation}
f \equiv \frac{k/a_0}{2 \pi} 
= 1.55\times10^{-15} \l(\f{k/a_0}{1\, \mathrm{Mpc}^{-1}}\r)\,\mathrm{Hz},
\label{eq:f}
\end{equation}
where $a_0$ is the scale factor of the 
Friedmann-Lema\^itre-Robertson-Walker~(FLRW) universe today.
As convenient, we shall describe the PS of GWs either in terms of
wave numbers or frequencies.
Note that the CMB constrains the primordial GWs only over large scales 
corresponding to wave numbers $10^{-4} \lesssim k/a_0\lesssim 1\,\rm{Mpc}^{-1}$ 
or, equivalently, over very small frequencies in the range of $10^{-19} 
\lesssim f\lesssim 10^{-15}\,\rm{Hz}$.
During the past decade, the direct detection of GWs from astrophysical 
sources, such as coalescing compact binary black holes and neutron stars, 
by the Ligo-Virgo-Kagra~(LVK) collaboration has opened a new window to the
universe (in this context, see, for instance, Refs.~\cite{LIGOScientific:2016aoc,
LIGOScientific:2016dsl,LIGOScientific:2016wyt,LIGOScientific:2017bnn,
LIGOScientific:2017ycc,LIGOScientific:2017vox}).
The GWs generated in the primordial universe are expected to be stochastic 
in nature.
The detection of GWs from astrophysical sources by the LVK collaboration has
led to the expectation that the direct detection of a stochastic GW 
background~(SGWB) may be around the corner.
But, despite a few rounds of observations, the LVK collaboration has not yet
detected a SGWB over the range of frequencies (say, $1 \lesssim f \lesssim 
10^2\, \mathrm{Hz}$) that its detectors are sensitive 
to~\cite{LIGOScientific:2016jlg,LIGOScientific:2019vic,KAGRA:2021kbb}.
However, about two years ago, the Pulsar Timing Arrays~(PTAs)---such as 
NANOGrav~\cite{NANOGrav:2023gor,NANOGrav:2023hde}, EPTA (including the 
data from InPTA)~\cite{EPTA:2023sfo,EPTA:2023fyk}, PPTA~\cite{Zic:2023gta,
Reardon:2023gzh}, and CPTA~\cite{Xu:2023wog}---reported the detection of 
a SGWB in the range of frequencies $10^{-9} \lesssim f \lesssim 10^{-6}\, 
\mathrm{Hz}$\footnote{We should mention that the detection of the SGWB by 
the PTAs also constitute an indirect detection~(for a discussion in this 
context, see, for instance, Ref.~\cite{Yokoyama:2021hsa}).}.
Many ongoing and forthcoming GW observatories, such as the Square
Kilometer Array (SKA) ($10^{-9} \lesssim f \lesssim 10^{-6}\, 
\mathrm{Hz}$)~\cite{Janssen:2014dka}, 
Laser Interferometer Space Antenna (LISA) ($10^{-4} \lesssim f \lesssim 0.1\, 
\rm{Hz}$)~\cite{Bartolo:2016ami,
LISACosmologyWorkingGroup:2022jok}, 
TianQin and Taiji ($10^{-4} \lesssim f \lesssim 0.1\, 
\rm{Hz}$)~\cite{Hu:2017mde,TianQin:2020hid,Gong:2021gvw},
Big Bang Observer (BBO) ($0.1 \lesssim f \lesssim 1\, 
\rm{Hz}$)~\cite{Crowder:2005nr,Corbin:2005ny,Harry:2006fi,Baker:2019pnp}, 
Decihertz Interferometer Gravitational wave Observatory (DECIGO)
($0.1 \lesssim f \lesssim 10\, \rm{Hz}$)~\cite{Kawamura:2019jqt,
Kawamura:2020pcg}, 
Matter-wave Atomic Gradiometer Interferometric Sensor (MAGIS)
($0.1 \lesssim f \lesssim 10\, \rm{Hz}$)~\cite{Coleman:2018ozp}, 
Cosmic Explorer (CE) ($1 \lesssim f \lesssim 10^2\, 
\rm{Hz}$)~\cite{Evans:2021gyd,Evans:2023euw}
and Einstein Telescope (ET) ($1 \lesssim f \lesssim 10^3\, 
\rm{Hz}$)~\cite{Sathyaprakash:2012jk,Branchesi:2023mws,Abac:2025saz},
are expected to operate over the indicated range of frequencies. 
At high frequencies ($10^3 \lesssim f \lesssim 10^{20}\,\rm{Hz}$), we 
already have some upper bounds on the characteristic strain of GWs 
from Bulk Acoustic Wave devices~(BAW)~\cite{Goryachev:2013fcc, 
Galliou:2013fvz}, Absolute Radiometer for Cosmology, Astrophysics 
and Diffuse Emission~(ARCADE)~\cite{Fixsen:2009xn}, 
Optical Search for QED Vacuum Birefringence, Axions and Photon 
Regeneration~(OSQAR)~\cite{OSQAR:2015qdv}, and the
CERN Axion Solar Telescope~(CAST)~\cite{CAST:2004gzq}.
In addition, a set of detectors have been proposed to observe GWs 
at these high frequencies (for broad discussions in this context, see, 
Refs.~\cite{Tong:2008rz,Domcke:2022rgu,Bringmann:2023gba,
Kahn:2023mrj,Kanno:2023whr}).
Specific detectors include Axion Dark Matter eXperiment (ADMX)~\cite{ADMX:2021nhd},
Superconducting Quantum Materials and Systems Center (SQMS)~\cite{Berlin:2021txa},
International Axion Observatory-heterodyne+single photon 
detectors (IAXO-HET+SPD)~\cite{Ringwald:2020ist},
electromagnetic Gaussian beams (GB)~\cite{Li:2003tv},
Joint Undertaking on the Research for Axion-like 
particles~(JURA)~\cite{Beacham:2019nyx}, 
and International Axion Observatory~(IAXO)~\cite{Ruz:2018omp}.
During the coming decade or two, these observatories are anticipated 
to provide further constraints on the amplitude and shape of the PS of 
primordial GWs (in this context, also see Ref.~\cite{Lasky:2015lej}).

In this work, we shall be interested in understanding the behavior of the 
PS of PGWs over small scales or, equivalently, at high frequencies.
Specifically, we shall be interested in wave numbers that never leave the
Hubble radius during the course of cosmological evolution.
Let~$\ke$ denote the wave number that leaves the Hubble radius at the end
of inflation.
We shall now briefly discuss the expected behavior of the PS of PGWs over 
small scales such that~$k\gtrsim \ke$. 
Since these wave numbers remain inside the Hubble radius from the early 
stages of inflation until today, the Fourier mode functions describing the
PGWs associated with these wave numbers will always behave in the same manner 
(barring an overall time-dependent function that depends inversely on the
scale factor, say, $a$) as the mode functions in Minkowski spacetime. 
Then, it can be immediately shown that, over such wave numbers, at any 
time, the PS of PGWs will behave as~$k^2$.
Moreover, as we shall illustrate in due course, when the PS of PGWs is 
evaluated today, the $k^2$ rise will suggest that their strengths at high 
frequencies can become comparable to the sensitivities of the ongoing and 
forthcoming GW observatories (in this context, see Fig.~\ref{fig:PT-hf}).
It has been argued that the PS of PGWs has to be regularized,
which can truncate the~$k^2$ rise in the quantity at large
wave numbers (see Ref.~\cite{Wang:2015zfa}; for related discussions on 
the scalar power spectrum, see Refs.~\cite{Parker:2007ni,Agullo:2008ka,
Agullo:2009vq,Agullo:2009zi,Urakawa:2009xaa,delRio:2014aua,Pla:2024xsv}; 
in this context, also see Ref.~\cite{Finelli:2007fr}).
As we shall discuss, in the context of PGWs, the process of regularization 
is essential to avoid the $k^2$ rise in the PS on small scales.

There is another aspect of the background cosmological dynamics that is 
expected to modify the PS of PGWs around the wave number~$\ke$.
It has been suggested in the literature (we should say with inadequate
proof) that the PS of GWs will exhibit an exponential suppression over 
wave numbers $k \gtrsim \ke$ (see, for instance, Refs.~\cite{Kuroyanagi:2008ye,
Ringwald:2022xif,Ito:2020neq,Giovannini:2024vei}).
Barring an occasional exception, the PGWs are typically evolved from their 
origin during inflation to the present epoch assuming that the transitions 
from one epoch to another (say, from inflation to reheating or the epoch of
radiation domination) are instantaneous or abrupt.
Realistically, these transitions should be infinitely smooth and recent 
arguments in the literature suggest that smoothing the transitions can 
lead to a sharp cut-off in the PS of PGWs for~$k\gtrsim \ke$ (in this 
context, see Refs.~\cite{Pla:2024xsv,Pi:2024kpw}).

Our aim in this work is to carefully examine the effects due to 
regularization as well as smoothing of the transitions on the PS 
of PGWs.
To understand the behavior around~$\ke$, we use the method of adiabatic 
regularization to calculate the observed PS of PGWs today.
We also  smooth the transition from inflation to the epoch of radiation 
domination in a specific manner and explicitly evaluate the corresponding
modifications to the PS of PGWs.
We show that, in the case of an instantaneous transition, regularization 
leads to a PS which oscillates with a fixed amplitude about a vanishing
mean value over~$k\gtrsim \ke$.
We further illustrate that, at such high wave numbers, the smoother
transition leads to a power-law suppression of the amplitude of the 
oscillations in the regularized PS of PGWs.

This paper is organized as follows.
In the following section, we shall describe some of the essential properties
of PGWs.
We shall describe the equation of motion satisfied by the PGWs, their quantization,
and introduce the PS associated with them.
We shall also discuss the generic form of the PS expected in the post-inflationary
universe.
In Sec.~\ref{sec:ds-rd-it}, we shall discuss the popular case of instantaneous 
transitions from de Sitter inflation to the epochs of radiation and matter 
domination, and evaluate the resulting PS of PGWs.
In Sec.~\ref{sec:r-ps-sed-gws}, we shall initially discuss the need for the
regularization of the PS of PGWs.
We shall also describe the manner in which the PS of PGWs can be regularized 
using the method of adiabatic subtraction.
Moreover, in these sections, we shall clarify a few points related to the 
amplitude and shape of the actual as well as the regularized PS of PGWs.
In particular, we shall show that, the regularized PS of PGWs, evaluated 
during the epochs of radiation and matter domination, oscillates with a 
constant amplitude about zero over $k \gtrsim \ke$.
In Sec.~\ref{sec:i-rd-st-b}, we shall smooth the transition from power-law 
inflation (in fact, we shall smooth the `effective potential' that governs 
the dynamics of PGWs) to the epoch of radiation domination with the aid of 
a linear function and determine the scale factor during the transition.
In Sec.~\ref{sec:i-rd-st-p}, we shall evolve the mode functions of the PGWs
across the smooth linear transition and illustrate that, during radiation
domination, over $k \gtrsim \ke$, 
the regularized PS of PGWs contains oscillations about a vanishing mean and 
that the amplitude of the oscillations exhibits a power-law suppression.
Finally, in Sec.~\ref{sec:so}, after a quick summary of the results we have
obtained, we shall outline a few related issues that need to be examined 
further.

At this point in our discussion, we should make a few clarifying remarks
concerning the conventions and notations that we shall work with.
We shall work with natural units such that $\hbar=c=1$ and set the reduced 
Planck mass to be $\Mpl=\l(8 \pi G\r)^{-1/2}$.
We shall adopt the signature of the metric to be~$(-,+,+,+)$.
Note that Latin indices shall represent the spatial coordinates, except 
for~$k$ which shall be reserved for denoting the wave number. 
We shall assume the background to be the spatially flat, FLRW universe 
described by the following line-element:
\begin{equation}
\d s^2=-\d t^2+a^2(t) \d {\bm x}^2
=a^2(\eta) \l(-\d \eta^2+\d {\bm x}^2\r),\label{eq:flrw-le}
\end{equation}
where~$t$ and~$\eta$ denote the cosmic and conformal time coordinates, 
and~$a$ denotes the scale factor.
The overdots and overprimes shall denote differentiation with respect to 
the cosmic and conformal times, respectively.


\section{Essentials of PGWs}


In this section, we shall describe aspects of PGWs that are 
essential for our discussion.
We shall first introduce the equation of motion governing PGWs.
We shall also describe the quantization of PGWs and discuss the 
associated PS. 


\subsection{Equation of motion of PGWs and quantization}\label{sec:quan}

Let $h_{ij}$ denote the tensor perturbations characterizing the GWs 
in a FLRW universe.
When these tensor perturbations are taken into account, the
line-element~\eqref{eq:flrw-le} describing the spatially 
flat, FLRW universe is modified to be (see, for instance, 
Refs.~\cite{Mukhanov:1990me,Weinberg:2008zzc})
\begin{align}
\d s^2
&= -\d t^2+a^2(t) \l[\delta_{ij}+h_{ij}(t,{\bm x})\r]
\d x^i \d x^j\nn\\
&=a^2(\eta) \l\{ -\d \eta^2+\l[\delta_{ij}+h_{ij}(\eta,{\bm x})\r]
\d x^i \d x^j\r\}.\label{eq:flrw-wgws}
\end{align}
Since the tensor perturbations $h_{ij}$ are transverse and traceless, they
satisfy the conditions 
\begin{equation}
\delta^{ij}\pa_ih_{jl} =0,\; \delta^{ij}h_{ij}=0.\label{eq:ttc-rs} 
\end{equation}
Also, in our discussion, we shall assume that no sources with an anisotropic
stress-energy tensor are present.
Under these conditions, the Einstein's equations at the first order in the 
perturbations lead to the following equation of motion for the PGWs (for a 
derivation, see App.~\ref{app:eom-gws}):
\begin{equation}
h_{ij}'' + 2 {\cal H} h_{ij}'-\pa^2h_{ij}=0,\label{eq:eom-h}
\end{equation}
where ${\cal H}=a'/a=a H$ is the conformal Hubble parameter, with $H=\dot{a}/a$
being the Hubble parameter, and $\pa^{2}=\delta^{ij}\pa_i\pa_j$.

As we mentioned earlier, the PGWs originate from the quantum vacuum.
On quantization, the tensor perturbations $h_{ij}$ can be decomposed in 
terms of the rescaled mode functions, say, $\mu_k(\eta)$, 
as follows~\cite{Mukhanov:1990me,Martin:2003bt,Martin:2004um,Bassett:2005xm,
Sriramkumar:2009kg,Baumann:2008bn,Baumann:2009ds,Sriramkumar:2012mik,
Linde:2014nna,Martin:2015dha}:
\begin{equation}
\hat{h}_{ij}(\eta, {\bm x}) 
= \frac{\sqrt{2}}{\Mpl a(\eta)}\sum_{\lambda={(+,\times)}}
\int \f{\d^{3}{\vk}}{(2\pi)^{3/2}} 
\varepsilon^{\lambda}_{ij}(\hat{\bm n}) \l[\hat{a}^{\lambda}_{\bm k}
\mu_{k}(\eta) \mathrm{e}^{i{\bm k}\cdot{\bm x}}
+\hat{a}^{\lambda\dag}_{\vk} \mu ^{\ast}_{k}(\eta)
\mathrm{e}^{-i {\bm k}\cdot{\bm x}}\r],\label{eq:dh}
\end{equation}
where ${\bm k}=k\hat{\bm n}$ and $\hat{\bm n}$ is a unit vector.
The quantity $\varepsilon^{\lambda}_{ij}(\hat{\bm n})$ is real and
it represents the polarization tensor, with the index~$\lambda$ 
denoting the polarizations~$+$ and~$\times$ of the GWs.
The polarization tensors $\varepsilon_{ij}^+(\hat{\bm n})$ and 
$\varepsilon_{ij}^\times(\hat{\bm n})$ can be expressed in terms of 
the set of orthonormal unit vectors $(\hat{\bm e}_1(\hat{\bm n}),
\hat{\bm e}_2(\hat{\bm n}), \hat{\bm n})$ in the following 
manner~\cite{Maggiore:1999vm,Maggiore:2007ulw}:
\begin{subequations}\label{eq:ptd}
\begin{align}
\varepsilon_{ij}^+(\hat{\bm n})
&=\hat{\bm e}_{1i}(\hat{\bm n}) \hat{\bm e}_{1j}(\hat{\bm n})
-\hat{{\bm e}}_{2i}(\hat{\bm n}) \hat{{\bm e}}_{2j}(\hat{\bm n}),\\
\varepsilon_{ij}^\times(\hat{\bm n})
&=\hat{\bm e}_{1i}(\vk) \hat{{\bm e}}_{2j}(\hat{\bm n})
+\hat{{\bm e}}_{2i}(\vk) \hat{\bm e}_{1j}(\hat{\bm n}).
\end{align}
\end{subequations}
The polarization tensor $\varepsilon_{ij}^\lambda(\hat{\bm n})$ obeys 
the transverse and traceless conditions [cf. Eqs.~\eqref{eq:ttc-rs}], viz.
\begin{equation}
\delta^{ij}k_{i}\varepsilon_{jl}^\lambda(\hat{\bm n})=0,\; 
\delta^{ij} \varepsilon_{ij}^{\lambda}(\hat{\bm n})=0.\label{eq:ttc}
\end{equation}
Also, the orthonormal nature of the unit vectors $\hat{\bm e}_1(\hat{\bm n})$ 
and $\hat{\bm e}_2(\hat{\bm n})$ lead to the normalization condition
\begin{equation}
\delta^{im} \delta^{jn}
\varepsilon_{mn}^{\lambda}(\hat{\bm n})
\varepsilon_{ij}^{\lambda'}(\hat{\bm n})=2\, \delta^{\lambda\lambda'}.
\label{eq:n}
\end{equation}
In the Fourier decomposition~\eqref{eq:dh} of the tensor perturbations, 
the operators $\hat{a}_{\bm k}^{\lambda}$ and $\hat{a}^{\lambda\dagger}_{\bm k}$ 
denote the annihilation and creation operators corresponding to the tensor 
modes associated with the wave vector~${\bm k}$ and polarization~$\lambda$.
The annihilation and creation operators obey the standard commutation
relations, viz.
\begin{equation}
\left[\hat{a}^{\lambda}_{\bm k},\hat{a}_{\bm k'}^{\lambda'}\right]
=\left[\hat{a}^{\lambda\dag}_{\bm k},
\hat{a}_{\bm k'}^{\lambda'\dagger}\right]=0,\quad
\left[\hat{a}^{\lambda}_{\bm k},\hat{a}_{\bm k'}^{\lambda'\dagger}\right]
=\delta^{(3)}\left({\bm k}-{\bm k}'\right) \delta^{\lambda\lambda'}. 
\end{equation}
In the absence of sources with anisotropic stresses, the equation of 
motion~\eqref{eq:eom-h} for $h_{ij}$ leads to the following differential
equation for the rescaled mode function $\mu _{k}(\eta)$~\cite{Mukhanov:1990me,
Martin:2003bt,Martin:2004um,Bassett:2005xm,Sriramkumar:2009kg,
Baumann:2008bn,Baumann:2009ds,Sriramkumar:2012mik,Linde:2014nna,
Martin:2015dha}:
\begin{equation}
\mu_{k}''+ \l(k^2 - \f{a''}{a}\r) \mu_{k} = 0.\label{eq:mse}
\end{equation}
Clearly, in terms of the variable $\mu_{k}(\eta)$, the tensor mode
functions correspond to solutions of a time-dependent oscillator whose
frequency is determined by the behavior of the scale factor.


\subsection{PS of PGWs}

In this work, we shall focus on the PS of the primordial tensor perturbations,
which is one of the quantities of observational interest.
The PS is a measure of the two-point correlation of the tensor perturbations 
in Fourier space.
The PS of PGWs at the conformal time $\eta$, say, $\pt(k,\eta)$, 
is defined through the relation~\cite{Mukhanov:1990me,Martin:2003bt,
Martin:2004um,Bassett:2005xm,Sriramkumar:2009kg,Baumann:2008bn,
Baumann:2009ds,Sriramkumar:2012mik,Linde:2014nna,Martin:2015dha}
\begin{equation}
\delta^{im}\delta^{jn}
\left \langle 0 \left \vert \hat{h}_{mn}(\eta,{\bm x})
\hat{h}_{ij}(\eta,{\bm x})\right \vert 0 \right \rangle 
=\int _0^{\infty} \frac{\mathrm{d}k}{k} \pt(k,\eta),\label{eq:twopoint}
\end{equation}
where, as is usually done, we have assumed that the expectation value on 
the left hand side is evaluated in the quantum vacuum $\vert 0\rangle$,
which is defined as ${\hat a}_{\bm k}^{\lambda}\vert 0\rangle=0$ for all 
${\bm k}$ and $\lambda$.
On utilizing the decomposition~\eqref{eq:dh} of the quantized tensor
perturbations~$\hat{h}_{ij}$ in terms of the rescaled mode 
functions~$\mu_{k}(\eta)$, the PS of PGWs at any time can be expressed as
\begin{equation}
\pt(k,\eta)=\f{8}{\Mpl^2} \f{k^3}{2\pi^2}
\f{\vert \mu_{k}(\eta)\vert^2}{a^2(\eta)}.\label{eq:tps}
\end{equation}


\subsection{PGWs in the post-inflationary universe}

Evidently, to arrive at the PS of PGWs {\it today},\/ we need to evolve the 
tensor perturbations in the post-inflationary universe. 
After inflation, the different epochs of the universe can be described by 
an equation-of-state parameter, say, $w$, that is a constant and is greater 
than~$-1/3$.
In such situations, the scale factor describing the spatially flat,
FLRW universe can be written as
\begin{align}
\label{eq:scalefactorpl}
a(\eta)=a_w\left(\eta-\eta_w\right)^{2/(1+3w)}.
\end{align}
The two constants $a_w$ and $\eta_w$ are to be determined by 
matching the scale factor and its time derivative at the onset of the 
epoch, with those of the previous epoch.
For the above $a(\eta)$, we have 
\begin{equation}
\f{a''}{a}=\f{2(1-3w)}{(1+3w)^2(\eta-\eta_w)^2}.
\end{equation}
In such a case, the general solution to the rescaled mode function
$\mu_{k}(\eta)$ that describes the PGWs [cf. Eq.~\eqref{eq:mse}] can be
written as
\begin{align}
\label{eq:modefunctionrad}
\mu_{k}(\eta)=\alpha_{k}^{w} m_{k}(\eta)
+\beta_{k}^{w} n_{k}(\eta),
\end{align}
where $(\alpha_k^{w},\beta_k^{w})$ are the so-called Bogoliubov
coefficients.
As in the case of the constants describing the scale 
factor~\eqref{eq:scalefactorpl}, 
the coefficients $(\alpha_k^{w},\beta_k^{w})$ are to be determined 
by the boundary conditions on the rescaled mode function~$\mu_k(\eta)$ 
imposed at the beginning of the epoch.
The functions $m_{k}(\eta)$ and $n_{k}(\eta)$ are given by
\begin{subequations}\label{eq:gs}
\begin{align}
m_{k}(\eta) &=x^{1/2}\mathrm{H}_{\nu(w)}^{(2)}(x),\\
n_{k}(\eta) &=x^{1/2}\mathrm{H}_{\nu(w)}^{(1)}(x)=m_{k}^*(\eta),
\end{align}
\end{subequations}
where $x=k(\eta-\eta_w)$ and $\nu(w)=3(1-w)/[2(1+3w)]$, while 
$\mathrm{H}_\nu^{(1)}(x)$ and $\mathrm{H}_\nu^{(2)}(x)$ are the 
Hankel functions of types one and two, respectively, and of 
order~$\nu$.

During the era where the universe is dominated by radiation, $w=1/3$ 
and $\nu=1/2$. 
In such a case, $a''/a=0$, and the function $m_{k}(\eta)$ has the 
following simple form~\cite{Gradshteyn:2007}:
\begin{equation}
m_{k}(\eta)
=i \sqrt{\f{2}{\pi}}\e^{-ix}=n_k^{\ast}(\eta).\label{eq:mn-rd}
\end{equation}
On substituting the corresponding $\mu_{k}(\eta)$ in the
expression~\eqref{eq:tps}, we obtain the PS of PGWs during the 
radiation-dominated epoch to be
\begin{align}
\pt(k,\eta)=\frac{8k^3}{\pi^3 \Mpl^2 a^2}
\l[\vert \alpha_k^\mathrm{r}\vert^2 +\vert \beta_k^\mathrm{r}\vert^2
-2 \Re(\alpha_k^\mathrm{r}\beta_k^\mathrm{r*}) \cos(2x)
-2 \Im (\alpha_k^\mathrm{r}\beta_k^{\mathrm{r}*}) \sin(2x)\r].
\label{eq:ptrad}
\end{align}
This formula is general and it has a distinct advantage.
To predict $\pt(k,\eta)$, all we need to do is determine the form of the
coefficients $(\alpha_k^\mathrm{r},\beta_k^\mathrm{r})$ and substitute 
them in the above expression.

As we indicated above, the coefficients $(\alpha_k^\mathrm{r},
\beta_k^\mathrm{r})$ are determined by the boundary conditions
imposed on the rescaled mode function~$\mu_k(\eta)$ at the 
beginning of the epoch of radiation domination.
In other words, the coefficients depend on the history of the 
universe prior to the era of radiation domination.
Let us now consider the special case wherein the universe, before 
the radiation-dominated era, underwent a phase of cosmic inflation. 
During inflation, the scale factor is typically of the form
\begin{equation}
a(\eta)=\ell_0(-\eta)^q,\label{eq:a-pli}
\end{equation}
where $q < -1$ in power-law inflation and $q=-1-\epsilon_1$ (with 
$\epsilon_1 = -\dot{H}/H^2\ll 1$ being the first slow roll parameter) 
in slow roll inflation.
Also, $\ell_0$ determines the scale of inflation.
Moreover, note that exact de Sitter inflation corresponds to $q=-1$.
Further, in the inflationary scenario, it is common to impose the 
so-called Bunch-Davies initial conditions on the perturbations when 
the modes are in the sub-Hubble domain during the early 
stages~\cite{Bunch:1977sq,Bunch:1978yq,Bunch:1978yw}.
This corresponds to assuming that the tensor perturbations are in 
the vacuum state. 
Under such assumptions, the solution for the rescaled mode 
function~$\mu_k(\eta)$ during inflation can be written as
\begin{align}
\label{eq:infmodefunction}
\mu_{k}(\eta)=A_{k} f_{k}(y),
\end{align}
where $y\equiv -k\eta>0$, and the function $f_k(y)$ is given by
\begin{equation}
f_k(\eta) =y^{1/2}\, \mathrm{H}_{\mu}^{(1)}(y),\label{eq:fk}
\end{equation}
with $\mu=1/2-q$.
The constant $A_k$ is given by
\begin{align}
A_{k} =\sqrt{\frac{\pi}{2}}\frac{1}{\sqrt{2k}}
\e^{-i\pi q/2+i\pi/2+ik\eta_\mathrm{i}}=\sqrt{\frac{\pi}{2}}
\frac{1}{\sqrt{2k}}\e^{i\pi \mu/2+i \pi/4+ik \eta_\mathrm{i}},
\end{align}
where $k\eta_{\mathrm{i}}$ is an arbitrary phase factor.
We should clarify that inflation could have occurred either immediately 
before the epoch of radiation domination or there could have been other 
phases between inflation and the radiation-dominated era, such as the
phase of reheating. 
Whatever the details of the scenario that one considers, the coefficients 
$(\alpha_k^\mathrm{r},\beta_k^\mathrm{r})$ will always be proportional 
to $A_{k}$, and they will satisfy the normalization condition 
\begin{align}  
\vert \alpha_k^{\mathrm{r}}\vert^2
-\vert \beta_k^{\mathrm{r}}\vert^2=\vert A_k\vert^2.\label{eq:wc}
\end{align}
Therefore, the power spectrum~\eqref{eq:ptrad} can be expressed as
\begin{align}
\label{eq:generalpt}
\pt(k,\eta) &=\frac{2}{\pi^2}\left(\frac{H_\mathrm{e}}{\Mpl}\right)^2
\left(\frac{a_\mathrm{e}}{a}\right)^2\,\bar{y}_{\e}^2
\Biggl[1+2\left \vert
\frac{\beta_k^\mathrm{r}}{A_{k}}\right\vert^2
-2\Re \left(\frac{\alpha_k^\mathrm{r}\beta_k^{\mathrm{r}*}}{\vert 
A_{k}\vert^2}\right)\cos(2x)\nn\\ 
&\quad-2\Im \left(\frac{\alpha_k^\mathrm{r}
\beta_k^{\mathrm{r}*}}{\vert A_{k}\vert^2}\right)\sin(2x)\Biggr]
\end{align}
and, in this expression, we have set 
\begin{align}
\bar{y}_\mathrm{e}=\frac{k}{\ke}\label{eq:yb}
\end{align}
where $\ke=\ae \He$, with $\ae$ and $\He$ denoting the scale factor
and the Hubble parameter at the end of inflation.
We should emphasize again that the above result is very general and, to
arrive at explicit expressions for the PS of PGWs, we need to evaluate 
the coefficients $(\alpha_k^{\mathrm{r}},\beta_k^{\mathrm{r}})$.
In order to calculate these coefficients, we need to specify the manner 
in which inflation and the epoch of radiation domination are related.
There is a point we wish to make before we proceed further.
Note that, if we assume that inflation is of the de Sitter form (so that 
$q=-1$), then the rescaled mode function [cf. Eqs.~\eqref{eq:infmodefunction}
and~\eqref{eq:fk}]
simplifies to be 
\begin{align}
\mu_k(\eta)=\f{1}{\sqrt{2k}} \l(1-\f{i}{k\eta}\r) \e^{-ik\eta}.
\end{align}
In such a case, at the end of inflation, say, at the conformal time~$\ee$, 
the PS of PGWs can be expressed as
\begin{align}
\pt(k,\ee)=\f{2 \He^2}{\pi^2\Mpl^2}\l(1+y_{\e}^2\r),
\end{align}
where $y_{\e}=-k\ee=k/\ke$.
This spectrum is scale invariant over $y_{\e} \ll 1$, whereas it behaves 
as~$y_{\e}^2$ over~$y_{\e} \gg 1$.
We shall see that, over $y_{\e}\gg 1$, the PS continues to behave as $y_{\e}^2$
even when we evolve the modes post-inflation.


\section{Instantaneous transition from de Sitter inflation to 
radiation and matter domination}\label{sec:ds-rd-it}

In this section, we shall consider the simple case of instantaneous
transition from de Sitter inflation to the epoch of radiation domination
and calculate the coefficients~$(\alpha_k^{\mathrm{r}},\beta_k^{\mathrm{r}})$.
Thereafter, assuming instantaneous transition from radiation to matter 
domination, we shall further evolve the mode functions describing the PGWs
through the epoch of matter domination.
We shall also take the opportunity to highlight a few points related to 
the resulting PS of PGWs evaluated during the epochs of radiation and 
matter domination.


\subsection{PS during radiation domination}

We shall now discuss the simplest of scenarios which allows us to calculate 
the coefficients~$(\alpha_k^{\mathrm{r}},\beta_k^{\mathrm{r}})$ easily. 
We shall consider the popular model wherein de Sitter inflation ends at the 
conformal time~$\ee$, and is immediately followed by the epoch of radiation 
domination.
In this scenario, the transition between inflation and the radiation-dominated 
epoch is sharp and reheating is instantaneous. 
Recall that, during de Sitter inflation and radiation domination, the scale
factors are given by [cf. Eqs.~\eqref{eq:a-pli} and~\eqref{eq:scalefactorpl}] 
\begin{subequations}
\begin{align}
a(\eta)&=-\f{\ell_0}{\eta},\label{eq:a-dsi}\\
a(\eta)&=a_\mathrm{r}(\eta-\eta_\mathrm{r}).\label{eq:a-rd}
\end{align}
\end{subequations}
During de Sitter inflation, the conformal Hubble parameter is given by
${\cal H}=-1/\eta$, while during the radiation dominated era, we have
${\cal H}=1/(\eta-\eta_\mathrm{r})$.
Then, the continuity of ${\cal H}$ implies that $\eta_\mathrm{r}=2
\eta_\mathrm{e}$, whereas the continuity of the scale factor leads to
$a_\mathrm{r}=\ell_0/\eta_\mathrm{e}^2$. 
In such a situation, the coefficients~$(\alpha_k^{\mathrm{r}},
\beta_k^{\mathrm{r}})$ are arrived at by matching the rescaled mode 
function~$\mu_k(\eta)$ and its time derivative at the time $\ee$, i.e. at the
end of inflation. 
We find that the coefficients~$(\alpha_k^{\mathrm{r}},
\beta_k^{\mathrm{r}})$ can be written as
\begin{subequations}
\begin{align}
\alpha_k^{\mathrm{r}}
&=\frac{A_{k}}{W_k^\mathrm{r}}\left[f_{k}(y_\mathrm{e})
n_{k}'(x_\mathrm{e})
-f_{k}'(y_\mathrm{e})
n_{k}(x_\mathrm{e})\right],\\
\beta_k^{\mathrm{r}}
&=-\f{A_{k}}{W_k^{\mathrm{r}}}\left[f_{k}(y_\mathrm{e})
m_{k}'(x_\mathrm{e})
-f_{k} '(y_\mathrm{e})m_{k}(x_\mathrm{e})\right],
\end{align}
\end{subequations}
where $x_\mathrm{e}\equiv k(\eta_{\e}-\eta_\mathrm{r})
=-k\eta_{\e}\equiv y_{\e}=k/\ke=\bar{y}_{\e}$ and $\ke=-1/\ee$. 
The quantity $W_k^{\mathrm{r}}$ is the Wronskian defined as
\begin{align}
W_k^{\mathrm{r}}(\eta)\equiv m_k(\eta) n_k'(\eta)-m_k'(\eta)n_k(\eta).
\label{eq:wkr}
\end{align}
The equation of motion for $\mu_{k}(\eta)$ implies that the Wronskian is 
a constant and, using the forms of $m_k(\eta)$ and $n_k(\eta)$ in  the
radiation-dominated epoch [cf. Eq.~\eqref{eq:mn-rd}], it can be 
established that $W_k^\mathrm{r}=4ik/\pi$. 
Then, straightforward manipulations lead to
\begin{subequations}\label{eq:CDabrupt-o}
\begin{align}
\label{eq:CDabrupt}
\alpha_k^{\mathrm{r}}
&=-\frac{iA_k}{2y_\mathrm{e}^2}
(1-2iy_\mathrm{e}-2y_\mathrm{e}^2)
\e^{2iy_\mathrm{e}},\\
\beta_k^{\mathrm{r}}
&=-\frac{iA_k}{2y_\mathrm{e}^2},\label{eq:bk-abrupt}
\end{align}
\end{subequations}
and it can be easily verified that $\vert \alpha_k^{\mathrm{r}}\vert^2
-\vert\beta_k^{\mathrm{r}}\vert^2=\vert A_k\vert^2$, as required.
On substituting the coefficients $(\alpha_k^{\mathrm{r}},
\beta_k^{\mathrm{r}})$ in the expression~\eqref{eq:generalpt} for the
PS of PGWs during radiation domination, we obtain that 
\begin{align}
\pt(k,\eta)
&=\frac{2}{\pi^2}\left(\frac{H_\mathrm{e}}{\Mpl}\right)^2
\left(\frac{a_\mathrm{e}}{a}\right)^2\frac{1}{y_\mathrm{e}^2}
\biggl\{y_\mathrm{e}^4+\frac12-\frac12 \left(1-2y_\mathrm{e}^2\right)\cos
\left[2 (x-y_\mathrm{e})\right]\nn\\
&\quad+y_\mathrm{e}\sin \left[2(x-y_\mathrm{e})\right]\biggr\},
\label{eq:ps-ds-rd-it}
\end{align}
where, recall that, $x=k(\eta-\eta_\mathrm{r})=ay_\mathrm{e}/a_\mathrm{e}$.

The above PS is valid at any time during the epoch of radiation domination.
Let us now discuss in detail the shape of the PS when it is evaluated at 
the time of radiation-matter equality, say, $\eeq$. 
Note that $\pt(k,\eeq)$ is controlled by two quantities, $y_\mathrm{e}=k/
k_\mathrm{e}$, where $k_\mathrm{e}$ is the wave number that leaves the Hubble 
radius at the end of inflation, and $x_\mathrm{eq}=k/k_\mathrm{eq}$, where 
$k_\mathrm{eq}=a_\mathrm{eq}H_\mathrm{eq}$ is the wave number that re-enters
the Hubble radius at the time of radiation-matter equality.
For wave numbers that are on super-Hubble scales at the end of inflation
$y_\mathrm{e} \ll 1$, and in such a limit, we find that 
the PS given by Eq.~\eqref{eq:ps-ds-rd-it} reduces to
\begin{align}
\label{eq:ptlargescale}
\pt(k,\eta)\simeq \frac{2}{\pi^2}\left(\frac{H_\mathrm{e}}{\Mpl}\right)^2
\left(\frac{a_\mathrm{e}}{a}\right)^2\frac{1}{y_\mathrm{e}^2}
\sin^2(x).
\end{align}
In this domain, i.e. over $y_\mathrm{e} \ll 1$, let us focus on wave 
numbers that have remained in the super-Hubble domain in the subsequent 
radiation-dominated era. 
This corresponds to considering the limit $x\ll 1$. 
At the time of radiation-matter equality, such a limit corresponds to $x_\mathrm{eq}
\ll 1$ or, equivalently, $k\ll k_\mathrm{eq}$. 
In such a case, we obtain the PS to be
\begin{align}
\pt(k,\eta)=\frac{2}{\pi^2}\left(\frac{H_\mathrm{e}}{\Mpl}\right)^2
\left(\frac{a_\mathrm{e}}{a}\right)^2\frac{x^2}{y_\mathrm{e}^2}
=\frac{2}{\pi^2}\left(\frac{H_\mathrm{e}}{\Mpl}\right)^2,\label{eq:ps-si}
\end{align}
since $x=a y_\mathrm{e}/a_\mathrm{e}$. 
In other words, the PS of PGWs is scale-invariant in this regime. 
For modes that were in the super-Hubble domain at the end of inflation but 
have then re-entered the Hubble radius during the radiation-dominated epoch, 
we have~$x\gg 1$. 
At the time of radiation-matter equality, such a limit corresponds to 
$k_\mathrm{eq} \ll k < k_\mathrm{e}$. 
In such a case, the PS is still given by Eq.~\eqref{eq:ptlargescale} above.
It should be clear that, over this regime, the PS behaves as $k^{-2}$ with 
superimposed oscillations. 
Finally, over wave numbers that are in the sub-Hubble regime at the end of inflation, 
i.e. $y_\mathrm{e}\gg 1$ or $k\gg k_\mathrm{e}$, and, therefore, always remain in the
sub-Hubble domain subsequently, we find that the PS behaves as 
\begin{align}
\label{eq:ptsmallscale}
\pt(k,\eta)\simeq \frac{2}{\pi^2}\left(\frac{H_\mathrm{e}}{\Mpl}\right)^2
\left(\frac{a_\mathrm{e}}{a}\right)^2y_\mathrm{e}^2.
\end{align}
In other words, the PS of PGWs rises as~$k^2$ over wave numbers such that 
$k \gg \ke$.

\begin{figure}[!t]
\centering
\includegraphics[width=1.0\textwidth]{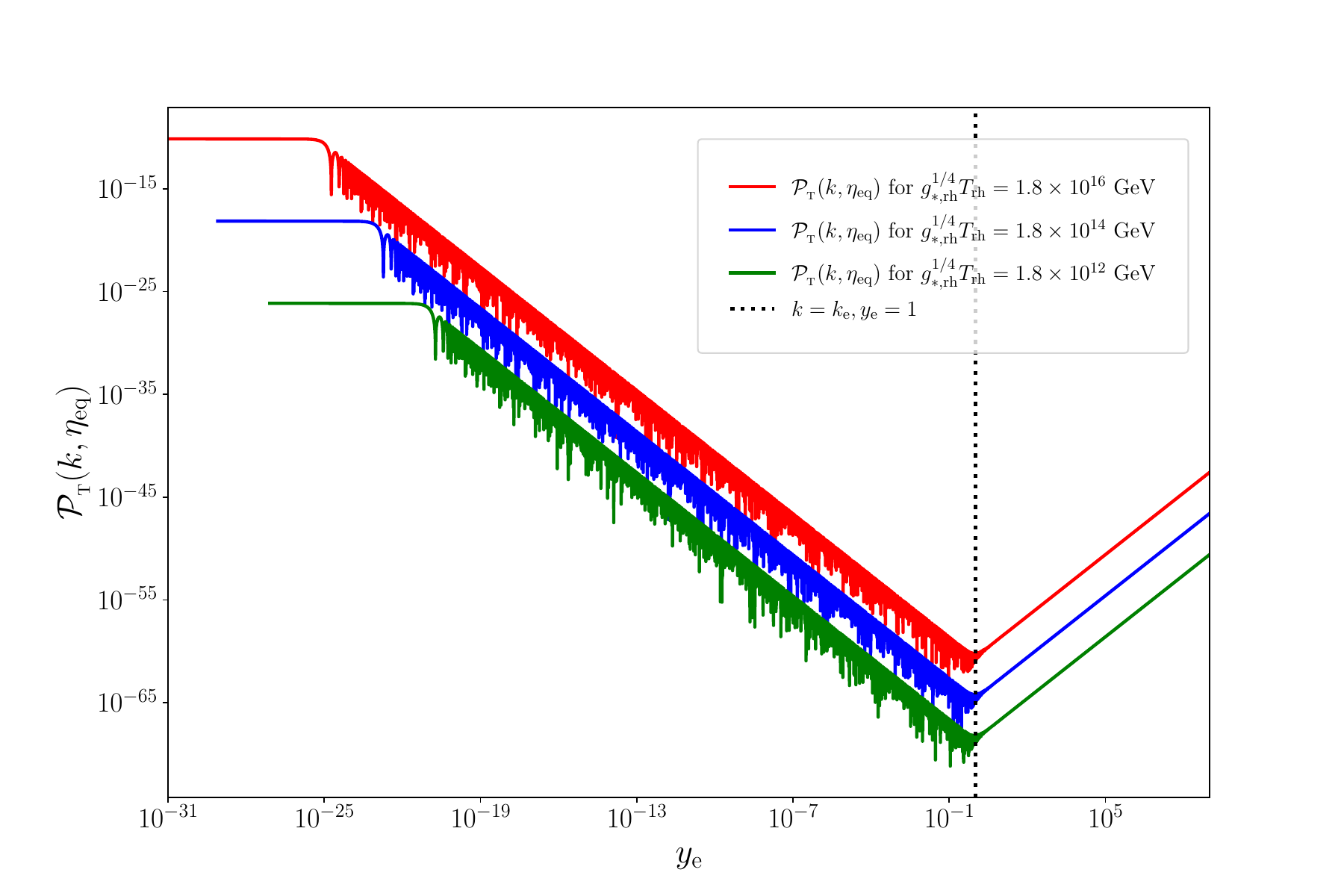}
\caption{The PS of PGWs $\pt(k,\eta)$ evaluated at the time of radiation-matter 
equality~$\eeq$ has been plotted for the case of instantaneous transition from 
de Sitter inflation to the epoch of radiation domination 
[cf. Eq.~\eqref{eq:ps-ds-rd-it}].
If we assume the standard $\Lambda$CDM model of cosmology, the PS depends 
{\it only}\/ on the reheating temperature~$\Tre$.
For $g_{\ast,\mathrm{rh}}^{1/4} \Tre =1.8\times10^{16}\,\mathrm{GeV}$, 
we find that the tensor-to-scalar ratio proves to be $r\simeq 0.034$ over 
large scales, which is roughly the current upper bound from the 
CMB~\cite{Planck:2018jri,BICEP:2021xfz}.
Keeping this constraint in mind, we have plotted the PS
for $g_{\ast,\mathrm{rh}}^{1/4} \Tre=1.8\times(10^{16}, 
10^{14},10^{12})\, \mathrm{GeV}$ (in red, blue and green).
Evidently, the PS is scale invariant over large scales such that $y_{\e}\ll 1$
and $x\ll 1$ [cf. Eq.~\eqref{eq:ps-si}].
These limits correspond to wave numbers $k\lesssim \keq$, which are outside the
Hubble radius at the time of radiation-matter equality.
Also, the fact that $\pt(k,\eeq)\propto \He^2 \propto \Tre^4$ on large scales
should be clear from the comparison of the scale-invariant amplitudes for the 
different~$\Tre$.
The PS behaves as $k^{-2}$ with superimposed oscillations over the intermediate 
scales such that $y_{\e}\ll 1$ and $x\gg 1$ [cf. Eq.~\eqref{eq:ptlargescale}].
These limits correspond to wave numbers $k$ such that $\keq \lesssim k \lesssim 
\ke$, which renter the Hubble radius during the radiation-dominated era.
The sharp dips in the PS over this domain correspond to wave numbers where the  
PS vanishes.
Importantly, note that, when $y_{\e} > 1$ (with $y_{\e}=1$
indicated by the vertical dotted line), $\pt(k,\eeq) \propto k^2$ 
[cf. Eq.~\eqref{eq:ptsmallscale}].
These correspond to wave numbers $k > \ke$ which never leave the Hubble radius.
We should add that PS is scale invariant until~$\keq/a_0 
\simeq 0.007\,\mathrm{Mpc}^{-1}$.
The change from the scale-invariant behavior to the~$k^{-2}$ behavior occurs
at different~$y_{\e}$ because~$\ke$ is different for different~$\Tre$.}
\label{fig:ps-sed-ds-rd-it1} 
\end{figure}

To understand the amplitude and the complete shape of the PS of PGWs as
well as to confirm the behavior in three different regimes identified
above, let us plot the quantity~$\pt(k,\eeq)$, i.e. the PS at the time of 
radiation-matter equality.
In order to do so, we need to choose values for the parameters involved. 
In the standard $\Lambda$CDM model of cosmology, it can be easily shown that 
\begin{align}
\left(\frac{H_\mathrm{e}}{\Mpl}\right)^2
\left(\frac{a_\mathrm{e}}{a_\mathrm{eq}}\right)^2
&=\frac{\pi}{\sqrt{90}} \l(\frac{H_0/h}{\Mpl}\r)
\l(\frac{g_{s,\mathrm{rh}}}{g_{s,\mathrm{eq}}}\r)^{-2/3}
\l(\frac{g_{\ast,\mathrm{rh}}}{g_{\ast,\mathrm{eq}}}\r)^{1/2}\,
\l(\frac{g_{\ast,\mathrm{rh}}^{1/4} T_{\mathrm{rh}}}{\Mpl}\r)^2
\frac{\l(\Omega_\mathrm{c} h^2+\Omega_\mathrm{b} h^2\r)^2}
{\left(\Omega_\gamma h^2+\Omega_\nu h^2\right)^{3/2}}\nn\\
&\simeq 3.725 \times 10^{-75} \l(\frac{g_{*,\mathrm{rh}}^{1/4} 
T_{\mathrm{rh}}}{10^9\, \mathrm{GeV}}\r)^2,
\end{align}
where $a_\mathrm{eq}$ denotes the scale factor at the time of 
radiation-matter equality.
Also, $T_{\mathrm{rh}}$ represents the reheating temperature and $H_0=2.133\,
h \times 10^{-42}\, \mathrm{GeV}$ is the value of the Hubble parameter 
{\it today}.\/
Moreover, $(g_{\ast,\mathrm{rh}},g_{\ast,\mathrm{eq}})$ and 
$(g_{s,\mathrm{rh}},g_{s,\mathrm{eq}})$ denote the number of 
effective relativistic degrees of freedom that contribute to 
the energy density and entropy of radiation at the time of 
reheating and radiation-matter equality, respectively.
The quantities $\Omega_\mathrm{c}\, h^2=0.12011$, $\Omega_\mathrm{b}\,h^2
=0.0223$, $\Omega_\gamma\,h^2=2.47\times 10^{-5}$, $\Omega _{\nu}\,h^2
=1.68 \times 10^{-5}$ are the present day values for the dimensionless
densities of cold dark matter, baryons, photons and neutrinos, 
respectively~\cite{Planck:2018vyg}. 
Further, to evaluate the argument of the trigonometric functions 
appearing in the PS, we need to calculate the quantity~$x$. 
Recall that
\begin{align}
x=k(\eta-\eta_\mathrm{r})=\frac{k}{aH}
=\frac{k}{a_\mathrm{e}H_\mathrm{e}}\frac{a_\mathrm{e}H_\mathrm{e}}{aH}
= y_\mathrm{e} \frac{a}{a_\mathrm{e}},\label{eq:expressionx}
\end{align}
where we have used the fact that $a H={\cal H}=(\eta-\eta_\mathrm{r})^{-1}$ in
the radiation-dominated era.
Therefore, in order to calculate~$x_\mathrm{eq}$, we need to evaluate the 
ratio, $a_\mathrm{eq}/a_\mathrm{e}$ which can be expressed as
\begin{align}\label{eq:aeq/ae}
\frac{a_\mathrm{eq}}{a_\mathrm{e}}
&=\left(\frac{\pi^2}{90}\right)^{1/4}
\left(\frac{H_0/h}{\Mpl}\right)^{-1/2}
\l(\frac{g_{s,\mathrm{rh}}}{g_{s,\mathrm{eq}}}\r)^{1/3}
\l(\frac{g_{\ast,\mathrm{eq}}}{g_{\ast,\mathrm{rh}}}\r)^{1/4}\,
\left(\frac{g_{\ast,\mathrm{rh}}^{1/4} T_{\mathrm{rh}}}{\Mpl}\right)
\frac{(\Omega _\gamma h^2+\Omega_\nu
h^2)^{3/4}}{(\Omega_\mathrm{c}h^2+\Omega_\mathrm{b} h^2)}\nn\\ 
&\simeq  9.188\times 10^{17}
\l(\f{g_{\ast,\mathrm{rh}}^{1/4} T_{\mathrm{rh}}}{10^9\, \mathrm{GeV}}\r).
\end{align}
Using these results, in Fig.~\ref{fig:ps-sed-ds-rd-it1}, we have plotted the 
PS of PGWs evaluated at the time of radiation-matter equality as a function 
of~$y_\mathrm{e}$. 
The figure clearly reflects the behavior of the PS in the three different limits 
we discussed above [as given by Eqs.~\eqref{eq:ptlargescale}, \eqref{eq:ps-si} 
and~\eqref{eq:ptsmallscale})].


\subsection{PS during matter domination}\label{sec:PS_MD}

Let us now arrive at the PS during the epoch of matter domination that 
follows the epoch of radiation domination. 
In order to do so, we need to evolve the rescaled mode functions $\mu_k(\eta)$
post the epoch of radiation-matter equality.
During matter domination, $w=0$ and hence the scale factor can be expressed 
as [cf. Eq.~\eqref{eq:scalefactorpl}] 
\begin{align}
a(\eta)=\am (\eta-\ema)^2,\label{eq:sf-md}
\end{align}
where $\am$ and $\ema$ are constants.
We shall assume that the transition from radiation to matter domination
is instantaneous.
On matching the above scale factor and its time derivative at~$\eeq$ 
with the values from the earlier epoch of radiation domination 
[cf. Eq.~\eqref{eq:a-rd}], we obtain that 
\begin{subequations}
\begin{align}
\am &= \f{\ar}{2(\eeq-\ema)},\\
\ema &= -\eeq+2 \er=-\eeq + 4\ee.
\end{align}
\end{subequations}

During the epoch of matter domination, the general solution for the rescaled 
mode function $\mu_k(\eta)$ can be expressed as [cf.~Eq.~\eqref{eq:gs}]
\begin{align}\label{eq:muk_matter}
\mu_k(\eta)=\alpha_k^\mathrm{m} p_k(\eta)+\beta_k^\mathrm{m} q_k(\eta),
\end{align}
where the functions $p_k(\eta)$ and $q_k(\eta)$ are given by
\begin{align}
p_k(\eta) & =z^{1/2} H_{3/2}^{(2)}(z)=\sqrt{\frac{2}{\pi}}
\e^{-iz}\left(-1+\frac{i}{z}\right)=q_k^\ast,\label{eq:pq-md}
\end{align}
with $z=k(\eta-\ema)$. 
On substituting the above mode function in the expression~\eqref{eq:tps}, we 
find that the PS of PGWs during the epoch of matter domination can be written
as
\begin{align}
\pt(k,\eta) 
&=\frac{8k^3}{\pi^3\Mpl^2 a^2}
\Biggl\{\left(\left \vert \alpha_k^\mathrm{m}\right \vert^2
+\left \vert \beta_k^\mathrm{m}\right\vert ^2\right) \left(1+\frac{1}{z^2}\right)
+2\left(1-\frac{1}{z^2}\right)
\biggl[\Re(\alpha_k^\mathrm{m}\beta_k^\mathrm{m*}) \cos(2z)\nn\\ 
&\quad+\Im (\alpha_k^\mathrm{m}\beta_k^{\mathrm{m}*}) \sin(2z)\biggr]
-\frac{4}{z}
\biggl[\Re(\alpha_k^\mathrm{m}\beta_k^\mathrm{m*}) \sin(2z)
-\Im (\alpha_k^\mathrm{m}\beta_k^{\mathrm{m}*}) \cos(2z)\biggr]
\Biggr\}.\label{eq:PTaccurate}
\end{align}
As in the case of the Bogoliubov coefficients~$(\alpha_k^{\mathrm{r}},
\beta_k^{\mathrm{r}})$ [cf. Eq.~\eqref{eq:wc}], it can be established that 
the coefficients $(\alpha_k^{\mathrm{m}},\beta_k^{\mathrm{m}})$ satisfy 
the condition
\begin{equation}  
\vert \alpha_k^{\mathrm{m}}\vert^2
-\vert \beta_k^{\mathrm{m}}\vert^2=\vert A_k\vert^2.
\end{equation}
Using this Wronskian condition, we can rewrite the PS above as
\begin{align}
\label{eq:generalptmatunregulated}
\pt(k,\eta) 
&=\f{2}{\pi^2}\l(\frac{H_\mathrm{e}}{\Mpl}\r)^2
\l(\frac{a_\mathrm{e}}{a}\r)^2 \bar{y}_\mathrm{e}^2
\Biggl\{\left(1+2\,\left \vert\f{\beta_k^\mathrm{m}}{A_{k}}\right\vert ^2\right) 
\left(1+\frac{1}{z^2}\right)\nn\\
&\quad +2\left(1-\frac{1}{z^2}\right)
\l[\Re\l(\f{\alpha_k^\mathrm{m}\beta_k^\mathrm{m*}}{\vert A_{k}\vert^2}\r) \cos(2z)
+\Im \l(\f{\alpha_k^\mathrm{m}\beta_k^{\mathrm{m}*}}{\vert A_{k} \vert ^2}\r)  
\sin(2z)\r]\nn\\
&\quad-\f{4}{z} \l[\Re\l(\f{\alpha_k^\mathrm{m}\beta_k^\mathrm{m*}}{\vert A_{k} 
\vert ^2}\r) \sin(2z)
-\Im \l(\f{\alpha_k^\mathrm{m}\beta_k^{\mathrm{m}*}}{\vert A_{k} \vert^2}\r) 
\cos(2z)\r]\Biggr\},
\end{align}
where, recall that, $\bar{y}_{\e}=k/(\ae \He)$.

As before, the Bogoliubov coefficients $(\alpha_k^\mathrm{m},\beta_k^\mathrm{m})$
are to be determined by matching the mode function~$\mu_k(\eta)$ and its time 
derivative with those of the previous epoch at the time of transition from one 
epoch to another.
On matching the mode function and its time derivative in the radiation and
matter-dominated epochs at the time of radiation-matter equality, we find that 
the coefficients~$(\alpha_k^{\mathrm{m}},\beta_k^{\mathrm{m}})$ can be written 
as
\begin{subequations}
\begin{align}
\alpha_k^{\mathrm{m}}
&=\frac{1}{W_k^\mathrm{m}}
\biggl\{-\l[\alpha_k^{\mathrm{r}} m_k'(\eeq)
+\beta_k^{\mathrm{r}} n_k'(\eeq)\r] q_{k}(\eeq)
+\l[\alpha_k^{\mathrm{r}} m_k(\eeq)
+\beta_k^{\mathrm{r}} n_k(\eeq)\r] q_{k}'(\eeq)\biggr\},\\
\beta_k^{\mathrm{m}}
&=\frac{1}{W_k^\mathrm{m}}
\biggl\{\l[\alpha_k^{\mathrm{r}} m_k'(\eeq)
+\beta_k^{\mathrm{r}} n_k'(\eeq)\r] p_{k}(\eeq)
-\l[\alpha_k^{\mathrm{r}} m_k(\eeq)
+\beta_k^{\mathrm{r}} n_k(\eeq)\r] p_{k}'(\eeq)\biggr\},
\end{align}
\end{subequations}
where $W_k^\mathrm{m}$ is the Wronskian defined as
\begin{align}
W_k^{\mathrm{m}}(\eta)\equiv m_k(\eta) n_k'(\eta)-m_k'(\eta)n_k(\eta).
\end{align}
As in the radiation-dominated epoch, the Wronskian is a constant, which
can be determined to be $W_k^\mathrm{m}=4ik/\pi$.
Using the forms of the functions $(m_k,n_k)$ and~$(p_k,q_k)$ from 
Eqs.~\eqref{eq:mn-rd} and~\eqref{eq:pq-md} in the above expressions,
we obtain~$(\alpha_k^{\mathrm{m}},\beta_k^{\mathrm{m}})$ to be
\begin{subequations}\label{eq:alpham_betam}
\begin{align}
\alpha_k^\mathrm{m}
&=\frac{\e^{ix_\mathrm{eq}}}{2}
\left[\alpha_k^\mathrm{r}\left(-2i+\frac{1}{x_\mathrm{eq}}
+\frac{i}{4x^2_\mathrm{eq}}\right)
-\frac{i\beta_k^\mathrm{r}}{4x_\mathrm{eq}^2}\e^{2ix_\mathrm{eq}}\right],\\
\beta_k^\mathrm{m} 
&=\frac{\e^{-ix_\mathrm{eq}}}{2}
\left[\frac{i\alpha_k^\mathrm{r}}{4x^2_\mathrm{eq}}\e^{-2ix_\mathrm{eq}}
+\beta_k^\mathrm{r}\left(2i+\frac{1}{x_\mathrm{eq}}
-\frac{i}{4x^2_\mathrm{eq}}\right)\right],
\end{align}
\end{subequations}
where $x_\mathrm{eq}=k(\eta_\mathrm{eq}-\eta_\mathrm{r})$, and we have 
used the fact that $z=k(\eta+\eta_\mathrm{eq}-2\eta_\mathrm{r})$ and 
hence $z_\mathrm{eq}=2x_\mathrm{eq}$. 
The quantities involving the Bogoliubov coefficients $(\alpha_k^\mathrm{m},
\beta_k^\mathrm{m})$ that appear in the PS~\eqref{eq:PTaccurate} above can 
be expressed as 
\begin{subequations}\label{eq:alphamod2_matter_exact}
\begin{align}
\vert \alpha_k^\mathrm{m}\vert ^2 
&=\vert \alpha_k^\mathrm{r}\vert ^2
\left(1+\frac{1}{64 x^4_\mathrm{eq}}\right)
+\frac{\vert \beta_k^\mathrm{r}\vert ^2}{64 x^4_\mathrm{eq}}
+\frac{\alpha_k^\mathrm{r}\beta_k^{\mathrm{r}*}}{16x^2_\mathrm{eq}}
\left(2+\frac{i}{x_\mathrm{eq}}
-\frac{1}{4x^2_\mathrm{eq}}\right)\e^{-2ix_\mathrm{eq}}\nn\\ 
&\quad+\frac{\alpha_k^{\mathrm{r}*}\beta_k^\mathrm{r}}{16x^2_\mathrm{eq}}
\left(2-\frac{i}{x_\mathrm{eq}}-\frac{1}{4x^2_\mathrm{eq}}\right)
\e^{2ix_\mathrm{eq}},\\
\vert \beta_k^\mathrm{m}\vert ^2 
&=\frac{\vert \alpha_k^\mathrm{r}\vert ^2}{64 x^4_\mathrm{eq}}
+\vert \beta_k^\mathrm{r}\vert ^2
\left(1+\frac{1}{64 x^4_\mathrm{eq}}\right)
+\frac{\alpha_k^\mathrm{r}\beta_k^\mathrm{r}{}^*}{16x^2_\mathrm{eq}}
\left(2+\frac{i}{x_\mathrm{eq}}
-\frac{1}{4x^2_\mathrm{eq}}\right)\e^{-2ix_\mathrm{eq}}\nn\\ 
&\quad+\frac{\alpha_k^\mathrm{r}{}^*\beta_k^\mathrm{r}}{16x^2_\mathrm{eq}}
\left(2-\frac{i}{x_\mathrm{eq}}
-\frac{1}{4x^2_\mathrm{eq}}\right)\e^{2ix_\mathrm{eq}},\\
\Re(\alpha_k^\mathrm{m}\beta_k^\mathrm{m*})
&=\left(\vert \alpha_k^\mathrm{r}\vert ^2
+\vert \beta_k^\mathrm{r}\vert ^2\right)
\Biggl[\frac{1}{16x^2_\mathrm{eq}}
\left(-2+\frac{1}{4x^2_\mathrm{eq}}\right)\cos(4x_\mathrm{eq})
+\frac{1}{16x^3_\mathrm{eq}}\sin(4x_\mathrm{eq})\Biggr]\nn\\ 
&\quad+\Re(\alpha_k^\mathrm{r}\beta_k^\mathrm{r*})
\Biggl[\frac14 \left(-4+\frac{2}{x^2_\mathrm{eq}}
-\frac{1}{16x^4_\mathrm{eq}}\right)\cos(2x_\mathrm{eq})
-\frac14 \left(-\frac{4}{x_\mathrm{eq}}
+\frac{1}{2x^3_\mathrm{eq}}\right)\sin(2x_\mathrm{eq})\nn\\ 
&\quad-\frac{1}{64x^4_\mathrm{eq}}\cos(6x_\mathrm{eq})\Biggr]
+\Im(\alpha_k^\mathrm{r}\beta_k^\mathrm{r*})
\Biggl[-\frac14 \left(-4+\frac{2}{x^2_\mathrm{eq}}
-\frac{1}{16x^4_\mathrm{eq}}\right)\sin(2x_\mathrm{eq})\nn\\ 
&\quad-\frac14 \left(-\frac{4}{x_\mathrm{eq}}
+\frac{1}{2x^3_\mathrm{eq}}\right)\cos(2x_\mathrm{eq})
-\frac{1}{64x^4_\mathrm{eq}}\sin(6x_\mathrm{eq})\Biggr],\\
\Im(\alpha_k^\mathrm{m}\beta_k^\mathrm{m*})
&=\left(\vert \alpha_k^\mathrm{r}\vert ^2
+\vert \beta_k^\mathrm{r}\vert ^2\right)
\Biggl[\frac{1}{16x^2_\mathrm{eq}}\left(-2
+\frac{1}{4x^2_\mathrm{eq}}\right)\sin(4x_\mathrm{eq})
-\frac{1}{16x^3_\mathrm{eq}}\cos(4x_\mathrm{eq})\Biggr]\nn\\ 
&\quad+\Re(\alpha_k^\mathrm{r}\beta_k^\mathrm{r*})
\Biggl[\frac14 \left(-4+\frac{2}{x^2_\mathrm{eq}}
-\frac{1}{16x^4_\mathrm{eq}}\right)\sin(2x_\mathrm{eq})
+\frac14 \left(-\frac{4}{x_\mathrm{eq}}
+\frac{1}{2x^3_\mathrm{eq}}\right)\cos(2x_\mathrm{eq})\nn\\ 
&\quad-\frac{1}{64x^4_\mathrm{eq}}\sin(6x_\mathrm{eq})\Biggr]
+\Im(\alpha_k^\mathrm{r}\beta_k^\mathrm{r*})
\Biggl[\frac14 \left(-4+\frac{2}{x^2_\mathrm{eq}}
-\frac{1}{16x^4_\mathrm{eq}}\right)\cos(2x_\mathrm{eq})\nn\\ 
&\quad-\frac14 \left(-\frac{4}{x_\mathrm{eq}}
+\frac{1}{2x^3_\mathrm{eq}}\right)\sin(2x_\mathrm{eq})
+\frac{1}{64x^4_\mathrm{eq}}\cos(6x_\mathrm{eq})\Biggr].
\end{align}
\end{subequations}
In the case of instantaneous transitions from de Sitter inflation to 
the epoch of radiation and matter domination, the PS of PGWs today can
be arrived at by substituting these expressions, along with the 
expressions~\eqref{eq:CDabrupt-o} for the Bogoliubov 
coefficients $(\alpha_k^\mathrm{r},\beta_k^\mathrm{r})$, in 
Eq.~\eqref{eq:generalptmatunregulated}.
The complete expression proves to be rather long and cumbersome to write
down explicitly.
Therefore, let us discuss the shape of the PS in the different limits of 
interest.

The shape of the PS during the matter-dominated era is controlled by
the three quantities (in contrast to two during the radiation-dominated 
era):~$x_\mathrm{e}=y_\mathrm{e}$, 
$x_\mathrm{eq}=y_\mathrm{e}a_\mathrm{eq}/a_\mathrm{e}$ and $z_0=2x_\mathrm{eq}
(a_0/a_\mathrm{eq})^{1/2}$, where, recall that, $a_0$ is the scale factor today. 
Note that these relations imply the inequality $z_0>x_\mathrm{eq}>y_\mathrm{e}$. 
Let us now consider the behavior of the PS of PGWs in the two limits
$x_\mathrm{eq}\gg 1$ and  $x_\mathrm{eq}\ll 1$.
In the limit $x_\mathrm{eq}\gg 1$, it is easy to show that the 
quantities in the expressions~\eqref{eq:alphamod2_matter_exact} 
reduce to the following forms:
\begin{subequations}    
\begin{align}
\vert \alpha_k^\mathrm{m}\vert ^2 
&\simeq \vert \alpha_k^\mathrm{r}\vert ^2,\\
\vert \beta_k^\mathrm{m}\vert ^2 
&\simeq \vert \beta_k^\mathrm{r}\vert ^2,\\
\Re(\alpha_k^\mathrm{m}\beta_k^\mathrm{m*})
&\simeq 
-\Re(\alpha_k^\mathrm{r}\beta_k^\mathrm{r*})
\cos(2x_\mathrm{eq})
+\Im(\alpha_k^\mathrm{r}\beta_k^\mathrm{r*})
\sin(2x_\mathrm{eq}),\\
\Im(\alpha_k^\mathrm{m}\beta_k^\mathrm{m*})
&\simeq 
-\Re(\alpha_k^\mathrm{r}\beta_k^\mathrm{r*})
\sin(2x_\mathrm{eq})
-\Im(\alpha_k^\mathrm{r}\beta_k^\mathrm{r*})
\cos(2x_\mathrm{eq}).
\end{align}\label{eq:alphamod2_matter_approx_abrupt}
\end{subequations}
It then follows that, in the limit, the PS of PGWs in the matter-dominated
era reduces to
\begin{align}
\pt(k,\eta) 
&=\frac{2}{\pi^2}\left(\frac{H_\mathrm{e}}{\Mpl}\right)^2
\left(\frac{a_\mathrm{e}}{a}\right)^2\frac{1}{y_\mathrm{e}^2}
\Biggl[y_\mathrm{e}^4+\frac12-\frac12\left(1-2y_\mathrm{e}^2\right)
\cos\left(2z-2y_\mathrm{e}-2x_\mathrm{eq}\right)\nn\\ 
&\quad+y_\mathrm{e} \sin\left(2z-2y_\mathrm{e}-2x_\mathrm{eq}\right)\Biggr].
\label{eq:ps-md-xeqgg1}
\end{align}
Apart from the difference in the phase factors of the trigonometric
functions, this PS has the same form as the PS in Eq.~\eqref{eq:ps-ds-rd-it} 
we had obtained in the radiation-dominated era.
From the above expression, we can easily arrive at the PS in the limits 
$x_\mathrm{eq}\gg 1$ and $y_{\rm e}\gg1$ or, equivalently, when $k\gg \ke$.
In these limits, we obtain that 
\begin{equation}\label{eq:ptsmall}
\pt(k,\eta) \simeq\frac{2}{\pi^2}\left(\frac{H_\mathrm{e}}{\Mpl}\right)^2
\left(\frac{a_\mathrm{e}}{a}\right)^2\left[y_\mathrm{e}^2+\cos\left(2z-
2y_\mathrm{e}-2x_\mathrm{eq}\right)\right].
\end{equation}
On the other hand, in the limits $x_\mathrm{eq}\gg 1$ and $y_{\rm e}\ll1$ or, 
equivalently, when $\keq\ll k\ll\ke$, we obtain that
\begin{align}\label{eq:ptradenter}
\pt(k,\eta) 
\simeq\frac{2}{\pi^2}\left(\frac{H_\mathrm{e}}{\Mpl}\right)^2
\left(\frac{a_\mathrm{e}}{a}\right)^2\f{1}{y_\mathrm{e}^2}
\sin^2(x_{\rm eq}-z).
\end{align}

We shall now discuss the behavior of the PS over large scales such that
$x_\mathrm{eq}\ll 1$. 
In the limit $x_\mathrm{eq}\ll 1$, from Eqs.~\eqref{eq:alpham_betam},
we obtain that
\begin{subequations}\label{eq:matxeqsmall}
\begin{align}
\alpha_k^\mathrm{m}&\simeq \f{i}{8x_\mathrm{eq}^2}(\alpha_k^\mathrm{r}
-\beta_k^\mathrm{r})+\f{3}{8x_\mathrm{eq}}(\alpha_k^\mathrm{r}+\beta_k^\mathrm{r})
-\f{9i}{16}(\alpha_k^\mathrm{r}-\beta_k^\mathrm{r})
+\f{x_\mathrm{eq}}{48}(37\alpha_k^\mathrm{r}-27\beta_k^\mathrm{r}),\\
\beta_k^\mathrm{m}&\simeq \f{i}{8x_\mathrm{eq}^2}(\alpha_k^\mathrm{r}
-\beta_k^\mathrm{r})+\f{3}{8x_\mathrm{eq}}(\alpha_k^\mathrm{r}+\beta_k^\mathrm{r})
-\f{9i}{16}(\alpha_k^\mathrm{r}-\beta_k^\mathrm{r})+\f{x_\mathrm{eq}}{48}
(-27\alpha_k^\mathrm{r}+37\beta_k^\mathrm{r}).
\end{align}
\end{subequations}
Since $a_0$ and $H_0$ are the scale factor and the Hubble parameter today, 
$k_0/a_0= H_0$ is the smallest wave number of observational interest.
On substituting the above expressions for $(\alpha_k^\mathrm{m},\beta_k^\mathrm{m})$
in the PS~\eqref{eq:PTaccurate} and considering the limit $z\gg 1$, which corresponds
to $k_0\ll k\ll\keq$, we obtain that
\begin{equation}
\pt(k,\eta) \simeq\frac{9}{8\pi^2}\left(\frac{H_\mathrm{e}}{\Mpl}\right)^2
\left(\frac{a_\mathrm{e}}{a}\right)^2\f{1}{y_\mathrm{e}^2x_\mathrm{eq}^2}\cos^2(z).
\label{eq:pteqtodaycale}
\end{equation}
In contrast, if we take the limits $\xeq\ll 1$, $y_{\e} \ll 1$ and $z\ll 1$
one after the another, we find that the PS~\eqref{eq:PTaccurate} simplifies
to the form
\begin{equation}\label{eq:PTlarge}
\pt(k,\eta) \simeq \frac{1}{8\pi^2}\left(\frac{H_\mathrm{e}}{\Mpl}\right)^2
\left(\frac{a_\mathrm{e}}{a}\right)^2\f{z^4}{y_\mathrm{e}^2x_\mathrm{eq}^2}
=\f{2}{\pi^2}\left(\frac{H_\mathrm{e}}{\Mpl}\right)^2,
\end{equation}
which is the standard scale-invariant behavior of the PS of PGWs on the 
largest scales.

\begin{figure}[!t]
\centering
\includegraphics[width=1.0\textwidth]{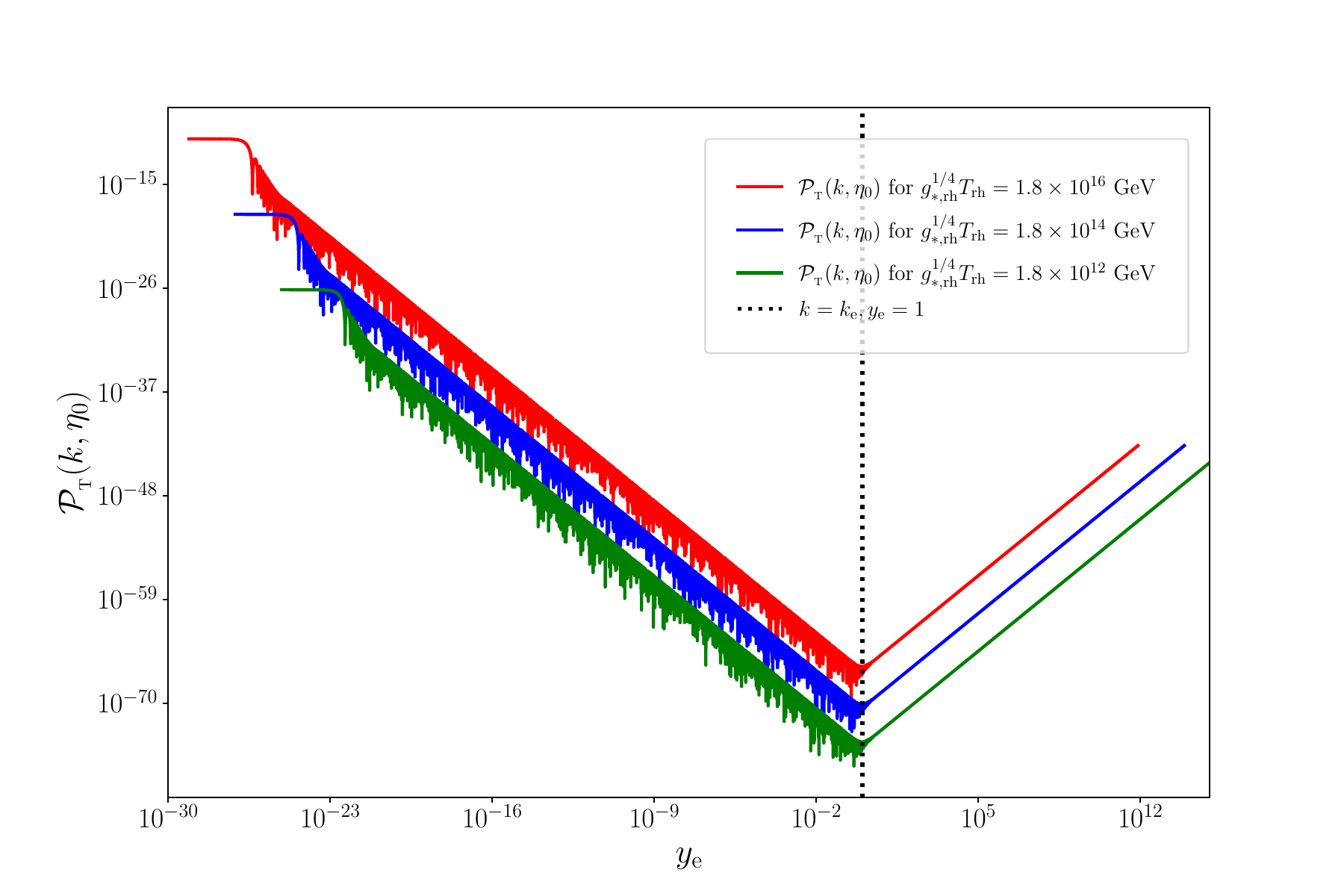}
\caption{The PS of PGWs~$\pt(k,\eta)$ evaluated today, i.e. at $\eta_0$,
has been plotted for the case of instantaneous transitions from de Sitter 
inflation to the epoch of radiation domination and subsequently to the 
epoch of matter domination [cf. Eq.~\eqref{eq:generalptmatunregulated}].
We have plotted the PS for the three values of~$g_{\ast,\mathrm{rh}}^{1/4} 
\Tre$ we had plotted in the previous figure.
Note that, in addition to the different behavior of the PS in the three 
regimes mentioned in the previous figure, there arises a fourth regime 
when $\xeq \ll 1$ and $z\gg 1$, where the PS behaves as $k^{-4}$ with 
superimposed oscillations [cf. Eq.~\eqref{eq:pteqtodaycale}].
These wave numbers correspond to $k_0 \lesssim  k \lesssim \keq$ which 
re-enter the Hubble radius during the epoch of matter domination.}
\label{fig:ps-sed-ds-rd-it} 
\end{figure}

We can plot the PS of PGWs today using Eq.~\eqref{eq:generalptmatunregulated},
along with the expressions~\eqref{eq:alpham_betam} and~\eqref{eq:CDabrupt-o} 
for the Bogoliubov coefficients~$(\alpha_k^{\mathrm{m}},\beta_k^{\mathrm{m}})$ 
and~$(\alpha_k^{\mathrm{r}},\beta_k^{\mathrm{r}})$. 
To plot the PS, we also require the ratios $\He/\Mpl$ and~$\ae/a_0$.
We can express the ratio~$\ae/a_0$ as 
\begin{equation}
\f{\ae}{a_0}=\f{\ae}{a_\mathrm{eq}}\f{a_\mathrm{eq}}{a_0},
\end{equation}
and we have already evaluated the ratio $\ae/a_\mathrm{eq}$  [cf. 
Eq.~\eqref{eq:aeq/ae}].
The ratio of scale factors at the time of radiation-matter equality and today is given by
\begin{equation}\label{eq:aeq/a0}
\frac{a_\mathrm{eq}}{a_0}=\frac{\Omega_\gamma h^2
+\Omega_\nu h^2}{\Omega_\mathrm{c} h^2+\Omega_\mathrm{b} h^2}.
\end{equation}
In Fig.~\ref{fig:ps-sed-ds-rd-it}, we have plotted the
PS~\eqref{eq:generalptmatunregulated} 
evaluated today as a function of~$y_{\e}$ for the same set of parameters as 
in the previous figure.
It should be clear from the figure that the PS exhibits four different 
regimes, viz. a scale-invariant behavior (over $k\lesssim k_0$), $k^{-4}$
and $k^{-2}$ behavior  with superimposed oscillations (over $k_0 \lesssim k 
\lesssim \keq$ and  $\keq \lesssim k \lesssim \ke$), and a $k^2$ behavior (over 
$k\gtrsim \ke$), as described by Eqs.~\eqref{eq:PTlarge}, \eqref{eq:pteqtodaycale}, 
\eqref{eq:ptradenter} and~\eqref{eq:ptsmall}, respectively.


\subsection{Difficulties with the PS}

In the previous section, we had obtained the expression for the PS of PGWs
in the case of instantaneous transitions from de Sitter inflation to the 
epochs of radiation and matter domination.
We shall now examine the expression closely and highlight some of the
behavior which can be considered as physically unsatisfactory.

\begin{figure}[!t]
\centering
\includegraphics[width=1.0\textwidth]{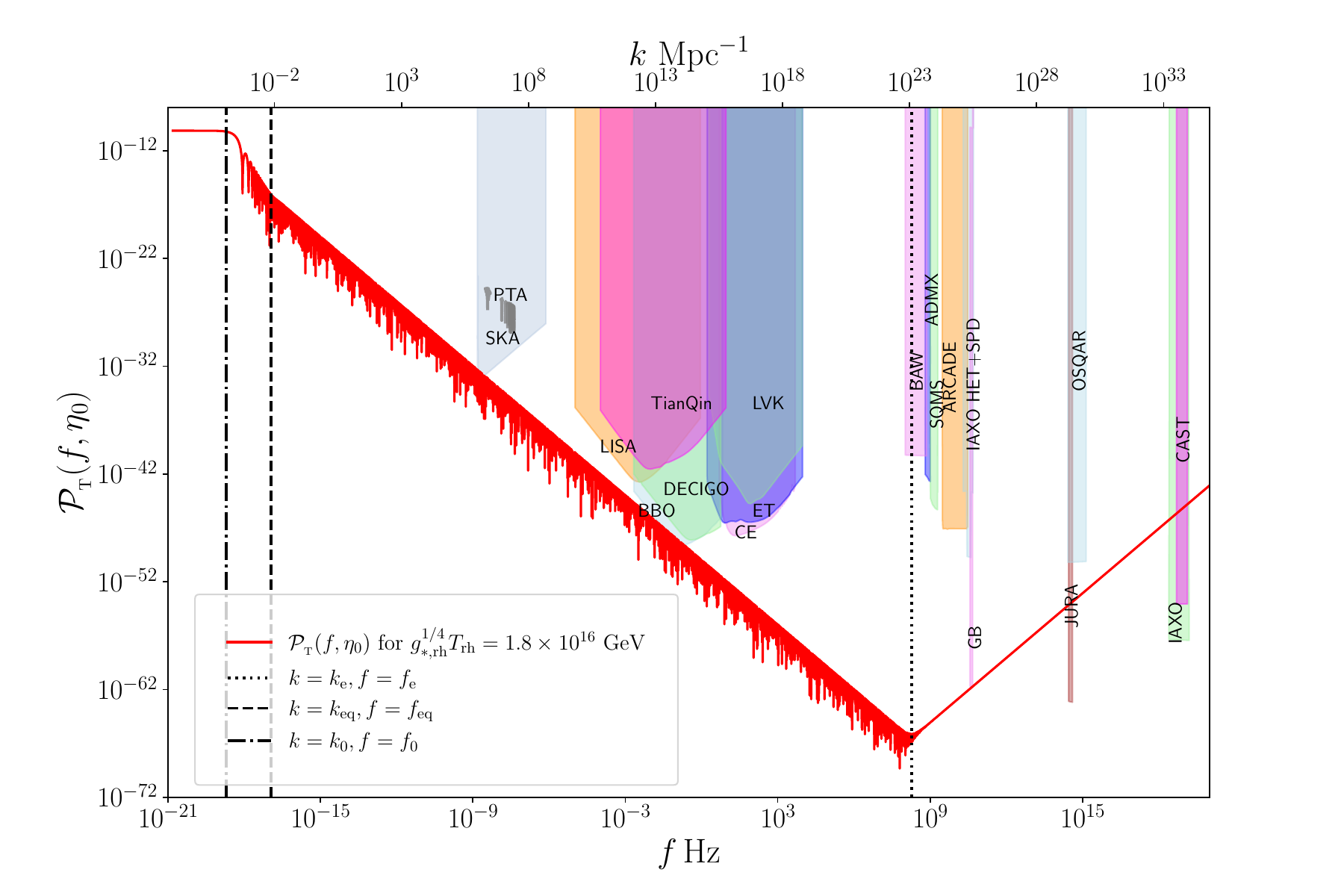}
\caption{The PS of PGWs~$\pt(k,\eta)$ evaluated today, i.e. at the conformal 
time $\eta_0$, has been plotted~(in red) 
assuming~$g_{\ast,\mathrm{rh}}^{1/4} \Tre=1.8 \times 10^{16}\,\mathrm{GeV}$ 
(corresponding to a tensor-to-scalar ratio of $r\simeq 0.034$, which is 
consistent with the current constraints) for the case of instantaneous 
transitions from de Sitter inflation to the epochs of radiation and matter
domination.
In contrast to the previous two figures, we have plotted the PS as a function
of frequency~$f$ rather than the dimensionless ratio~$y_{\e}$.
For the above-mentioned scenario and reheating temperature, the different 
wave numbers can be estimated to be~$(k_0/a_0,\keq/a_0,\ke/a_0)\simeq 
(1.2 \times 10^{-4}, 0.007,
1.24\times 10^{23})\,\mathrm{Mpc}^{-1}$, which correspond to the frequencies 
$(f_0,f_\mathrm{eq},f_{\e})=(1.94\times10^{-19},1.14\times10^{-17}, 
1.92\times10^{8})\,\mathrm{Hz}$ (indicated by the vertical dotted-dashed, 
dashed and dotted lines).
In the figure, we have also included the sensitivity curves of various present 
and future GW observatories operating over a wide range of frequencies (in this 
context, see the discussion in the introductory section as well as 
Refs.~\cite{Moore:2014lga,Kanno:2023whr,Franciolini:2022htd}). 
While the curves for ADMX, SQMS, GB, JURA and IAXO represent projected future 
sensitivities, it is important to point out that the sensitivity curves of some
detectors at high frequencies, such as BAW, ARCADE, OSQAR and CAST, are actually 
current upper bounds, not mere projections. 
The fact that CAST has not observed the signal implies that the PS of PGWs 
that we have plotted is ruled out. 
It should also be clear from the figure that, unless the $k^2$ rise in the PS 
of PGWs is truncated, the signal can, in principle, be observed by one or 
more of the detectors operating at high frequencies.}\label{fig:PT-hf}
\end{figure}

Previously, in Fig.~\ref{fig:ps-sed-ds-rd-it}, we had plotted the PS 
of PGWs today as a function of the dimensionless variable~$y_{\e}$.
Instead, in Fig.~\ref{fig:PT-hf}, we have plotted the same PS as a 
function of frequency~$f$\footnote{For simplicity, we shall ignore 
the effects due to late-time acceleration since this affects only 
the largest scales or the smallest frequencies.}.  
We have plotted the PS for the highest value of the reheating temperature
$\Tre$ that is consistent with the current bounds on the tensor-to-scalar
ratio $r$ from the CMB~\cite{Planck:2015sxf,BICEP:2021xfz}.
Specifically, we shall be interested in the behavior of the PS of PGWs
over wave numbers such that $k \gtrsim \ke =\ee^{-1}$.
Note that these wave numbers are always inside the Hubble radius, from 
the initial stages of inflation until today.
Therefore, clearly, the PS of PGWs over this range of wave numbers reflects 
the unprocessed Bunch-Davies vacuum. 
It is then interesting to ask whether such a signal can be detected.

To answer the question, in Fig.~\ref{fig:PT-hf}, we have also included 
the sensitivity curves of the various ongoing and forthcoming GW 
observatories such as SKA, LISA, BBO, TianQin, DECIGO, CE, ET and LVK 
over the smaller range of frequencies (for a brief discussion 
on the relation between the characteristic strain in terms of which 
the  sensitivity curves are often indicated and the sensitivity curves 
for the PS, see App.~\ref{app:strain}). 
In addition, we have also included the sensitivity curves of detectors 
that operate (or are set to operate) at high frequencies such as BAW, 
ADMX, SQMS, ARCADE, GB, JURA, IAXO HET + SPD, OCSAR, JURA, IAXO and CAST.
As should be evident from the figure, since the PS of PGWs rises as $k^2$
over $k \gtrsim \ke$, the PS intersects the sensitivity curves of some 
of the detectors.
In other words, it should be possible to detect the signal by one or more
of the detectors at high frequencies.
However, the $k^2$ rise in the PS on small scales is actually problematic. 
The reason being that, while the curves for ADMX, SQMS, GB, JURA and IAXO 
are projected future sensitivities, the sensitivity curves for BAW, ARCADE,
OSQAR and CAST actually represent upper bounds. 
The fact that, contrary to what the figure suggests, no signal has been seen
by CAST casts a shadow on the validity of the calculations that leads to the 
$k^2$ rise in the PS of PGWs at high frequencies. 
Moreover, as can be seen from Eq.~\eqref{eq:twopoint}, the $k^2$ rise would
lead to a divergence in the two-point correlation function in real space. 
These arguments indicate that PS of PGWs needs to be regularized, a topic that 
we shall turn to in the next section.


\section{Regularization of the PS of PGWs}\label{sec:r-ps-sed-gws}

In the previous section, assuming the Bunch-Davies initial conditions,  
we calculated the PS of PGWs at the time of radiation-matter equality
as well as today.
We found that, over small scales such that $y_\mathrm{e}\gg 1$, the PS 
behaves as~$k^2$.
As is well known, it is such a rise that leads to ultraviolet divergences 
in the coincident limit of the two-point functions of quantum fields in 
real space. 
In quantum field theory, these divergences are usually dealt with the 
help of regularization and renormalization procedures (for discussions in 
this context, see the books~\cite{Weinberg:1995mt,Weinberg:1996kr,
Birrell:1982ix,Parker:2009uva}).
For this reason, we shall also regularize the two-point function in Fourier
space, i.e. the~PS (for related discussions, see Refs.~\cite{Parker:2007ni,
Finelli:2007fr,Agullo:2008ka,Agullo:2009vq,Agullo:2009zi,Urakawa:2009xaa,
delRio:2014aua,Pla:2024xsv}).
We should emphasize that this last statement is non-trivial.
Note that, in real space, divergences arise in the coincident limit of 
two-point functions of quantum fields because they involve an integral
over {\it all}\/ the wave numbers.
In contrast, the two-point function of quantum fields in Fourier space, 
i.e. the PS, is evaluated {\it at a given wave number}.\/
Therefore, actually, the PS {\it does not contain any divergence}.\/ 
Nevertheless, the PS has to be regularized in order to ensure that 
the unphysical~$k^2$ behavior does not arise at small scales.


\subsection{Employing the method of adiabatic subtraction}\label{adiabatic_method}

There exist a variety of regularization procedures to handle the divergences
that are encountered when examining the behavior of quantum fields in curved
spacetimes (in this context, see the books~\cite{Birrell:1982ix,Parker:2009uva}).
These include dimensional regularization (see, for instance, 
Refs.~\cite{Brown:1976wc,Weinberg:1996kr}), 
zeta-function regularization (for the initial discussions, see
Refs.~\cite{Dowker:1975xj,Dowker:1975tf,Hawking:1976ja}; for specific 
applications, see Ref.~\cite{Elizalde:1995hck}),
proper-time regularization (for the original discussion, see 
Ref.~\cite{Schwinger:1951nm}; for additional discussions in electromagnetic
backgrounds, see Refs.~\cite{Dittrich:1985yb,Dittrich:2000zu}; for early 
applications in curved spacetimes, see, for instance,  
Refs.~\cite{DeWitt:1975ys,Dowker:1975xj}; for further
discussions, see Refs.~\cite{Birrell:1982ix,Parker:2009uva}),
point-splitting regularization (for the initial discussions, see 
Refs.~\cite{Christensen:1976vb,Davies:1976ei,Davies:1977ze,Bunch:1977sq,Wald:1977up,
Wald:1978pj,Bunch:1978yq,Bunch:1978yw,Christensen:1978yd}; 
for further discussions, see Refs.~\cite{Birrell:1982ix,Parker:2009uva}), 
Hadamard regularization~(see, for example, Refs.~\cite{Tadaki:1987dq,Decanini:2005eg}
and, for a discussion related to the topic of our interest, see Refs.~\cite{Negro:2024bbf,Negro:2024iwy}),
and adiabatic regularization (for the original discussions, see Refs.~\cite{Fulling:1974zr,
Parker:1974qw,Fulling:1974pu,Bunch:1978gb,Bunch:1980vc,Anderson:1987yt}).
Among these regularization procedures, the approach that is the most convenient
to implement in time-dependent situations such as the FLRW universe is the 
method of adiabatic regularization.
Since we shall be interested in regularizing quantities at a given wave number,
the procedure turns out to be relatively simple to employ.
We should also mention that the different methods of regularization are expected
to lead to the same results.
In due course, it will be interesting to reproduce the results we shall obtain 
below using, say, the method of point-splitting.

The method of adiabatic regularization involves identifying the contribution 
to the two-point function at a given adiabatic order and subtracting it from 
the complete contribution.
In order to parametrize the slowness of the changes in the background and calculate
the corresponding contributions, it proves to be convenient to introduce a parameter,
say, $T$, often referred to as the adiabatic parameter.
Thereafter, an expansion in the inverse powers of~$T$ is carried out and the 
terms at the order~$T^{-n}$ is referred to as the contributions at $n$\/th 
adiabatic order~\cite{Birrell:1982ix}.
Eventually, the parameter~$T$ is set to unity at the end of the calculations.
As we shall see, the regularization of the PS is carried out by subtracting 
terms up to the second adiabatic order.

Consider a field propagating in a FLRW universe. 
The homogeneity of the background allows us to decompose the field in terms
of the Fourier modes and the isotropy of the background implies that the mode
functions will depend only on the wave number~$k$.
In general, a suitably rescaled Fourier mode function of the field, say, 
$\chi_k$, will satisfy an equation of motion of the following form:
\begin{equation}
\chi_k''(\eta)+ \l[\omega_k^2(\eta)+\sigma^2(\eta)\r]\chi_k(\eta)=0,
\label{eq:eom-chik}
\end{equation}
where $\omega_k^2(\eta)$ and $\sigma^2(\eta)$ are functions of time.  
For instance, in the case of a massive scalar field, the
quantity $\omega_k^2(\eta)$ will depend on the wave number, the mass 
of the field and the scale factor  of the FLRW universe, while the
quantity $\sigma^2(\eta)$ will depend on the scale factor and its time 
derivatives.
To solve such an equation, let us assume the WKB ansatz
\begin{equation}
\chi_k^\mathrm{ad}(\eta)=\frac{1}{\sqrt{2 W_k(\eta)}}
\exp[-i \int^{\eta}_{\ei} \d\tilde{\eta}\,  W_k(\tilde{\eta})],\label{eq:wkb-a}
\end{equation}
where $\ei$ is some initial time. 
On substituting this ansatz into the equation of motion~\eqref{eq:eom-chik}, 
we obtain that
\begin{equation}
W_k^2(\eta) =\omega_k^2(\eta)+\f{\sigma^2(\eta)}{T^2}
-\f{1}{2 T^2}\l[\f{W_k''}{W_k}-\f{3}{2}\l(\f{W_k'}{W_k}\r)^2\r].\label{eq:Wk}
\end{equation}
where we have introduced powers of $1/T$ to facilitate the adiabatic expansion. 
We can now expand $W_k(\eta)$ in adiabatic order, i.e. we can write
\begin{equation} 
W_k(\eta)=\sum_{n=0}^{\infty}\f{W^{(n)}_k(\eta)}{T^n},\label{eq:exp}
\end{equation}
where the index ${(n)}$ refers to the adiabatic order of the expansion. 
In the following, we shall assume that $\omega_{k}(\eta)$ and $\sigma^2(\eta)$ 
are of adiabatic orders~zero and~two, respectively. 
On substituting this expansion into Eq.~\eqref{eq:Wk} and, on separating the 
terms by the adiabatic order (i.e. equating the coefficients of $T^{-n}$), we 
obtain that (in this context, see, for instance, Ref.~\cite{Pla:2024xsv})
\begin{subequations}\label{eq:adia}
\begin{align}
W_k^{(0)}&=\omega_k,\\
W_k^{(1)}&=0,\\
W_k^{(2)}&=\f{\sigma^2}{2\omega_k} 
+ \f{3 \omega_k'^2}{8 \omega_k^3}
-\f{\omega_k''}{4 \omega_k^2}.
\end{align}
\end{subequations}
For instance, note that, the adiabatic solution~\eqref{eq:eom-chik} leads to 
\begin{equation}\label{eq:uk2}
\vert \chi_k^{\mathrm{ad}}\vert^2=\f{1}{2 W_k}.
\end{equation}
According to Eq.~\eqref{eq:exp}, when $T$ is eventually set to unity, we have
\begin{equation}\label{eq:Winv}
W_k^{-1}\equiv(W_k^{-1})^{(0)}+(W_k^{-1})^{(2)}+\cdots,
\end{equation}
with the terms at the different adiabatic orders $(W_k^{-1})^{(n)}$ defined as
\begin{subequations}\label{eq:adiaWinv}
\begin{align}
(W_k^{-1})^{(0)} &=\f{1}{\omega_k},\\
(W_k^{-1})^{(2)} &=-\f{W_k^{(2)}}{\omega_k^2}.
\end{align}
\end{subequations}


\subsection{Regularization of the PS of PGWs}

In the case of the rescaled mode function~$\mu_k(\eta)$ that characterizes 
the PGWs [cf. Eq.~\eqref{eq:mse}], we have
\begin{subequations}
\begin{align}
\omega_k &= k,\\
\sigma^2(\eta) &= -\f{a''}{a}.
\end{align}
\end{subequations}
To regularize the PS of PGWs, we need to calculate the quantity
$\vert \mu_k^\mathrm{ad}\vert^2$ up to the second adiabatic order. 
We have  
\begin{subequations}
\begin{align}
\label{eq:wadzero}
W_k^{(0)} &=k,\\
\label{eq:wadtwo}
W_k^{(2)} &=-\f{1}{2 k} \f{a''}{a}.
\end{align}
\end{subequations}
and, on using Eqs.~\eqref{eq:uk2}--\eqref{eq:adiaWinv}, we obtain 
the contribution to the PS of PGWs due to the adiabatic solution
up to the second order to be
\begin{align}
\ptad(k,\eta)
&=\f{8}{\Mpl^2} \f{k^3}{2 \pi^2}
\f{\vert \mu_k^\mathrm{ad}\vert^2}{a^2}
=\f{8}{\Mpl^2} \f{k^3}{2 \pi^2 a^2} 
\f{1}{2 W_k}\nn\\
&=\f{8}{\Mpl^2} \f{k^3}{2 \pi^2 a^2}\,
\f{1}{2 W_k^{(0)}}
\l(1-\f{W_k^{(2)}}{W_k^{(0)}}\r)
=\f{8}{\Mpl^2} \f{k^3}{2 \pi^2 a^2} \f{1}{2 k}
\l(1+\f{a''}{2k^2 a}\r).\label{eq:ptad1}
\end{align}
We should mention that the final expression is valid for arbitrary 
scale factor.
The above adiabatic contribution should be subtracted from the original
PS to lead to the following regularized PS of PGWs:
\begin{equation}
\ptr(k,\eta)\equiv \pt(k,\eta)-\ptad(k,\eta).\label{eq:rps-d}
\end{equation}
Note that, in the post-inflationary universe, we can express 
the adiabatic contribution as follows:
\begin{align}
\ptad(k,\eta)
&=\frac{2}{\pi^2} \left(\frac{H_\mathrm{e}}{\Mpl}\right)^2
\left(\frac{a_\mathrm{e}}{a}\right)^2 \bar{y}_\mathrm{e}^2
\left(1+\f{1}{2 k^2}\f{a''}{a}\right),\label{eq:ptad}
\end{align}
where, recall that, $\bar{y}_{\e}=k/(\ae \He)$.

Let us now apply this procedure to the PS of PGWs we obtained during 
the radiation domination era. 
During radiation domination, we have $a''=0$.
Hence, in the expression for~$\ptad(k,\eta)$ in Eq.~\eqref{eq:ptad} 
above, the second term vanishes. 
Upon subtracting the remaining contribution from the PS of PGWs 
as given by Eq.~\eqref{eq:generalpt}, we obtain the 
regularized PS during the epoch of radiation domination to be
\begin{align}
\label{eq:generalptregulated}
\ptr(k,\eta)
=\frac{2}{\pi^2}\left(\frac{H_\mathrm{e}}{\Mpl}\right)^2
\left(\frac{a_\mathrm{e}}{a}\right)^2 2\bar{y}_\mathrm{e}^2
\Biggl[\left \vert\frac{\beta_k^\mathrm{r}}{A_{k}}\right\vert^2
-\Re \left(\frac{\alpha_k^\mathrm{r}\beta_k^{\mathrm{r}*}}{\vert 
A_{k}\vert^2}\right)\cos(2x)-\Im \left(\frac{\alpha_k^\mathrm{r}
\beta_k^{\mathrm{r}*}}{\vert A_{k}\vert^2}\right)\sin(2x)\Biggr],
\end{align}
which is valid for any coefficients $(\alpha_k^{\mathrm{r}},
\beta_k^{\mathrm{r}})$.

Similarly, during the epoch of matter domination, the contribution 
to the PS of PGWs at the second adiabatic order can be written 
as [cf. Eqs.~\eqref{eq:ptad} and~\eqref{eq:sf-md}] 
\begin{align}
\ptad(k,\eta)
=\frac{2}{\pi^2} \left(\frac{H_\mathrm{e}}{\Mpl}\right)^2
\left(\frac{a_\mathrm{e}}{a}\right)^2 \bar{y}_\mathrm{e}^2
\left(1+\f{1}{ z^2}\right).\label{ptmatad}
\end{align}
On subtracting this contribution from the PS of PGWs as given by
Eq.~\eqref{eq:generalptmatunregulated}, we obtain the regularized 
PS during the epoch of matter domination to be
\begin{align}
\label{eq:generalptmatregulated}
\ptr(k,\eta)
&=\frac{2}{\pi^2}\left(\frac{H_\mathrm{e}}{\Mpl}\right)^2
\left(\frac{a_\mathrm{e}}{a}\right)^2 2\bar{y}_\mathrm{e}^2
\Biggl\{\,\left \vert \f{\beta_k^\mathrm{m}}{A_{k}}\right\vert ^2\left(1+\frac{1}{z^2}\right)
+\left(1-\frac{1}{z^2}\right)
\biggl[\Re\left(\f{\alpha_k^\mathrm{m}\beta_k^\mathrm{m*}}{\vert A_{k} \vert ^2}\right) \cos(2z)\nn\\ 
&\quad+\Im \left(\f{\alpha_k^\mathrm{m}\beta_k^{\mathrm{m}*}}{\vert A_{k} \vert ^2}\right) \sin(2z)\biggr]
-\frac{2}{z}
\biggl[\Re\left(\f{\alpha_k^\mathrm{m}\beta_k^\mathrm{m*}}{\vert A_{k} \vert ^2}\right) \sin(2z)
-\Im \left(\f{\alpha_k^\mathrm{m}\beta_k^{\mathrm{m}*}}{\vert A_{k} \vert ^2}\right) \cos(2z)\biggr]
\Biggr\},
\end{align}
which is valid for any coefficients $(\alpha_k^{\mathrm{m}},
\beta_k^{\mathrm{m}})$.


\subsection{Regularized PS in the case of the instantaneous transitions}

\begin{figure}[!t]
\centering
\includegraphics[width=1.0\textwidth]{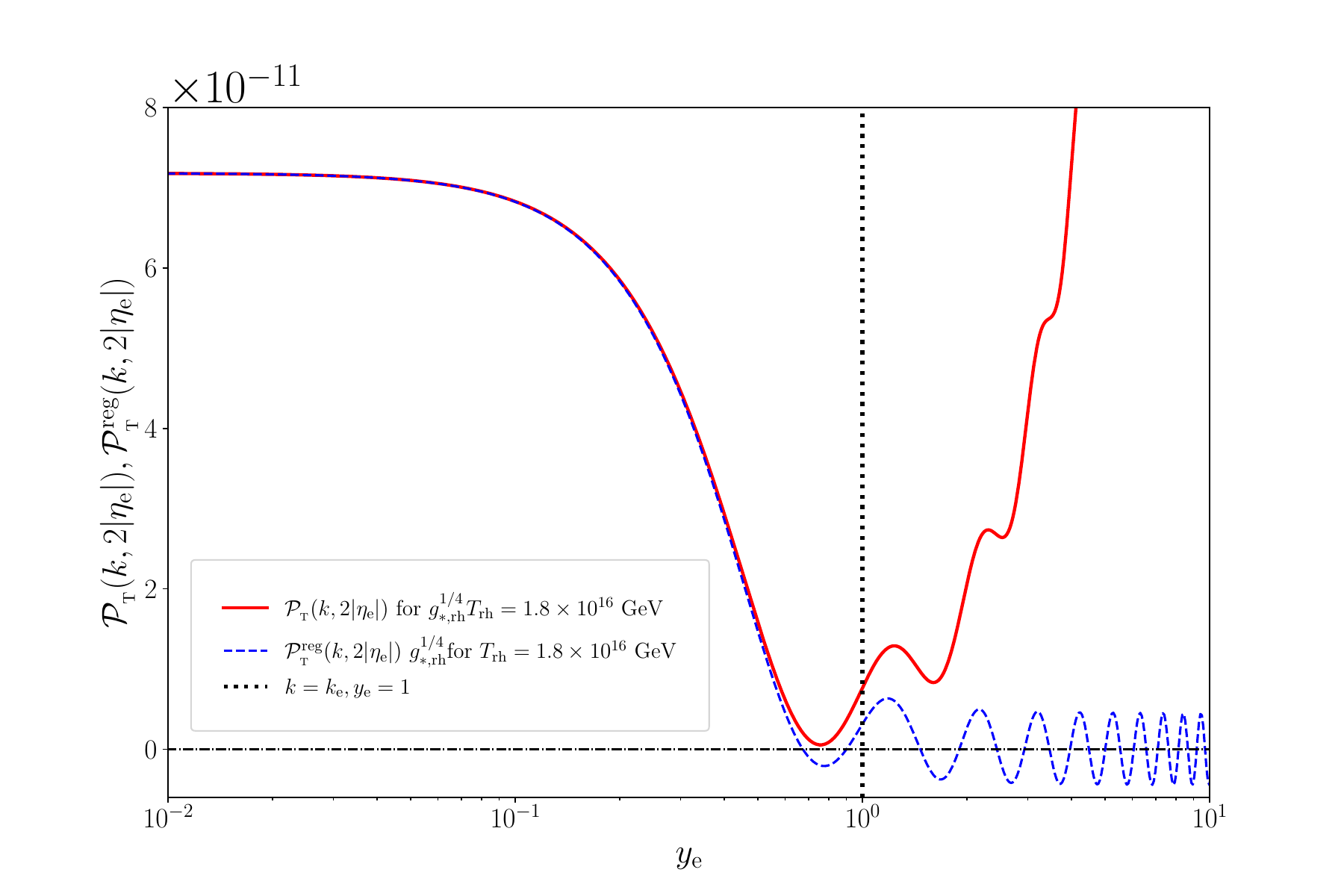}
\caption{The actual and the regularized PS of PGWs, viz. $\pt(k,\eta)$ and
$\ptr(k,\eta)$, evaluated during the early stages of the radiation-dominated 
epoch at $\eta=-2\ee=2\vert\ee\vert$ have been plotted~(in red and blue) for 
the case of instantaneous transition from de Sitter inflation
[cf. Eq.~\eqref{eq:ps-ds-rd-it-r}].
We have plotted the PS for the same value of $g_{\ast,\mathrm{rh}}^{1/4} 
\Tre$ as in the previous figure. 
Note that, over $y_{\e}>1$ (demarcated by the vertical dotted line), while 
the original PS rises as~$k^2$ (with oscillations superimposed upon it), 
the regularized PS oscillates about zero with a constant amplitude.
In other words, over $k\gtrsim \ke$, $\ptr(k,\eta)$ can turn negative.}
\label{fig:neg-PS}
\end{figure}

In the case of the instantaneous transition from de Sitter inflation
to radiation domination, the regularized PS above takes the form (with 
$\bar{y}_\mathrm{e}=y_\mathrm{e}$)
\begin{align}
\ptr(k,\eta)=\frac{2}{\pi^2}\left(\frac{H_\mathrm{e}}{\Mpl}\right)^2
\left(\frac{a_\mathrm{e}}{a}\right)^2
\frac{1}{2y_\mathrm{e}^2}\biggl\{1-\left(1-2y_\mathrm{e}^2\right)
\cos\left[2(x-y_\mathrm{e})\right]
+2y_\mathrm{e}\sin\left[2(x-y_\mathrm{e})\right]\biggr\}.
\label{eq:ps-ds-rd-it-r}
\end{align}
We find that, for wavenumbers that are in the super-Hubble domain  
at the end of inflation, i.e. when $y_\mathrm{e}\ll 1$, the above
expression for $\ptr(k,\eta)$ exactly matches the 
PS~\eqref{eq:ptlargescale} on large scales that we had obtained. 
On the contrary, for wavenumbers that are in the sub-Hubble domain
at the end of inflation, i.e. when $y_\mathrm{e}\gg 1$, we obtain 
that
\begin{align}
\ptr(k,\eta)\simeq \frac{2}{\pi^2}\left(\frac{H_\mathrm{e}}{\Mpl}\right)^2
\l(\frac{a_\mathrm{e}}{a}\r)^2\cos\l[2(x-y_\mathrm{e})\r].\label{eq:osc}
\end{align}
This should be contrasted with the result~\eqref{eq:ptsmallscale} for the 
unregularized PS in this limit.
Clearly, there is no $k^2$ rise anymore for $k \gtrsim \ke$.
In fact, $\ptr(k,\eta)$ oscillates about a vanishing mean value in this regime. 
In Fig.~\ref{fig:neg-PS}, we have plotted the original and the regularized PS 
of PGWs viz. $\pt(k,\eta)$ and $\ptr(k,\eta)$ [as given by 
Eqs.~\eqref{eq:ps-ds-rd-it} and~\eqref{eq:ps-ds-rd-it-r}] evaluated during the
early stages of the radiation-dominated epoch.
The figure highlights the points we have made above.
As expected, the regularization curtails the~$k^2$ rise that occurs in 
the original PS for $k\gtrsim \ke$.
However, as should be clear from the figure, the regularized PS can be 
negative in this regime (in this context, also see the discussion
and figure in Ref.~\cite{Pla:2024xsv}). 
Moreover, the process of regularization does not alter the conventional
(i.e. the unregularized) PS over wave numbers~$k \lesssim \ke$.  
The last point is reassuring since the regularized PS will not alter the 
standard imprints on the CMB.

\begin{figure}[!t]
\centering
\includegraphics[width=1.0\textwidth]{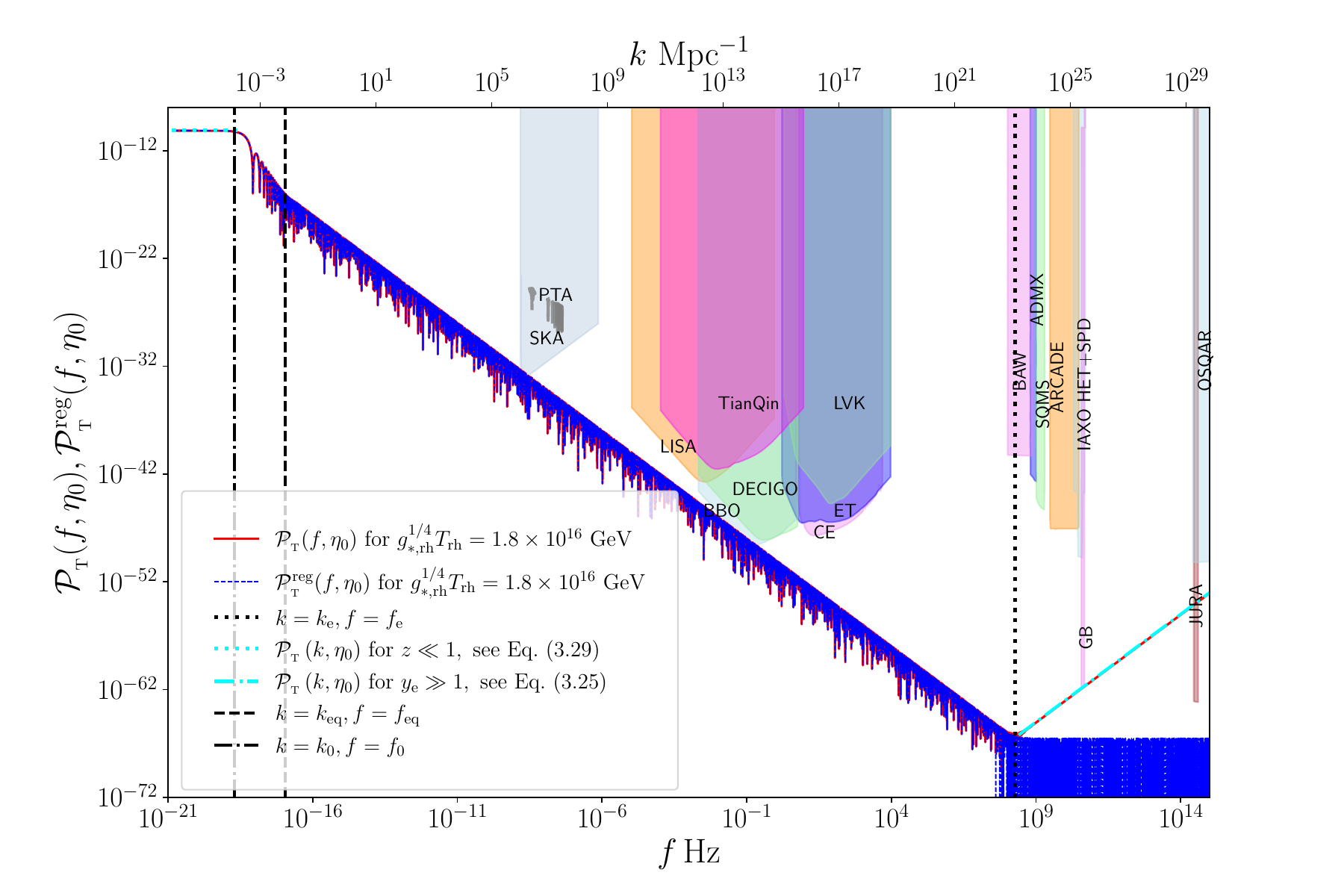}
\caption{The actual and the regularized PS of PGWs, viz. $\pt(k,\eta)$ 
and $\ptr(k,\eta)$, evaluated today at the conformal time $\eta_0$,
have been plotted (in red and blue) as a function of frequency $f$ for 
the case of instantaneous transitions from de Sitter inflation to the 
epochs of radiation and matter domination [cf. Eqs.~\eqref{eq:generalptmatunregulated}
and~\eqref{eq:generalptmatregulated}].
We have chosen to work with~$g_{\ast,\mathrm{rh}}^{1/4} \Tre=1.8\times 
10^{16}\, \mathrm{GeV}$ as in the previous two figures.
Also, as in Fig.~\ref{fig:PT-hf}, the vertical lines indicate the frequencies
corresponding to the wave numbers $k_0/a_0$, $\keq/a_0$ and~$\ke/a_0$.
Moreover, as in Fig.~\ref{fig:PT-hf}, we have included the sensitivity 
curves of the various GW observatories.
It is clear that the process of regularization truncates the~$k^2$ rise 
for~$k \gtrsim \ke$.
Also, note that the regularization procedure does not alter the PS over 
scales $k \lesssim \ke$.
We find that, over $k \gtrsim \ke$, the regularized PS $\ptr(k,\eta)$ oscillates
about zero as illustrated in the previous figure.
However, the amplitude of the oscillations is significantly suppressed during
the late stages of matter domination when compared to the values during the 
early stages of radiation domination.}
\label{fig:ps-rps-ds-rd-it} 
\end{figure}

Let us now turn to the regularized PS of PGWs in the matter-dominated epoch.
In the case of instantaneous transitions from de Sitter inflation to radiation
domination and further to matter domination, the regularized PS of PGWs can
be arrived at by substituting the expressions~\eqref{eq:alpham_betam} for the 
Bogoliubov coefficients $(\alpha_k^\mathrm{m},\beta_k^\mathrm{m})$, along
with the expressions~\eqref{eq:CDabrupt-o} for  
$(\alpha_k^\mathrm{r},\beta_k^\mathrm{r})$,
in Eq.~\eqref{eq:generalptmatregulated}.
As in the case of the original PS, the expression for the regularized PS also 
turns out to be rather lengthy.
However, as we had seen earlier, the PS reduces to a simpler form in the 
limit $x_\mathrm{eq}\gg 1$ [cf. Eq.~\eqref{eq:ps-md-xeqgg1}].
The corresponding regularized PS can be obtained to be (since, in this 
limit, $z\gg 1$ as well)
\begin{align}
\ptr(k,\eta) 
&=\frac{2}{\pi^2}\left(\frac{H_\mathrm{e}}{\Mpl}\right)^2
\left(\frac{a_\mathrm{e}}{a}\right)^2\frac{1}{y_\mathrm{e}^2}
\Biggl[\frac12-\frac12\left(1-2y_\mathrm{e}^2\right)
\cos\left(2z-2y_\mathrm{e}-2x_\mathrm{eq}\right)\nn\\ 
&\quad+y_\mathrm{e} \sin\left(2z-2y_\mathrm{e}-2x_\mathrm{eq}\right)\Biggr].
\end{align}
In the limit $y_{\e}\gg 1$, this regularized PS simplifies to the form
\begin{align}
\ptr(k,\eta) 
\simeq \frac{2}{\pi^2}\left(\frac{H_\mathrm{e}}{\Mpl}\right)^2
\left(\frac{a_\mathrm{e}}{a}\right)^2
\cos\left(2z-2y_\mathrm{e}-2x_\mathrm{eq}\right),\label{eq:ps-md-lwn}
\end{align}
which, barring the phase factor, has the same behavior as in the 
radiation-dominated epoch [cf. Eq.~\eqref{eq:osc}].
In Fig.~\ref{fig:ps-rps-ds-rd-it}, we have plotted the original PS
and the regularized PS of PGWs evaluated today.

\subsection{Adiabatic regularization in real space}

The fact that the regularized PS of PGWs $\ptr(k,\eta)$ is negative over some
scales (as we saw in the previous section) may be viewed as problematic and, 
in this section, we shall discuss this issue in some detail. 
To begin with, note that the original PS $\pt(k,\eta)$ is a positive definite
quantity, as should be evident from Eq.~\eqref{eq:tps}.
However, since the regularized PS $\ptr(k,\eta)$ is the difference between two
quantities, it does not have to be positive definite.
Hence, in principle, there is no reason to expect that $\ptr(k,\eta)$ will remain
positive on all scales.
Secondly, it is the two-point correlation function of PGWs in real space [cf.
Eq.~\eqref{eq:twopoint}] that is an observable and, therefore, physical 
requirements should be imposed only on this quantity.
The PS can be compared to the sensitivities of various GW detectors, as we 
have done, for instance, in Fig.~\ref{fig:PT-hf}.
But, the PS of PGWs is only a derived quantity and it is inferred from the
physical constraints arrived at on the two-point correlation function in 
real space. 
These constraints can imply a non-standard behavior for the regularized PS
$\ptr(k,\eta)$.
However, provided the corresponding correlation function in real space behaves 
in a proper fashion, the non-standard behavior of $\ptr(k,\eta)$ need not be 
viewed as problematic.

Let us now highlight the reasons as to why the two-point correlation 
function in real space is the quantity that can be observationally probed. 
At a given spatial location, say, ${\bm x}=0$, the signal, say, 
$\mathcal{S}(t)$, observed by a GW detector can be 
expressed as~\cite{Maggiore:1999vm,Maggiore:2007ulw}
\begin{align}
\mathcal{S}(t)=D^{ij} h_{ij}(t),
\end{align}
where $D^{ij}$ depends on the properties of the detector and is known as the 
detector tensor.
Then, the mean-squared displacement of the detector can be expressed as
\begin{align}
\l\langle \mathcal{S}^2(t) \r\rangle
= D^{ij}D^{mn}\l\langle h_{ij}(t) h_{mn}(t)\r\rangle,
\end{align}
where the angular brackets denote averages over the random variable~$h_{ij}$.
On using the result~\eqref{eq:hijhijavg1} from App.~\ref{app:strain} in the 
above expression, we obtain that
\begin{align}
\l\langle \mathcal{S}^2(t) \r\rangle
=\f{F}{4\pi}\int_{0}^{\infty} \d f S_h(f)
=\f{F}{16\pi}\int_{0}^{\infty}\f{\d f}{f}\,\pt(f,t),\label{eq:msd-c}
\end{align}
where the quantity $F$ is a number that depends on the detector, and is 
given by
\begin{align}
F=\sum_{\lambda=(+,\times)}\int_0^{2\pi} \d \phi 
\int _0^{\pi} \d \theta \sin \theta \left[D^{ij}
\varepsilon_{ij}^\lambda(\hat{\bm n})\right]^2.
\end{align}

If the GWs are of quantum origin as in the case of the PGWs of our interest, 
then the quantities~$\mathcal{S}(t)$ and~$h_{ij}$ have to be treated as 
quantum operators instead of random variables.
In such a situation, we can define the mean-squared displacement as
\begin{align}
\l\langle \hat{\mathcal{S}}^2(\eta)\r\rangle
= D^{ij}D^{mn}\l\langle \hat{h}_{ij}(\eta,{\bm x}) 
\hat{h}_{mn}(\eta,{\bm x})\r\rangle,
\end{align}
where the averages are to be evaluated in the quantum vacuum.
Using the decomposition~\eqref{eq:dh} of the PGWs, it is straightforward 
to show that, in the quantum vacuum, the mean-squared displacement
can be expressed as
\begin{align}
\l\langle \hat{\mathcal{S}}^2(\eta) \r\rangle 
= \frac{2 F}{\Mpl^2a^2}
\int _0^{\infty} \f{\d k}{(2\pi)^3} k^2 \vert \mu_k\vert ^2
=\f{F}{16\pi} \int_0^{\infty} \frac{\dd k}{k} \pt(k,\eta),
\end{align}
which is the same as the expression~\eqref{eq:msd-c} in the case of 
stochastic and isotropic, classical GWs.
It should be clear that, if the PS $\pt(k,\eta)$ grows as $k^2$ at 
large $k$, then the quantity $\langle \hat{\mathcal{S}}^2(\eta) \rangle$ 
is not finite, which is not physically possible. 
Moreover, the observable $\langle \hat{\mathcal{S}}^2 (\eta)\rangle$ is 
clearly related to the two-point correlation function in real space
[cf. Eq.~\eqref{eq:twopoint}]. 
These arguments justify our claim at the beginning of this section, 
viz. that the observable signal is related to the correlation function 
in real space.

In light of the above discussion, it also becomes important to understand
the implications of the procedure of adiabatic regularization in real space.  
The two-point correlation function of the tensor perturbations, evaluated 
at the same time but at different spatial locations, can be written as 
[cf. Eq.~\eqref{eq:twopoint}]
\begin{equation}
\delta^{im}\delta^{jn}
\left \langle 0 \left \vert \hat{h}_{mn}(\eta,{\bm x})
\hat{h}_{ij}(\eta,{\bm x}')\right \vert 0 \right \rangle 
=\int _0^{\infty}  \frac{\mathrm{d}k}{k} 
\f{\sin\l(k\vert {\bm x}-{\bm x}'\vert\r)}{k\vert {\bm x}-{\bm x}'\vert}
\pt(k,\eta).\label{eq:twopoint2}
\end{equation}
As we have discussed before, to regularize the PS, it is the contribution 
up to the second adiabatic order that is to be subtracted from the complete 
PS.
In real space, the contribution to the two-point correlation function of the
tensor perturbations up to the second adiabatic order in Fourier space [cf. 
Eq.~\eqref{eq:ptad1}] can be expressed as 
\begin{align}
\delta^{im}\delta^{jn}
\l \langle 0 \left \vert \hat{h}^{ij}(\eta,{\bm x})
\hat{h}_{ij}(\eta,{\bm x}')\r \vert 0 \right \rangle^\mathrm{ad}
&=\int _0^{\infty} \f{\mathrm{d}k}{k} 
\frac{\sin (k\vert {\bm x}-{\bm x}'\vert)}{k\vert {\bm x}-{\bm x}'\vert}
\ptad(k,\eta)\nn\\
&=\f{2}{\pi^2\Mpl^2 a^2}
\l[\f{1}{\vert {\bm x}-{\bm x}'\vert^2}
\int_0^{\infty}\mathrm{d}\kappa\, \sin \kappa 
+\f{a''}{2a} \int_0^{\infty}\f{\d\kappa}{\kappa^2} \sin \kappa\r],
\end{align}
where we have defined $\kappa=k\vert {\bm x}-{\bm x}'\vert $. 
On suitably regulating the first integral, we can evaluate it to be
\begin{align}
\int_0^{\infty}\mathrm{d}\kappa \sin \kappa
=\lim_{\epsilon\to 0}\int_0^{\infty}
\mathrm{d}\kappa\, \mathrm{e}^{-\epsilon \kappa} \sin \kappa=1.
\end{align}
We can evaluate the second integral by introducing a comoving infrared cutoff, 
say, $k_{_\mathrm{IR}}$.
Upon doing so, we finally arrive at the adiabatic contribution up to the 
second order to the two-point function in real space to be
\begin{align}
\delta^{im}\delta^{jn}
\l \langle 0 \left \vert \hat{h}_{mn}(\eta,{\bm x})
\hat{h}_{ij}(\eta,{\bm x}')\right \vert 0 \r\rangle^\mathrm{ad}
&=\frac{2}{\pi^2\Mpl^2 a^2}
\biggl\{\frac{1}{\vert {\bm x}-{\bm x}\vert^2}
+\f{a''}{2a}\biggl[\f{\sin \l(k_{_\mathrm{IR}}\vert {\bm x}-{\bm x}'\vert\r)}
{k_{_\mathrm{IR}}\vert {\bm x}-{\bm x}'\vert}\nn\\
&\quad-\mathrm{Ci}(k_{_\mathrm{IR}}\vert {\bm x}-{\bm x}'\vert)\biggr]\biggr\},
\end{align}
where $\mathrm{Ci}(z)$ denotes the cosine integral function~\cite{Gradshteyn:2007}.
On Taylor expanding for small $\vert {\bm x}-{\bm x}'\vert$, we obtain that
\begin{align}
\delta^{im}\delta^{jn}
\l \langle 0 \left \vert \hat{h}_{mn}(\eta,{\bm x})
\hat{h}_{ij}(\eta,{\bm x}')\right \vert 0 \r \rangle^\mathrm{ad}
&=\f{2}{\pi^2\Mpl^2}\f{1}{\vert {\bm x}_\mathrm{p}-{\bm x}_\mathrm{p}'\vert^2}
\biggl\{1+\f{a''}{2 a^3} \vert {\bm x}_\mathrm{p}-{\bm x}'_\mathrm{p}\vert^2 
\biggl[1-\gamma_{_\mathrm{E}}\nn\\
&-\ln \left(k_{_\mathrm{IR}}\vert {\bm x}-{\bm x}'\vert\right)
+{\cal O}\left(k_{_\mathrm{IR}}^2\vert {\bm x}-{\bm x}'\vert^2\right)
\biggr]\biggr\},\label{eq:realspace_twopoint_adiabatic}
\end{align}
where ${\bm x}_\mathrm{p}=a {\bm x}$ denotes the physical distance. 
Note that, when $a''=0$, this quantity reduces to the correlation function 
of a free field in Minkowski spacetime, as expected. 
To regularize in real space, it is this quantity that should be subtracted
from the complete correlation function.

In fact, as we already discussed, the PS should be viewed as a derived quantity, 
obtained from the observable quantity, which is the correlation function in real 
space. 
Indeed, the two-point correlation function of the tensor perturbations in real
space can be inverted back to obtain the PS.
We find that the PS of PGWs can be expressed as
\begin{align}
\pt(k,\eta)=\f{2k^2}{\pi}\int_0^\infty\d r\, r \sin(kr)\,
\delta^{im}\delta^{jn}
\l\langle 0 \l \vert \hat{h}_{mn}(\eta,{\bm x})
\hat{h}_{ij}(\eta,{\bm x}')\r \vert 0 \r\rangle
\end{align}
and one can easily verify that the regularized PS can similarly be written as
\begin{align}
\ptr(k,\eta)
&=\f{2k^2}{\pi}\int_0^\infty\d r\, r \sin(kr)\nn\\
&\quad\times\delta^{im}\delta^{jn}
\l[\left \langle 0 \left \vert \hat{h}_{mn}(\eta,{\bm x})
\hat{h}_{ij}(\eta,{\bm x}')\right \vert 0 \right \rangle
-\left \langle 0 \left \vert \hat{h}_{mn}(\eta,{\bm x})
\hat{h}_{ij}(\eta,{\bm x}')\right \vert 0 \right \rangle^\mathrm{ad}\r].
\end{align}
From such a point of view, clearly, the fact that $\ptr(k,\eta)$ can turn
negative should not appear as problematic.

Finally, a last set of remarks are in order. 
Often, it has been suggested that the signal of GWs has to be averaged over
time (in this context, see, for instance, Ref.~\cite{Maggiore:2007ulw}).
We find that, if we average over time, the PS of PGWs, evaluated during the 
epoch of radiation domination [cf. Eq.~\eqref{eq:ps-ds-rd-it}], reduces to
the following simple form:
\begin{align}
\label{eq:ptwooscillations}
\overline{\mathcal{P}}_{_{\mathrm T}}(k)
=\frac{2}{\pi^2}\left(\frac{H_\mathrm{e}}{\Mpl}\right)^2
\left(\frac{a_\mathrm{e}}{a}\right)^2\frac{1}{y_\mathrm{e}^2}
\left(\frac12+y_\mathrm{e}^4\right).
\end{align}
However, note that the averaged PS $\overline{\mathcal{P}}_{_{\mathrm T}}(k)$ 
behaves as $k^{-2}$ on large scales and as $k^2$ on small scales. 
In particular, it no longer leads to the scale-invariant behavior on the largest 
scales.
Interestingly, if we average the regulated PS $\ptr(k,\eta)$ 
[cf. Eq.~\eqref{eq:ps-ds-rd-it-r}] during the 
radiation-dominated epoch, we obtain that 
\begin{align}
\overline{\mathcal{P}}_{_{\mathrm T}}^{\,\mathrm{reg}}(k)
=\frac{2}{\pi^2}\left(\frac{H_\mathrm{e}}{\Mpl}\right)^2
\left(\frac{a_\mathrm{e}}{a}\right)^2\frac{1}{y_\mathrm{e}^2},
\end{align}
which behaves as $k^2$ on {\it all}\/ scales.
These suggest that averaging should be performed only on small time scales
and not on cosmological time scales.
It also matters whether the limits of large and small scales are considered
before or after averaging.
If we first take the small scale limit (i.e. the limit $k\gg \ke$) of the 
regularized PS $\ptr(k,\eta)$ [see Eq.~\eqref{eq:osc}
and~\eqref{eq:ps-md-lwn}] and then carry out
the averaging, then the PS vanishes identically.
In other words, the resulting $\ptr(k,\eta)$ does not turn negative on such
small scales.


\section{Smoothing the transition from inflation to radiation 
domination}\label{sec:i-rd-st-b}

It should be clear from the discussions in the previous two sections that
the~$k^2$ rise in the PS of PGWs on small scales (over $k \gtrsim \ke$) 
can be circumvented by considering the regularized version of the quantity.
However, when transitions are involved, the process of regularization does
not prove to be sufficient to ensure that generic two-point functions in
real space are well behaved.
Note that, we have considered the transition from inflation to the 
post-inflationary, radiation-dominated epoch (and, later, to the 
matter-dominated epoch) to be instantaneous or abrupt.
Such instantaneous transitions can be considered to be unphysical.
Realistic transitions are expected to be smooth (to be precise, infinitely 
continuous) and, clearly, it would be interesting to investigate the effects 
of such smooth transitions on the PS of PGWs (for a recent discussion in
this context, see Ref.~\cite{Pi:2024kpw}).
In this section, we shall consider a smoother transition from inflation to 
the epoch of radiation domination and examine the corresponding behavior of 
the PS of PGWs.

\begin{figure}[!t]
\centering
\includegraphics[width=1.0\textwidth]{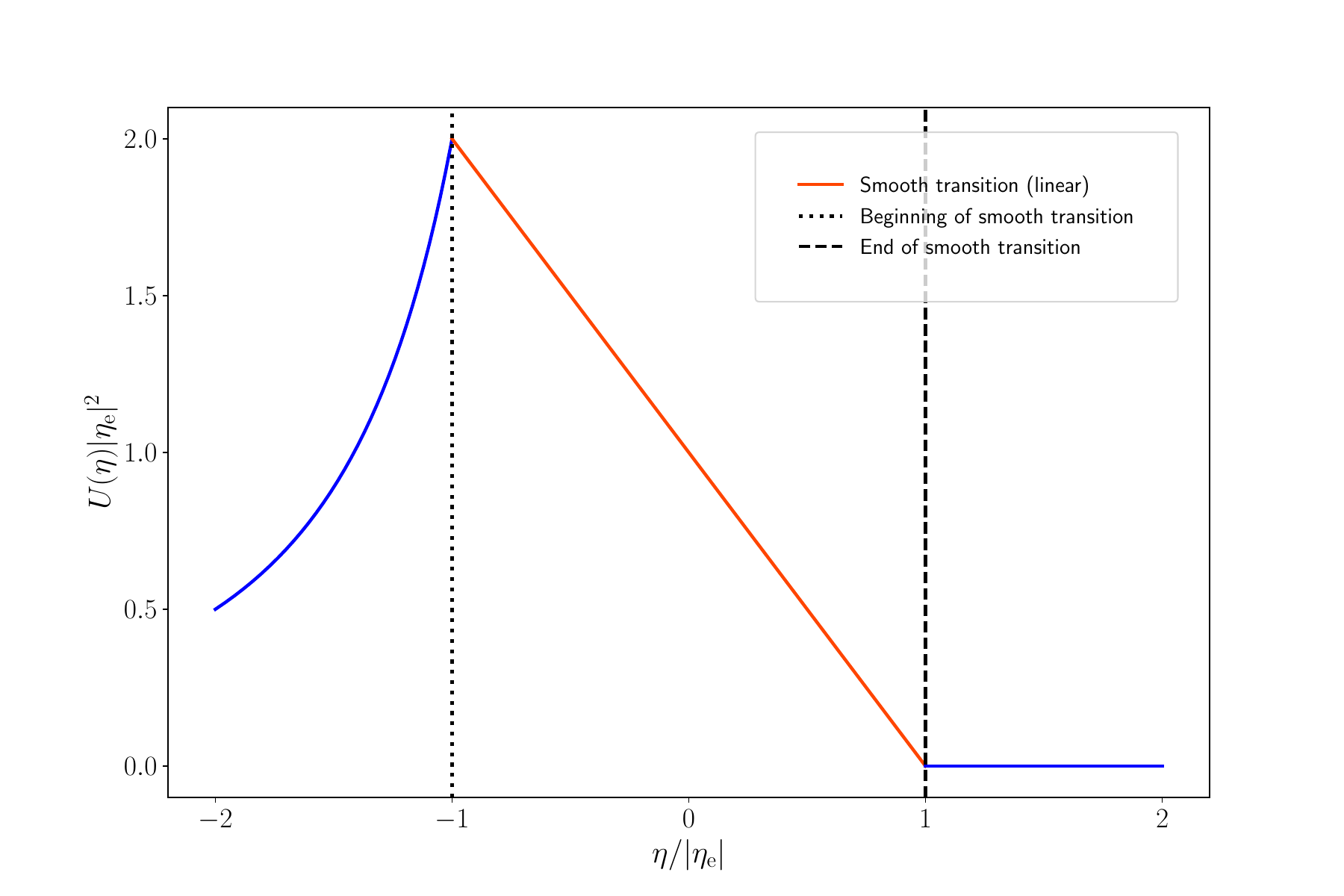}
\caption{The behavior of the `effective potential' $U(\eta)=a''/a$ has been 
plotted in the case of transition from de Sitter inflation (corresponding to $q=-1$)
to the epoch of radiation domination (corresponding to $w=1/3$), when 
$U(\eta)$ has been smoothed with a linear function [cf. Eq.~\eqref{eq:U-slt}].
Note that we have set $\Delta\eta=2\vert\ee\vert$ in plotting the figure.
We have also indicated the beginning and the end of the smooth
transition (as vertical dotted and dashed lines, respectively).}
\label{fig:U} 
\end{figure}

Instead of considering an instantaneous transition from inflation to the
epoch of radiation domination, let us introduce a smoother transition
which interpolates between these two phases of evolution of the universe. 
As we discussed, during inflation and the post-inflationary phase, the 
scale factors are given by Eqs.~\eqref{eq:a-pli} and~\eqref{eq:scalefactorpl}.
Soon, we will specialize our consideration to the case of~$w=1/3$ during the
post-inflationary phase, which corresponds to the radiation-dominated universe.
But, for the moment, let us present equations that are valid for any value
of the equation-of-state parameter~$w$.
Recall that inflation ends at~$\ee$.
Let us assume that the transition occurs over the time period $\Delta\eta$.
If we interpolate between the inflation and the post-inflationary phase
using a linear function for $U(\eta)=a''/a$, we obtain that
\begin{align}\label{eq:U-slt}
U(\eta)
=\l\{\begin{array}{ll}
\!\f{q\,(q-1)}{\eta^2}, &\mbox{for}~\eta \leq \ee, \\
\!\f{2}{\ee^2\,\Delta\eta}\!
\l\{\l[\f{\gamma(w)}{(1-\Delta\eta/\vert\ee\vert+\eta_w/\vert\ee\vert)^2}
-\f{q\,(q-1)}{2}\r](\eta-\ee)+\f{q\,(q-1)}{2}\,\Delta\eta\r\}, 
& \mbox{for}~\ee \leq \eta\leq \ee+\Delta\eta,\\
\!\f{2\,\gamma(w)}{(\eta-\eta_w)^2}, &\mbox{for}~\eta \geq \ee+\Delta\eta,
\end{array}\r.
\end{align}
where $\gamma(w)$ is given by
\begin{align}
\gamma(w)=\f{(1-3w)}{(1+3w)^2}.
\end{align}
It is straightforward to check that the quantity $U(\eta)$ is continuous 
at both $\ee$ and $\ee+\Delta\eta$.
We have plotted the $U(\eta)$ above for the case of transition from de Sitter
inflation to the epoch of radiation domination in Fig.~\ref{fig:U}.

Evidently, while we know the forms of the scale factor during inflation and the 
phase dominated by the fluid or field with the equation-of-state parameter~$w$, 
we do not know it during the intermediate transitory phase. 
However, since $U(\eta)=a''/a$, postulating a $U(\eta)$ actually corresponds
to specifying a scale factor. 
Let us determine the scale factor during the transition. 
The differential equation satisfied by the scale factor is given by
\begin{equation}
a''-U(\eta)a=0.
\end{equation}
which, for the above choice of $U(\eta)$, assumes the form
\begin{equation}
a''-\f{2}{\ee^2\,\Delta\eta}
\l\{\l[\f{\gamma(w)}{(1-\Delta\eta/\vert\ee\vert+\eta_w/\vert\ee\vert)^2}
-\f{q\,(q-1)}{2}\r](\eta-\ee)+\f{q\,(q-1)}{2}\,\Delta\eta\r\}\,a =0.
\end{equation}
If we define $\rho(\eta)=\alpha\, U(\eta)$, where $\alpha $ is a constant to 
be determined, then the above differential equation for $a(\eta)$ can be 
written as
\begin{align}
\alpha^3 \left(\frac{\mathrm{d}U}{\mathrm{d}\eta}\right)^2
\frac{\mathrm{d}^2 a}{\mathrm{d}\rho^2}-\rho a=0.
\end{align}
Therefore, if we choose $\alpha$ such that $\alpha^3 (\d U/\d\eta)^2=1$, 
i.e. $\alpha =\left \vert \mathrm{d}U/\mathrm{d}\eta\right \vert^{-2/3}$, 
then we find that the scale factor obeys the Airy differential equation. 
Notice that
\begin{align}
\left \vert
\frac{\mathrm{d}U}{\mathrm{d}\eta}\right \vert
=\f{1}{\vert\ee\vert^3} X(\eta_\mathrm{e},\Delta \eta,\eta_w),
\end{align}
where we have defined the quantity~$X$ to be
\begin{align}
X(\eta_\mathrm{e},\Delta \eta,\eta_w)
\equiv \left(\frac{2\vert \ee\vert}{\Delta \eta}{}\right)
\left \vert \frac{\gamma(w)}{(1-\Delta \eta/
\vert \eta_\mathrm{e}\vert+\eta_w/\vert \eta_\mathrm{e}\vert)^2}
-\frac{q(q-1)}{2}\right\vert  \label{eq:defbigX}
\end{align}
which, a priori, is a factor of order unity. 
As a result, the quantity $\alpha$ can be written as 
\begin{equation}
\alpha=\vert\ee\vert^2 X^{-2/3}.\label{eq:alpha}
\end{equation}
It follows from the above considerations that the solution for the scale 
factor can be written in terms of Airy functions $\mathrm{Ai}(x)$ and 
$\mathrm{Bi}(x)$ as follows:
\begin{align}
\frac{a(\eta)}{a_\mathrm{e}}&=A_1\, u(\eta)+A_2\, v(\eta),
\end{align}
where $u(\eta)=\mathrm{Ai}\left[\rho(\eta)\right]$,
$v(\eta)=\mathrm{Bi}\left[\rho(\eta)\right]$ and, recall that, $a_\mathrm{e}
=a(\eta_\mathrm{e})$.
It is useful to note that the above solution also implies that the 
conformal Hubble parameter~$\mathcal{H}$ during the transition is 
given by
\begin{align}
{\cal H}(\eta)
=\frac{\mathrm{d}\rho}{\mathrm{d}\eta}
\frac{A_1 (\mathrm{d}u/\mathrm{d}\rho)
+A_2 (\mathrm{d}v/\mathrm{d}\rho)}{A_1 u +A_2 v}.
\end{align}
The two constants $A_1$ and $A_2$ are to be determined by demanding that
the scale factor and its first time derivative are continuous at the end 
of inflation, i.e. at $\eta_\mathrm{e}$.
These conditions lead to
\begin{subequations}
\begin{align}
A_1 &=\pi \left[\frac{\mathrm{d}v}{\mathrm{d}\rho}\biggl \vert_{\ee}
-\left(\frac{\mathrm{d}\rho}{\mathrm{d}\eta}\right)^{-1}
{\cal H}_\mathrm{e}v(\eta_\mathrm{e})\right],\\
A_2 &=-\pi \left[\frac{\mathrm{d}u}{\mathrm{d}\rho}\biggl\vert_{\ee}
-\left(\frac{\mathrm{d}\rho}{\mathrm{d}\eta}\right)^{-1}
{\cal H}_\mathrm{e}u(\eta_\mathrm{e})\right],
\end{align}
\end{subequations}
where $\mathcal{H}_{\e}=\mathcal{H}(\ee)$.
We should mention that, in arriving at these expressions for $A_1$ and 
$A_2$, we have made use of the following Wronskian 
condition~\cite{Gradshteyn:2007}:
\begin{equation}
\Ai(x) \f{\d\Bi(x)}{\d x}-\f{\d \Ai(x)}{\d x} \Bi(x)=\f{1}{\pi}.
\label{eq:WAfns}
\end{equation}
Note that the factor $\mathrm{d}\rho/\mathrm{d}\eta$ can be expressed as
\begin{align}
\label{eq:dzdeta}
\frac{\mathrm{d}\rho}{\mathrm{d}\eta}
=\biggl \vert \frac{\mathrm{d}U}{\mathrm{d}\eta}
\biggr\vert ^{-2/3}\frac{\mathrm{d}U}{\mathrm{d}\eta}=\varepsilon
\biggl \vert \frac{\mathrm{d}U}{\mathrm{d}\eta}
\biggr\vert ^{1/3}
=\varepsilon \vert \ee\vert^{-1} X^{1/3},
\end{align}
where $\varepsilon$ is given by
\begin{equation}
\varepsilon
= \mathrm{sgn}\l[\f{\gamma(w)}{(1-\Delta\eta/\vert\ee\vert
+\eta_w/\vert\ee\vert)^2} -\f{q\,(q-1)}{2}\r].
\label{eq:varepsilon}
\end{equation}
In fact, $\varepsilon$ depends on whether
$\mathrm{d}U/\mathrm{d}\eta>0$ ($\varepsilon=1$) or
$\mathrm{d}U/\mathrm{d}\eta<0$ ($\varepsilon=-1$). 
Indeed, the sign of $\varepsilon$ in the space $(q,w,\ee,\Delta \eta)$ 
is rather non-trivial. 
Moreover, the coefficients $A_1$ and $A_2$ involve the functions $u(\eta)$ 
and $v(\eta)$ evaluated at the end of inflation which, in turn, requires 
the calculation of $\rho(\eta_\mathrm{e})$. 
Using the relation, $\rho(\ee)=\alpha U(\ee)=q(q-1)X^{-2/3}$, we can see 
that $\rho(\ee)$ is, a priori, a quantity of order unity. 
Therefore, we cannot use the asymptotic behavior of the Airy functions. 
Explicitly, the coefficients $A_1$ and $A_2$ can be expressed as
\begin{subequations}
\begin{align}
A_1 &=\pi \left\{\mathrm{Bi}'\left[q(q-1)X^{-2/3}\right] +\varepsilon
qX^{-1/3}\, \mathrm{Bi}\left[q(q-1)X^{-2/3}\right]\right\},\\
A_2 &=-\pi \left\{\mathrm{Ai}'\left[q(q-1)X^{-2/3}\right] 
+\varepsilon qX^{-1/3}\, \mathrm{Ai}\left[q(q-1)X^{-2/3}\right]\right\}.
\end{align}
\end{subequations}
where $\mathrm{Ai}'(x)=\d \mathrm{Ai}(x)/\d x$ and $\mathrm{Bi}(x)
=\d\mathrm{Bi}(x)/\d x$, {\it i.e. the overprimes on the Airy functions
in these two equations denote differentiation with respect to the 
arguments}.

The discussion above was valid for any value of~$w$. 
However, in this work, we are mainly interested in smoothing the transition
from inflation to the epoch of radiation domination. 
For this reason, from now on, we shall set $w=1/3$. 
In such a case $\gamma(w)=0$ and $\mathrm{d}U/\mathrm{d}\eta<0$, and we have
\begin{equation}
\varepsilon = \mathrm{sgn}\l[-\f{q\,(q-1)}{2}\r]=-1.
\end{equation}
Also, we have
\begin{equation}
X=\f{q\,(q-1)\,\vert \ee\vert}{\Delta \eta}\label{eq:Xrd}
\end{equation}
and, note that, $X$ no longer depends on $\eta_w$. 
As a consequence, the coefficients $A_1$ and $A_2$ take the form
\begin{subequations}
\begin{align}
\label{eq:A1}
\frac{A_1}{\pi} &=\mathrm{Bi}'\left\{\left[q(q-1)\right]^{1/3}
\left(\frac{\Delta \eta}{\vert \eta_\mathrm{e}\vert}\right)^{2/3}\right\}\nn\\ 
&\quad-q\left[q(q-1)\right]^{-1/3}
\left(\frac{\Delta \eta}{\vert \eta_\mathrm{e}\vert}\right)^{1/3}
\mathrm{Bi}\left\{\left[q(q-1)\right]^{1/3}
\left(\frac{\Delta \eta}{\vert \eta_\mathrm{e}\vert}\right)^{2/3}\right\},\\
\label{eq:A2}
\frac{A_2}{\pi} &=-\mathrm{Ai}'\left\{\left[q(q-1)\right]^{1/3}
\left(\frac{\Delta \eta}{\vert \eta_\mathrm{e}\vert}\right)^{2/3}\right\}\nn\\ 
&\quad+q\left[q(q-1)\right]^{-1/3}
\left(\frac{\Delta \eta}{\vert \eta_\mathrm{e}\vert}\right)^{1/3}
\mathrm{Ai}\left\{\left[q(q-1)\right]^{1/3}
\left(\frac{\Delta \eta}{\vert \eta_\mathrm{e}\vert}\right)^{2/3}\right\},
\end{align}
\end{subequations}
and we should stress again that the overprimes on the Airy functions
in these two equations denote differentiation with respect to the 
arguments. 

We must now match the scale factor and the conformal Hubble parameter at
the time $\eta_\mathrm{e}+\Delta \eta$, i.e. at the end of the smoother 
transition and the beginning of the epoch of radiation domination. 
These matching conditions lead to the determination of the constants 
$a_\mathrm{r}$ and $\eta_\mathrm{r}$ appearing in the expression~\eqref{eq:a-rd}
for the scale factor during the epoch of radiation domination. 
When $\gamma(w)=0$, $\rho(\ee+\Delta \eta)=0$ and the results~\cite{Gradshteyn:2007}
\begin{subequations}
\begin{align}
\Ai(0) &=\f{1}{3^{2/3}\,\Gamma(2/3)},\\
\Bi(0) &=\f{1}{3^{1/6}\,\Gamma(2/3)},\\
\Ai'(0) &=-\f{1}{3^{1/3}\,\Gamma(1/3)},\\
\Bi'(0) &=\f{3^{1/6}}{\Gamma(1/3)},
\end{align}
\end{subequations}
where $\Gamma(x)$ is the Gamma function, can be used to arrive at the 
following expressions for~$\er$ and~$\ar$:
\begin{subequations}
\begin{align}
\frac{\eta_\mathrm{r}}{\vert \eta_\mathrm{e}\vert}
&=\frac{\Delta \eta}{\vert \eta_\mathrm{e}\vert}-1
-\left[q(q-1)\right]^{-1/3}\left(\frac{\Delta \eta}
{\vert \eta_\mathrm{e}\vert}\right)^{1/3}\frac{\Gamma(1/3)}{3^{1/3}\Gamma(2/3)}
\left(\frac{A_1+\sqrt{3} A_2}{A_1-\sqrt{3}A_2}\right),\label{eq:etaw}\\
a_\mathrm{r} 
&=\frac{\left[q(q-1)\right]^{1/3}}{3^{1/3}\Gamma(1/3)
\vert \eta_\mathrm{e}\vert}
\left(\frac{\Delta \eta}{\vert \eta_\mathrm{e}\vert}\right)^{-1/3}
\left(A_1-\sqrt{3} A_2\right)a_\mathrm{e}.\label{eq:aw}
\end{align}
\end{subequations}
These two quantities depend only on the three parameters:~$q$,
$\eta_\mathrm{e}$ and $\Delta \eta/\vert \eta_\mathrm{e}\vert$.

It is also interesting to examine the manner in which the case of the 
instantaneous transition is recovered from the above expressions for 
$a_\mathrm{r}$ and $\eta_\mathrm{r}$. 
Clearly, the instantaneous transition corresponds to the limit $\Delta \eta 
\rightarrow 0$, while $\eta_\mathrm{e}$ remains finite. 
In such a limit, we find that the constants $A_1$ and $A_2$ simplify to
\begin{subequations}
\label{eq:lit-one}
\begin{align}
A_1 &\simeq\f{3^{1/6} \pi}{\Gamma(1/3)}
-\f{\pi q \Delta\eta^{1/3}}{3^{1/6} \Gamma(2/3) [q (q-1)]^{1/3} 
\vert\ee\vert^{1/3}},\\
\sqrt{3}\,A_2 
&\simeq \f{3^{1/6}\,\pi}{\Gamma(1/3)}
+\f{\pi q \Delta\eta^{1/3}}{3^{1/6} \Gamma(2/3) [q (q-1)]^{1/3}
\vert\ee\vert^{1/3}},
\end{align}
\end{subequations}
and, hence,
\begin{subequations}
\label{eq:lit-two}
\begin{align}
A_1+\sqrt{3} A_2
&\simeq \f{2\, 3^{1/6} \pi}{\Gamma(1/3)},\\
A_1-\sqrt{3} A_2
&\simeq -\f{2 \pi q \Delta\eta^{1/3}}{3^{1/6} \Gamma(2/3) [q(q-1)]^{1/3}
\vert\ee\vert^{1/3}}.
\end{align}
\end{subequations}
Since $\Gamma(1/3) \Gamma(2/3)=2 \pi/\sqrt{3}$, the above expressions immediately 
lead to $\ar=q \ae/\ee$ and $\er=(q-1)\ee/q$, as we had seen earlier [see our 
discussion following Eq.~\eqref{eq:a-rd}]. 


\section{Evolution of PGWs across the smoother transition}\label{sec:i-rd-st-p}

Having obtained the solution for the background across the smoother
linear transition, let us now construct the solution for the tensor 
mode functions.


\subsection{Evolution during the smoother transition}

We are now in a position to solve Eq.~\eqref{eq:mse} 
for the rescaled mode function~$\mu_k(\eta)$. 
The solutions during inflation and the radiation-dominated era 
are already known and, hence, let us focus on the solution during
the transition. 
If we define 
\begin{equation}
\tau(\eta)=-\alpha\left[k^2-U(\eta)\right]=-\alpha k^2 +\rho(\eta),
\end{equation}
then the equation governing the rescaled mode function during the 
transition takes the form
\begin{align}
\frac{\mathrm{d}^2\mu_{k}}{\mathrm{d}\tau ^2}
-\tau \mu_{k}=0,
\end{align}
which can be recognized again as the Airy differential equation. 
As a result, the general solution for the rescaled mode function 
can be expressed as~\cite{Gradshteyn:2007}
\begin{align}
\label{eq:modetransition}
\mu_{ k}(\eta)=B_1 i_{k}(\eta)+B_2 j_{k}(\eta)
=B_1 \mathrm{Ai}[\tau(\eta)]
+ B_2 \mathrm{Bi}[\tau(\eta)],
\end{align}
where $\mathrm{Ai}(x)$ and $\mathrm{Bi}(x)$ are the Airy functions. 
The two constants $B_1$ and $B_2$ are to be determined by
requiring the continuity of the rescaled mode function $\mu_k(\eta)$
and its time derivative at the end of inflation, i.e. at $\ee$. 
The constants can be expressed as
\begin{subequations}
\begin{align}
B_{1} &=\frac{A_{k}}{W_k^{\mathrm{slt}}}\left[f_{k}(\eta_\mathrm{e})
\frac{\mathrm{d}j_{k}}{\mathrm{d}\eta}\biggl \vert_{\eta_\mathrm{e}}
-\frac{\mathrm{d}f_{k}}{\mathrm{d}\eta}\biggl \vert_{\eta_\mathrm{e}}
j_{k}(\eta_{\mathrm{e}})\right]\nn\\ 
&=\frac{A_{k}}{W_k^{\mathrm{slt}}}\frac{\mathrm{d}\rho}{\mathrm{d}\eta}
\left[f_{k}(\eta_\mathrm{e})
\frac{\mathrm{d}j_{k}}{\mathrm{d}\tau}\biggl \vert_{\eta_\mathrm{e}}
+k\frac{\mathrm{d}f_{k}}{\mathrm{d}y} \biggl \vert_{\eta_\mathrm{e}}
\left(\frac{\mathrm{d}\rho}{\mathrm{d}\eta}\right)^{-1}
j_{k}(\eta_{\mathrm{e}})\right],\\
B_2 &=-\frac{A_{k}}{W_k^{\mathrm{slt}}}\left[f_{k}(\eta_\mathrm{e})
\frac{\mathrm{d}i_{ k}}{\mathrm{d}\eta}\biggl \vert_{\eta_\mathrm{e}}
-\frac{\mathrm{d}f_{k}}{\mathrm{d}\eta}\biggl \vert_{\eta_\mathrm{e}}
i_{k}(\eta_{\mathrm{e}})\right]\nn\\ 
&=-\frac{A_{k}}{W_k^{\mathrm{slt}}}\frac{\mathrm{d}\rho}{\mathrm{d}\eta}
\left[f_{k}(\eta_\mathrm{e})
\frac{\mathrm{d}i_{k}}{\mathrm{d}\tau}\biggl \vert_{\eta_\mathrm{e}}
+k\frac{\mathrm{d}f_{k}}{\mathrm{d}y}\biggl \vert_{\eta_\mathrm{e}}
\left(\frac{\mathrm{d}\rho}{\mathrm{d}\eta}\right)^{-1}
i_{k}(\eta_{\mathrm{e}})\right],
\end{align}
\end{subequations}
where $f_{k}(\eta)$ is the mode function during inflation [cf. 
Eqs.~\eqref{eq:infmodefunction} and~\eqref{eq:fk}], and we have 
made use of the fact that $\mathrm{d}\tau/\mathrm{d}\eta=\mathrm{d}\rho/\mathrm{d}\eta$,
since $\tau(\eta)$ and $\rho(\eta)$ differ only by a constant.
The quantity $W_k^{\mathrm{slt}}$ is the Wronskian given by
\begin{align}
\label{eq:Wronskiantransition}
W_k^{\mathrm{slt}}=i_k\f{\d j_k}{\d\eta}
-j_k\frac{\mathrm{d}i_k}{\mathrm{d}\eta}
=\frac{1}{\pi} \frac{\mathrm{d}\tau}{\mathrm{d}\eta}
=\frac{\alpha}{\pi} \frac{\mathrm{d}U}{\mathrm{d}\eta}=\frac{1}{\pi}
\left \vert \frac{\mathrm{d}U}{\mathrm{d}\eta}\right\vert^{-2/3}
\frac{\mathrm{d}U}{\mathrm{d}\eta}=-\frac{1}{\pi}
\left \vert \frac{\mathrm{d}U}{\mathrm{d}\eta}\right\vert^{1/3},
\end{align}
where we have made use of the Wronskian condition~\eqref{eq:WAfns}
for the Airy functions.

To proceed further and establish the detailed form of the
coefficients $B_1$ and $B_2$, we need the exact forms of 
the mode functions $f_{k}(\eta)$, $i_{k}(\eta)$ 
and~$j_{ k}(\eta)$.
Recall that the rescaled mode function during inflation 
can be expressed in terms of the Hankel functions [cf. 
Eqs.~\eqref{eq:infmodefunction} and~\eqref{eq:fk}].
It proves to be convenient to rewrite the mode function 
$f_k(\eta)$ as
\begin{align}
\label{eq:modeT}
f_{k}(\eta)=\sqrt{\frac{2}{\pi}} \e^{i\left[y-\pi(1-q)/2\right]}
T_{1/2-q}(y),
\end{align}
where the function $T_{\mu }(y)$ is defined by the following expression:
\begin{align}
T_{\mu}(y)=\sum_{m=0}^{\infty} \frac{(-1)^m}{m!}
\frac{\Gamma(1/2+\mu+m)}{\Gamma(1/2+\mu-m)}
\f{1}{(2iy)^m},
\end{align}
and we recall that $y=-k\eta >0$. 
As a consequence, we can also write that 
\begin{equation}
\f{\mathrm{d}f_{k}}{\mathrm{d}\eta}
=-k\f{\mathrm{d}f_{ k}}{\mathrm{d}y} 
=-k \l[i+T'_{1/2-q}(y)/T_{1/2-q}(y)\r]f_{k},
\end{equation}
{\it where $T'_{1/2-q}(y)$ denotes the derivative of the function 
$T_{1/2-q}(y)$ with respect to its argument}.\/ 
Note that the mode functions $i_{k}(\eta)$ and $j_{k}(\eta)$ are 
described by the Airy functions with the argument
$\tau(\eta)$ [see Eq.~\eqref{eq:modetransition}].  
The quantity $\tau(\ee)\equiv\tau_{\e}$ can be expressed as 
\begin{equation}
\tau_\mathrm{e}=-\alpha
\left[k^2-U(\eta_\mathrm{e})\right] =-\alpha
k^2+\rho(\eta_\mathrm{e})=-y_\mathrm{e}^2
X^{-2/3}\l[1-\f{q(q-1)}{y_\mathrm{e}^2}\r]\label{eq:tau-e}
\end{equation}
and, in the limit $y_{\e}\gg 1$, $\tau_\mathrm{e}$ is a large, 
negative quantity. 
As a consequence, it is convenient to use the following exact expressions 
for the Airy functions~\cite{Abramowitz:2008}:
\begin{subequations}\label{eq:afns}
\begin{align}
\label{eq:ai}
\mathrm{Ai}(s)
&=\frac{(-s)^{-1/4}}{\sqrt{\pi}}
\left[A(s)\sin\left(\zeta+\frac{\pi}{4}\right)
-B(s)\cos \left(\zeta+\frac{\pi}{4}\right)\right],\\
\label{eq:aip}
\mathrm{Ai}'(s)
&=\frac{(-s)^{1/4}}{\sqrt{\pi}}\left[-C(s)\cos\left(\zeta+
\frac{\pi}{4}\right)-D(s)\sin \left(\zeta+\frac{\pi}{4}\right)\right],\\
\label{eq:bi}
\mathrm{Bi}(s)
&=\frac{(-s)^{-1/4}}{\sqrt{\pi}}
\left[A(s)\cos\left(\zeta+\frac{\pi}{4}\right)
+B(s)\sin \left(\zeta+\frac{\pi}{4}\right)\right],\\
\label{eq:bip}
\mathrm{Bi}'(s)
&=\frac{(-s)^{1/4}}{\sqrt{\pi}}\left[C(s)\sin\left(\zeta+
\frac{\pi}{4}\right)-D(s)\cos \left(\zeta+\frac{\pi}{4}\right)\right],
\end{align}    
\end{subequations}
where $\zeta=(2/3)(-s)^{3/2}$, which is valid for negative arguments, 
i.e. when $s<0$. 
The four functions $A(s)$, $B(s)$, $C(s)$ and $D(s)$ are given 
by~\cite{Abramowitz:2008}
\begin{subequations}
\begin{align}
A(s) &= \sum_{m=0}^{\infty}(-1)^m \f{c_{2m}}{\zeta^{2m}},\\
B(s) &= \sum_{m=0}^{\infty}(-1)^m \f{c_{2m+1}}{\zeta^{2m+1}},\\
C(s)&= \sum_{m=0}^{\infty}(-1)^m \f{d_{2m}}{\zeta^{2m}},\\
D(s) &= \sum_{m=0}^{\infty}(-1)^m\, \f{d_{2m+1}}{\zeta^{2m+1}},
\end{align}
\end{subequations}
where the coefficients $c_m$ and $d_m$ are given by
\begin{subequations}
\begin{align}
c_m &=\frac{\Gamma(3m+1/2)}{54^m m! \, \Gamma(m+1/2)},\\
d_m &=-\frac{6m+1}{6m-1}c_m.
\end{align}    
\end{subequations}

On substituting these formulae into the expression for $B_1$ derived 
above, we obtain the following, exact result:
\begin{align}
B_1 &=\frac{A_{ k}}{W_{\mathrm{slt}}}\frac{\mathrm{d}\rho}{\mathrm{d}\eta}
\left\{f_{ k}(\eta_\mathrm{e})
\frac{\mathrm{d}j_{ k}}{\mathrm{d}\tau}\biggl\vert_{\eta_\mathrm{e}}
+k\left[i+\frac{T'_{1/2-q}(y_\mathrm{e})}{T_{1/2-q}(y_\mathrm{e})}\right]
f_{ k}(\eta_\mathrm{e})
\left(\frac{\mathrm{d}\rho}{\mathrm{d}\eta}\right)^{-1}
j_{ k}(\eta_{\mathrm{e}})\right\}\nn\\
&=\frac{A_{k}}{W_{\mathrm{slt}}}\frac{\mathrm{d}\rho}{\mathrm{d}\eta}
\frac{1}{\sqrt{\pi}}f_{ k}(\eta_\mathrm{e})
(-\tau_\mathrm{e})^{1/4}
\Biggl\{C(\tau_\mathrm{e})\sin\left(\zeta_\mathrm{e}
+ \frac{\pi}{4}\right)-D(\tau_\mathrm{e})
\cos \left(\zeta_\mathrm{e}+\frac{\pi}{4}\right)\nn\\ 
&\quad+k\left[i+\frac{T'_{1/2-q}(y_\mathrm{e})}{T_{1/2-q}(y_\mathrm{e})}\right]
\left(\frac{\mathrm{d}\rho}{\mathrm{d}\eta}\right)^{-1}
(-\tau_\mathrm{e})^{-1/2}
\left[A(\tau_\mathrm{e})\cos\left(\zeta_\mathrm{e}
+ \frac{\pi}{4}\right)+B(\tau_\mathrm{e})
\sin \left(\zeta_\mathrm{e}+\frac{\pi}{4}\right)\right]\Biggr\},
\end{align}
where we have used the expressions for the Airy functions given above. 
Then, if we use Eq.~\eqref{eq:dzdeta} and the expression for
$\tau_\mathrm{e}$ [see Eq.~\eqref{eq:tau-e}], it is easy to establish that 
\begin{equation}
\left(\f{\mathrm{d}\rho}{\mathrm{d}\eta}\right)^{-1}
(-\tau_\mathrm{e})^{-1/2}
=\f{\varepsilon}{k} \l[1-\f{q(q-1)}{y_\mathrm{e}^2}\r]^{-1/2},
\end{equation}
where $\varepsilon=\pm 1$ depending on the value of the equation-of-state
parameter $w$ in the subsequent epoch [cf. Eq.~\eqref{eq:varepsilon}].
(Recall that $\varepsilon=-1$ in the case of transition
to a radiation-dominated epoch.)
Finally, on using the expressions~\eqref{eq:Wronskiantransition} for
the Wronskian and the inflationary mode function~\eqref{eq:modeT}, we 
find that the coefficient $B_1$ can be written as
\begin{align}
\label{eq:exactB1}
B_1 &=A_{ k}\sqrt{2}\,
\e^{i\left[y_\mathrm{e}-\pi(1-q)/2\right]}
y_\mathrm{e}^{1/2}X^{-1/6}
\left[{\cal P}_{\mathrm{s}1}
\sin \left(\zeta_\mathrm{e}+\frac{\pi}{4}\right)
+{\cal P}_{\mathrm{c}1}
\cos \left(\zeta_\mathrm{e}+\frac{\pi}{4}\right)\right],
\end{align}
where the quantities $({\cal P}_{\mathrm{s}1},{\cal P}_{\mathrm{c}1})$
are given by
\begin{subequations}
\begin{align}
{\cal P}_{\mathrm{s}1} 
&=\varepsilon 
\left[iT_{1/2-q}(y_\mathrm{e})+T'_{1/2-q}(y_\mathrm{e})\right]
\left[1-\frac{q(q-1)}{y_\mathrm{e}^2}\right]^{-1/4}
B(\tau_\mathrm{e})\nn\\ 
&\quad +T_{1/2-q}(y_\mathrm{e})
\left[1-\frac{q(q-1)}{y_\mathrm{e}^2}\right]^{1/4}C(\tau_\mathrm{e}),\\ 
{\cal P}_{\mathrm{c}1} 
&= \varepsilon \left[iT_{1/2-q}(y_\mathrm{e})+T'_{1/2-q}(y_\mathrm{e})\right]
\left[1-\frac{q(q-1)}{y_\mathrm{e}^2}\right]^{-1/4}
A(\tau_\mathrm{e})\nn\\ 
&\quad-T_{1/2-q}(y_\mathrm{e})
\left[1-\frac{q(q-1)}{y_\mathrm{e}^2}\right]^{1/4}D(\tau_\mathrm{e}),
\end{align}
\end{subequations}
We should stress that that the result~\eqref{eq:exactB1} for $B_1$ is exact. 
Since
\begin{equation}
\zeta_\mathrm{e}=\f{2y_\mathrm{e}^3}{3X}
\l[1-\f{q(q-1)}{y_\mathrm{e}^2}\r]^{3/2},
\end{equation}
the quantity $B_1/A_{k}$ is in fact a function only of~$y_\mathrm{e}$. 
Moreover, note that the coefficients ${\cal P}_{\mathrm{c}1}$ and 
${\cal P}_{\mathrm{s}1}$, which multiply the cosine and sine functions, 
involve sums of negative powers of~$y_\mathrm{e}$.
Hence, they are easy to calculate at any order in the inverse powers of 
$y_\mathrm{e}$, which is exactly what is needed in order to evaluate $B_1$ 
in the small scale limit. 
We find that, in the limit $y_{\e}\gg 1$, the 
quantities~$({\cal P}_{\mathrm{s}1},{\cal P}_{\mathrm{c}1})$ can be
expressed as
\begin{subequations}
\begin{align}
{\cal P}_{\mathrm{s}1}
&=1+\f{iq(q-1)}{2y_\mathrm{e}}
-\f{q^2(q-1)^2}{8y_\mathrm{e}^2}\nn\\
&\quad+\f{i}{48 y_{\e}^3} 
\l[12 q-10 q^2- q^3 (3+q-3 q^2-q^3)+ 5 \varepsilon X\r]
+{\cal O}\l(\frac{1}{y_\mathrm{e}^4}\r),\\
{\cal P}_{\mathrm{c}1} &= i\varepsilon
-\frac{\varepsilon q(q-1)}{2y_\mathrm{e}}
-\frac{i \varepsilon q^2(q-1)^2}{8y_\mathrm{e}^2}\nn\\
&\quad-\frac{1}{48y_\mathrm{e}^3}
\l\{\l[12 q -14 q^2 +q^3 (3+q-3 q^2+q^3)\r]\varepsilon +7X\r\}
+{\cal O}\left(\frac{1}{y_\mathrm{e}^4}\right).
\end{align}
\end{subequations}
In these expressions, we have expanded the quantities~$({\cal P}_{\mathrm{s}1},
{\cal P}_{\mathrm{c}1})$ up to order~$y_\mathrm{e}^{-3}$ to show that the 
dependence on the details of the smoother transition only appear (through 
the parameter~$X$) at this order.

Of course, the same steps and considerations can be followed to derive
the expression of the coefficient $B_2$. 
We find that $B_2$ can be written as
\begin{align}
\label{eq:exactB2}
B_{2} &=A_{ k}\sqrt{2}\,
\e^{i\left[y_\mathrm{e}-\pi(1-q)/2\right]}
y_\mathrm{e}^{1/2}X^{-1/6}
\left[{\cal P}_{\mathrm{s}2}
\sin \left(\zeta_\mathrm{e}+\frac{\pi}{4}\right)
+{\cal P}_{\mathrm{c}2}
\cos \left(\zeta_\mathrm{e}+\frac{\pi}{4}\right)\right],
\end{align}
where $({\cal P}_{\mathrm{s}2},{\cal P}_{\mathrm{c}2})$
are given by
\begin{subequations}
\begin{align}
{\cal P}_{\mathrm{s}2} 
&=-\varepsilon \l[iT_{1/2-q}(y_\mathrm{e})+T'_{1/2-q}(y_\mathrm{e})\r]
\l[1-\frac{q(q-1)}{y_\mathrm{e}^2}\r]^{-1/4} A(\tau_\mathrm{e})\nn\\ 
&\quad+T_{1/2-q}(y_\mathrm{e})
\left[1-\frac{q(q-1)}{y_\mathrm{e}^2}\right]^{1/4}D(\tau_\mathrm{e}),\\
{\cal P}_{\mathrm{c}2} 
&=\varepsilon \left[iT_{1/2-q}(y_\mathrm{e})+T'_{1/2-q}(y_\mathrm{e})\right]
\left[1-\frac{q(q-1)}{y_\mathrm{e}^2}\right]^{-1/4} B(\tau_\mathrm{e})\nn\\ 
&\quad+T_{1/2-q}(y_\mathrm{e})
\left[1-\frac{q(q-1)}{y_\mathrm{e}^2}\right]^{1/4}C(\tau_\mathrm{e}).
\end{align}
\end{subequations}
In the limit $y_{\e}\gg 1$, the 
quantities~$({\cal P}_{\mathrm{s}2},{\cal P}_{\mathrm{c}2})$ can be
expressed as
\begin{subequations}
\begin{align}
{\cal P}_{\mathrm{s}2}&
=-i\varepsilon
+\f{\varepsilon q(q-1)}{2y_\mathrm{e}}
+\f{i \varepsilon q^2(q-1)^2}{8y_\mathrm{e}^2}\nn\\
&\quad-\f{1}{48y_\mathrm{e}^3}
\l\{\l[12 q -14 q^2 +q^3 (3+q-3 q^2+q^3)\r]\varepsilon +7X\r\}
+{\cal O}\left(\frac{1}{y_\mathrm{e}^4}\right),\\
{\cal P}_{\mathrm{c}2} 
&=1+\f{iq(q-1)}{2y_\mathrm{e}}
-\f{q^2(q-1)^2}{8y_\mathrm{e}^2}\nn\\
&\quad+\f{i}{48 y_{\e}^3}
\l[12 q -10 q^2 -q^3 (3+q-3 q^2+q^3) +5\varepsilon X\r]
+{\cal O}\l(\frac{1}{y_\mathrm{e}^4}\r).
\end{align}
\end{subequations}
Thus, we have completely determined the coefficients~$(B_1,B_2)$ during 
the transition. 
The next step consists of matching the solution for the recsaled mode
function~$\mu_k(\eta)$ during the transition to the solution during the 
radiation-dominated epoch.
Such a matching would lead to explicit expressions for the coefficients
$(\alpha^\mathrm{r}_{k},\beta^\mathrm{r}_{k})$. 
We shall now turn to the calculation.


\subsection{Evolution and PS after the smoother transition}

During the radiation-dominated era, the rescaled mode function~$\mu_k(\eta)$
is given by Eq.~\eqref{eq:modefunctionrad}, with the functions $m_k(\eta)$
and $n_k(\eta)$ being given by Eq.~\eqref{eq:mn-rd}. 
The coefficients $(\alpha^\mathrm{r}_k,\beta^\mathrm{r}_k)$ are obtained 
by matching the mode function and its derivative during the smoother linear
transition [cf. Eq.~\eqref{eq:modetransition}] with the mode function and 
its time derivative during the radiation-dominated era. 
The matching has to be carried out at the end of the period of the transition,
viz. at $\eta_\mathrm{tr}=\eta_\mathrm{e}+\Delta \eta$. 
On carrying out the matching, we find that the coefficients 
$(\alpha^\mathrm{r}_k,\beta^\mathrm{r}_k)$ can be written as
\begin{align}
\label{eq:matrix}
\begin{pmatrix}
\alpha^\mathrm{r}_k \\
\beta^\mathrm{r}_k
\end{pmatrix}
=\frac{1}{W_k^\mathrm{r}}
\begin{pmatrix}
T_{11} & T_{12}\\
T_{21} & T_{22}
\end{pmatrix}
\begin{pmatrix}
B_1 \\
B_2
\end{pmatrix},
\end{align}
where the matrix elements $(T_{11},T_{12},T_{21},T_{22})$ are given by
\begin{subequations}
\begin{align}
T_{11} &= n_{k}'(\etr) i_k(\etr)
-n_{k}(\etr) \f{\d\tau}{\d \eta}
\f{\d i_k}{\d\tau}\biggl\vert_{\etr},\\
T_{12} &= n_{k}'(\etr) j_k(\etr)
-n_{k}(\etr) \f{\d\tau}{\d \eta}
\f{\d j_k}{\d\tau}\biggl\vert_{\etr},\\
T_{21} &= -m_{k}'(\etr) i_k(\etr)
+m_{k}(\etr) \f{\d\tau}{\d \eta}
\f{\d i_k}{\d\tau}\biggl\vert_{\etr},\\
 T_{22} &= -m_{k}'(\etr) j_k(\etr)
 +m_{k}(\etr) \f{\d\tau}{\d \eta}
\f{\d j_k}{\d\tau}\biggl\vert_{\etr},
\end{align}
\end{subequations}
and $W_k^\mathrm{r}=4ik/\pi$ is the Wronskian of the functions~$m_{k}(\eta)$
and $n_{k}(\eta)$ [see Eq.~\eqref{eq:wkr} and the comment following it].
To proceed further, we make use of the 
results~$\mathrm{d}n_{ k}/\mathrm{d}\eta=ikn_{k}$ 
and $\mathrm{d}i_{ k}/\mathrm{d}\eta= (\mathrm{d}\tau/\mathrm{d}\eta)
\mathrm{d}i_{ k}/\mathrm{d}\tau =(\mathrm{d}\rho/\mathrm{d}\eta)
\mathrm{d}i_{ k}/\mathrm{d}\tau$. 
Together with Eq.~\eqref{eq:dzdeta}, these lead to the following 
expression for $T_{11}$:
\begin{align}
T_{11}=ik\, n_{k}(\eta_\mathrm{tr}) 
\l[i_{k}(\eta_\mathrm{tr})-\f{i X^{1/3}}{y_\mathrm{e}}
\f{\mathrm{d}i_{ k}}{\mathrm{d}\tau}\biggl\vert_{\eta_\mathrm{tr}}\r].
\end{align}
Note that the function $n_{k}(\eta)$ has to be evaluated at
\begin{align}
k(\eta_\mathrm{tr}-\eta_\mathrm{r})
=k(\eta_\mathrm{e}+\Delta
\eta-\eta_\mathrm{r}) =k\eta_\mathrm{e}
\left(1-\frac{\Delta\eta}{\vert \eta_\mathrm{e}\vert} 
+\frac{\eta_\mathrm{r}}{\vert \eta_\mathrm{e}\vert}\right)
=-y_{\e} Y,\label{eq:etatr}
\end{align}
where $Y$ is given by  [cf.~Eq.~\eqref{eq:etaw}]
\begin{equation}
Y=-\l[q(q-1)\r]^{-1/3}\l(\f{\Delta \eta}{\vert \eta_\mathrm{e}\vert}\r)^{1/3}
\f{\Gamma(1/3)}{3^{1/3}\Gamma(2/3)}
\l(\frac{A_1+\sqrt{3} A_2}{A_1-\sqrt{3}A_2}\r).\label{eq:defbigY}
\end{equation}
On the other hand, the Airy function appearing in $i_{k}(\eta)$ has
to be calculated at the following value of the argument:
\begin{align}
\tau(\eta_\mathrm{tr}) 
&\equiv \tau_\mathrm{tr} 
=-\alpha \left[k^2-U(\eta_\mathrm{e}+\Delta \eta)\right] 
=-\alpha k^2=-y_\mathrm{e}^2X^{-2/3},
\end{align}
where we have made use of the expression~\eqref{eq:alpha} for $\alpha$.
We find that the above expression for $T_{11}$ and, similar expressions
for $T_{12}$, $T_{21}$ and $T_{22}$, can be written as
\begin{subequations}
\begin{align}
T_{11} & = \sqrt{\f{2}{\pi}} k\, \e^{-i y_\mathrm{e}Y}
\l[\mathrm{Ai}(\tau_\mathrm{tr})
-\f{iX^{1/3}}{y_{\e}}\mathrm{Ai}'(\tau_\mathrm{tr})\r],\\
T_{12} &=\sqrt{\f{2}{\pi}} k\, \e^{-i y_\mathrm{e}Y}
\l[\mathrm{Bi}(\tau_\mathrm{tr})
-\f{iX^{1/3}}{y_{\e}}\mathrm{Bi}'(\tau_\mathrm{tr})\r],\\
T_{21} &=-\sqrt{\f{2}{\pi}} k\, \e^{i y_\mathrm{e}Y}
\l[\mathrm{Ai}(\tau_\mathrm{tr})
+\f{iX^{1/3}}{y_{\e}}\mathrm{Ai}'(\tau_\mathrm{tr})\r],\\
T_{22}
&=-\sqrt{\f{2}{\pi}} k\, \e^{i y_\mathrm{e}Y}
\l[\mathrm{Bi}(\tau_\mathrm{tr})
+\f{iX^{1/3}}{y_{\e}}\mathrm{Bi}'(\tau_\mathrm{tr})\r],
\end{align}
\end{subequations}
Finally, using the expressions~\eqref{eq:afns} for the Airy function 
and its derivatives, we can write the matrix elements $(T_{11},T_{12},T_{21},
T_{22})$ as follows:
\begin{subequations}
\begin{align}
T_{11} 
&=k\f{\sqrt{2}}{\pi}
\e^{-i y_\mathrm{e} Y}y_\mathrm{e}^{-1/2}X^{1/6}
\l[{\cal P}_{\mathrm{s11}}\sin \l(\zeta_\mathrm{tr}+\f{\pi}{4}\r)
+{\cal P}_{\mathrm{c11}}\cos \l(\zeta_\mathrm{tr}+\f{\pi}{4}\r)\r],\\
T_{12}
&=k\frac{\sqrt{2}}{\pi}
\e^{-i y_\mathrm{e} Y} y_\mathrm{e}^{-1/2}X^{1/6}
\left[{\cal P}_{\mathrm{s12}}\sin \left(\zeta_\mathrm{tr}
+\frac{\pi}{4}\right)
+{\cal P}_{\mathrm{c12}}\cos \left(\zeta_\mathrm{tr}+\frac{\pi}{4}\right)\right],\\
T_{21}& =-k\frac{\sqrt{2}}{\pi} 
\e^{i y\mathrm{e} Y}
y_\mathrm{e}^{-1/2}X^{1/6}
\left[{\cal P}_{\mathrm{s21}}\sin \left(\zeta_\mathrm{tr}
+\frac{\pi}{4}\right)
+{\cal P}_{\mathrm{c21}}\cos \left(\zeta_\mathrm{tr}+\frac{\pi}{4}\right)\right],\\
T_{22} &=-k\frac{\sqrt{2}}{\pi}
\e^{i y_\mathrm{e} Y}
y_\mathrm{e}^{-1/2}X^{1/6}
\left[{\cal P}_{\mathrm{s22}}\sin \left(\zeta_\mathrm{tr}+\frac{\pi}{4}\right)
+{\cal P}_{\mathrm{c22}}\cos \left(\zeta_\mathrm{tr}+\frac{\pi}{4}\right)\right],
\end{align}
\end{subequations}
where the quantities 
$({\cal P}_{\mathrm{s11}},{\cal P}_{\mathrm{c11}},{\cal P}_{\mathrm{s12}},
{\cal P}_{\mathrm{c12}},{\cal P}_{\mathrm{s21}},{\cal P}_{\mathrm{c21}},
{\cal P}_{\mathrm{s22}},{\cal P}_{\mathrm{c22}})$ are given by
\begin{subequations}
\begin{align}
{\cal P}_{\mathrm{s11}}
&\equiv A(\tau_\mathrm{tr})+iD(\tau_\mathrm{tr}),\\ 
{\cal P}_{\mathrm{c11}}
&\equiv -B(\tau_\mathrm{tr})+iC(\tau_\mathrm{tr})\\
{\cal P}_{\mathrm{s12}}
&\equiv B(\tau_\mathrm{tr})-iC(\tau_\mathrm{tr}),\\
{\cal P}_{\mathrm{c12}}
& \equiv A(\tau_\mathrm{tr})+iD(\tau_\mathrm{tr}),\\
{\cal P}_{\mathrm{s21}}
& \equiv A(\tau_\mathrm{tr})-iD(\tau_\mathrm{tr}),\\
{\cal P}_{\mathrm{c21}} 
&\equiv-B(\tau_\mathrm{tr})-iC(\tau_\mathrm{tr}),\\ 
{\cal P}_{\mathrm{s22}}
& \equiv B(\tau_\mathrm{tr})+iC(\tau_\mathrm{tr}),\\
{\cal P}_{\mathrm{c22}}
& \equiv A(\tau_\mathrm{tr})-iD(\tau_\mathrm{tr}).
\end{align}
\end{subequations}

The final step consists of making use of the above expressions for 
the matrix elements $(T_{11}, T_{12}, T_{21},T_{22})$ and the exact 
expressions for $(B_1,B_2)$ [cf. Eqs.~\eqref{eq:exactB1} 
and~\eqref{eq:exactB2}] in the matrix relation~\eqref{eq:matrix} 
to arrive at Bogoliubov coefficients during the epoch of radiation
domination, viz.~$(\alpha_{k}^{\mathrm{r}},\beta_k^\mathrm{r})$. 
We find that~$(\alpha_{k}^{\mathrm{r}},\beta_k^\mathrm{r})$ can be
written as
\begin{subequations}\label{eq:akbk-slt}
\begin{align}
\alpha^\mathrm{r}_k 
&=-\frac{iA_{ k}}{2} 
\e^{i[y_\mathrm{e}+(q-1)\pi/2]-i y_\mathrm{e} Y}
\Biggl[\left({\cal P}_{\mathrm{s}11}{\cal P}_{\mathrm{c}1}
+{\cal P}_{\mathrm{s}12}{\cal P}_{\mathrm{c}2}\right)
\sin\left(\zeta_\mathrm{tr}+\frac{\pi}{4}\right)
\cos\left(\zeta_\mathrm{e}+\frac{\pi}{4}\right)\nn\\ 
&\quad+\left({\cal P}_{\mathrm{s}11}{\cal P}_{\mathrm{s}1}
+{\cal P}_{\mathrm{s}12}{\cal P}_{\mathrm{s}2}\right)
\sin\left(\zeta_\mathrm{tr}+\frac{\pi}{4}\right)
\sin\left(\zeta_\mathrm{e}+\frac{\pi}{4}\right)
+\left({\cal P}_{\mathrm{c}11}{\cal P}_{\mathrm{c}1}
+{\cal P}_{\mathrm{c}12}{\cal P}_{\mathrm{c}2}\right)
\cos\left(\zeta_\mathrm{tr}+\frac{\pi}{4}\right)\nn\\ 
&\quad \times \cos\left(\zeta_\mathrm{e}+\frac{\pi}{4}\right)
+\left({\cal P}_{\mathrm{c}11}{\cal P}_{\mathrm{s}1}
+{\cal P}_{\mathrm{c}12}{\cal P}_{\mathrm{s}2}\right)
\cos\left(\zeta_\mathrm{tr}+\frac{\pi}{4}\right)
\sin\left(\zeta_\mathrm{e}+\frac{\pi}{4}\right)\Biggr],\\
\beta^\mathrm{r}_k 
&=\frac{iA_{k}}{2}
\e^{i[y_\mathrm{e}+(q-1)\pi/2]+iy_\mathrm{e} Y}
\Biggl[\left({\cal P}_{\mathrm{s}21}{\cal P}_{\mathrm{c}1}
+{\cal P}_{\mathrm{s}22}{\cal P}_{\mathrm{c}2}\right)
\sin\left(\zeta_\mathrm{tr}+\frac{\pi}{4}\right)
\cos\left(\zeta_\mathrm{e}+\frac{\pi}{4}\right)\nn\\ 
&\quad+\left({\cal P}_{\mathrm{s}21}{\cal P}_{\mathrm{s}1}
+{\cal P}_{\mathrm{s}22}{\cal P}_{\mathrm{s}2}\right)
\sin\left(\zeta_\mathrm{tr}+\frac{\pi}{4}\right)
\sin\left(\zeta_\mathrm{e}+\frac{\pi}{4}\right)
+\left({\cal P}_{\mathrm{c}21}{\cal P}_{\mathrm{c}1}
+{\cal P}_{\mathrm{c}22}{\cal P}_{\mathrm{c}2}\right)
\cos\left(\zeta_\mathrm{tr}+\frac{\pi}{4}\right)\nn\\ 
&\quad \times  \cos\left(\zeta_\mathrm{e}+\frac{\pi}{4}\right)
+\left({\cal P}_{\mathrm{c}21}{\cal P}_{\mathrm{s}1}
+{\cal P}_{\mathrm{c}22}{\cal P}_{\mathrm{s}2}\right)
\cos\left(\zeta_\mathrm{tr}+\frac{\pi}{4}\right)
\sin\left(\zeta_\mathrm{e}+\frac{\pi}{4}\right)\Biggr].
\end{align}
\end{subequations}
Note that $\zeta_\mathrm{tr}$ and~$\zeta_\mathrm{e}$ can be expressed 
in terms of $y_{\e}$ and $X$ as follows:
\begin{subequations}
\label{eq:ztrze}
\begin{align}
\zeta_\mathrm{tr} 
&\equiv \f{2}{3}(-\tau_\mathrm{tr})^{3/2}
=\f{2}{3} \f{y_\mathrm{e}^3}{X},\\
\zeta_\mathrm{e}
&\equiv \f{2}{3}(-\tau_\mathrm{e})^{3/2}
=\f{2}{3}\f{y_\mathrm{e}^3}{X} \l[1-\f{q(q-1)}{y_{\e}^2}\r]^{3/2}.
\end{align}
\end{subequations}
We should emphasize that the above expressions for the
coefficients~$(\alpha^\mathrm{r}_k,\beta^\mathrm{r}_k)$ are exact. 
In the limit $y_{\e}\gg 1$, it is easy to expand these expressions in 
inverse powers of~$y_\mathrm{e}$.
We find that, at the leading order, the 
coefficients~$(\alpha^\mathrm{r}_k,\beta^\mathrm{r}_k)$ are given by
\begin{subequations}\label{eq:cwdwsmallscales}
\begin{align}
\label{eq:cwsmallscales}
\alpha_k^\mathrm{r} 
&\simeq -iA_{k} \e^{i[y_\mathrm{e}+(q-1)\pi/2]-iy_\mathrm{e} Y}
\e^{i(\zeta_\mathrm{e}-\zeta_\mathrm{tr})},\\
\label{eq:dwsmallscales}
\beta_k^\mathrm{r}
&\simeq \frac{q(q-1)A_{ k}}{4y_\mathrm{e}^3}
\e^{i[y_\mathrm{e}+(q-1)\pi/2]+iy_\mathrm{e} Y}
\e^{-i(\zeta_\mathrm{e}-\zeta_\mathrm{tr})}
\l[1-\e^{i(\zeta_\mathrm{e}-\zeta_\mathrm{tr})}\frac{iX}{q(q-1)}
\sin \l(\zeta_\mathrm{e}-\zeta_\mathrm{tr}\r)\r].
\end{align}
\end{subequations}
We should highlight the fact that the coefficient $\beta^\mathrm{r}_{k}/A_k$ 
behaves as~$y_{\e}^{-3}$ in the limit $y_{\e}\gg 1$.
In other words, $\beta_k^\mathrm{r}/A_k \propto k^{-3}$ in the limit~$k \gg \ke$.
This should be contrasted with the result we had arrived at earlier in the
case of instantaneous transition from inflation to the epoch of radiation
domination.
Recall that, in the case of instantaneous transition we had obtained that 
$\beta^\mathrm{r}_k/A_k\propto k^{-2}$ in the limit $k\gg \ke$ 
[cf. Eq.~\eqref{eq:bk-abrupt}]. 
This implies that the smoother transition has modified the scale dependence 
of the coefficient $\beta_k^\mathrm{r}$. 
Such a modification will, in turn change the behavior of the PS at large~$k$. 
It should be clear that the behavior $\beta_k^\mathrm{r}/A_k\propto k^{-2}$ at 
large~$k$ is an artifact of the instantaneous transition from inflation to 
radiation domination.

Let us now consider the limit of instantaneous transition. 
Evidently, this corresponds to the limit $\Delta \eta\to 0$ or, 
equivalently, $X\to \infty$ [cf. Eq.~\eqref{eq:defbigX}].
Since we are also focusing on the limit $y_{\e}\gg 1$, we have
[cf. Eq.~\eqref{eq:ztrze}]
\begin{align}
\zeta_\mathrm{e}-\zeta_\mathrm{tr}=\frac{2}{3}\f{y_\mathrm{e}^3}{X}
\left\{\left[1-\frac{q(q-1)}{y_\mathrm{e}^2}\right]^{3/2}-1\right\}
\simeq -\frac{q(q-1)}{X}y_\mathrm{e}
\end{align}
and, hence, it follows that
\begin{align}
\frac{X}{q(q-1)}\sin(\zeta_\mathrm{e}-\zeta_\mathrm{tr})
\simeq -y_\mathrm{e}.
\end{align}
Moreover, in the limit $\Delta \eta \to 0$, on using the 
expressions~\eqref{eq:lit-two}, we obtain that
\begin{align}
\f{A_1+\sqrt{3} A_2}{A_1-\sqrt{3} A_2}
\simeq \f{3^{1/3}\Gamma(2/3)}{q \Gamma(1/3)}X^{1/3},
\end{align}
from which we can also deduce that, in this limit  [cf. Eq.~\eqref{eq:defbigY}]
\begin{align}
Y \simeq -q.
\end{align}
For simplicity, let us restrict ourselves to the case of de Sitter 
inflation wherein $q=-1$.
In such a case, in the limits, $y_{\e}\gg 1$ and $\Delta \eta\to 0$,
the expressions~\eqref{eq:cwdwsmallscales} for $(\alpha_k^\mathrm{r},
\beta_k^\mathrm{r})$ reduce to
\begin{subequations}
\begin{align}
\alpha^\mathrm{r}_{k} &\simeq iA_{k} \e^{2iy_\mathrm{e}},\\
\beta^\mathrm{r}_k &\simeq \frac{-iA_{ k}}{2y_\mathrm{e}^2}.
\end{align}
\end{subequations}
These are indeed the $y_{\e}\gg 1$ limits of the 
expressions~\eqref{eq:CDabrupt-o} for $(\alpha_k^\mathrm{r},
\beta_k^\mathrm{r})$ we obtained originally in the case
of the instantaneous transition from de Sitter inflation to the
epoch of radiation domination.

\begin{figure}[!t]
\centering
\includegraphics[width=0.975\linewidth]{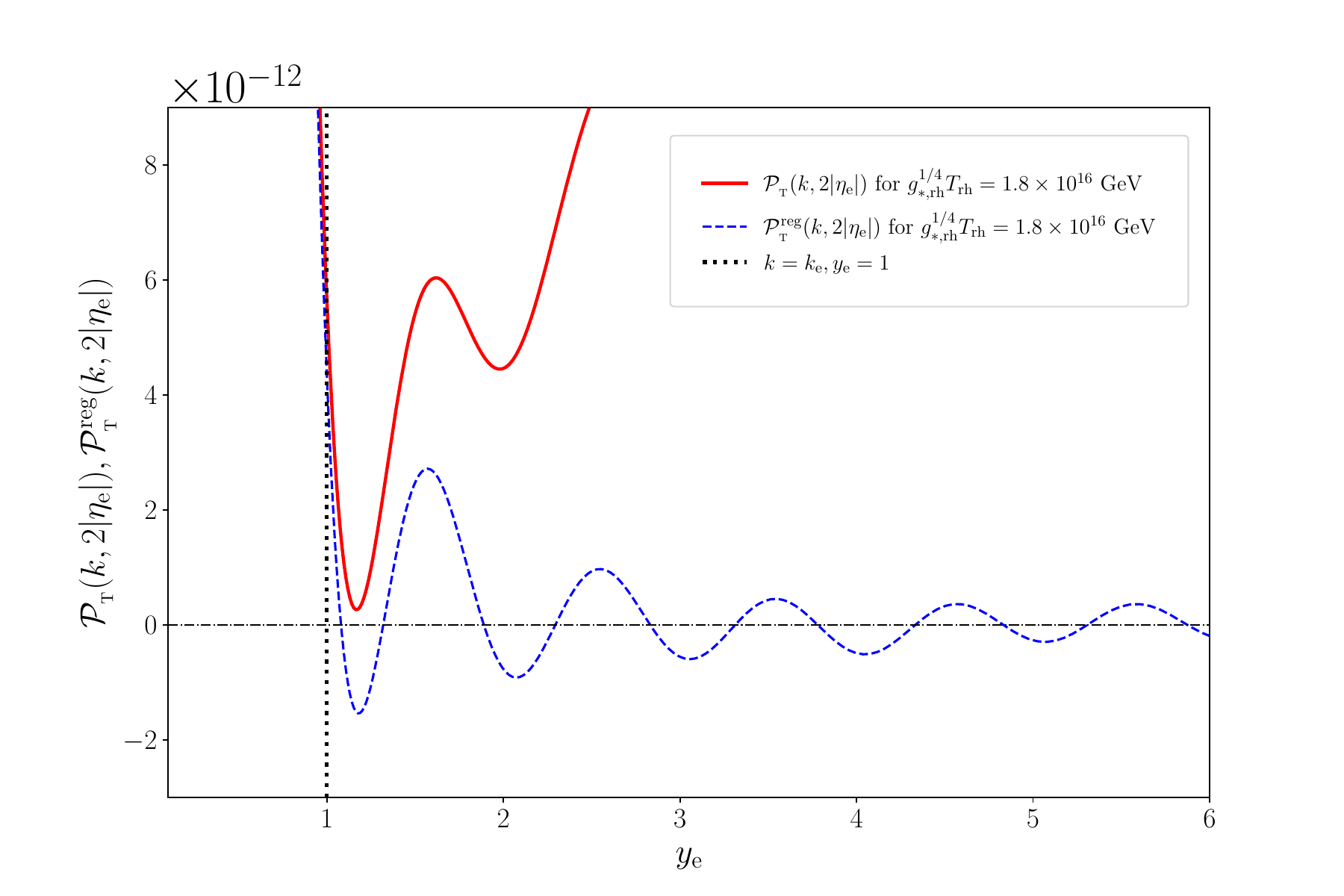}
\caption{The actual and the regularized PS of PGWs, viz. $\pt(k,\eta)$ and
$\ptr(k,\eta)$, evaluated during the early stages of the epoch of radiation 
domination at the conformal time $\eta=2\vert\ee \vert$ have been plotted~(in red
and blue) in the case of the smoother transition from inflation to radiation
domination. 
As in Figs.~\ref{fig:PT-hf}, \ref{fig:neg-PS} and~\ref{fig:ps-rps-ds-rd-it},
we have set $g_{\ast,\mathrm{rh}}^{1/4} \Tre=1.8\times 10^{16}\, \mathrm{GeV}$ 
in plotting this figure.
Also, we have chosen the width of the smooth transition to be~$\Delta\eta
=2\vert\ee\vert$.
The effects of the smoother transition from inflation to radiation domination 
modify the PS over wave numbers such that~$y_{\e} \gtrsim 1$. 
Hence, we have plotted the actual and regularized PS only over this domain.
Clearly, the regularized PS oscillates about zero (over $y_{\e}\gg 1$), as 
in the case of the instantaneous transition, which we had illustrated in 
Fig.~\ref{fig:neg-PS}. 
However, note that, in the case of the smoother transition, the amplitude of 
the oscillations decreases with~$k$ at large wave numbers.}
\label{fig:neg-PS_smooth}
\end{figure}

\begin{figure}[!t]
\centering
\includegraphics[width=0.975\linewidth]{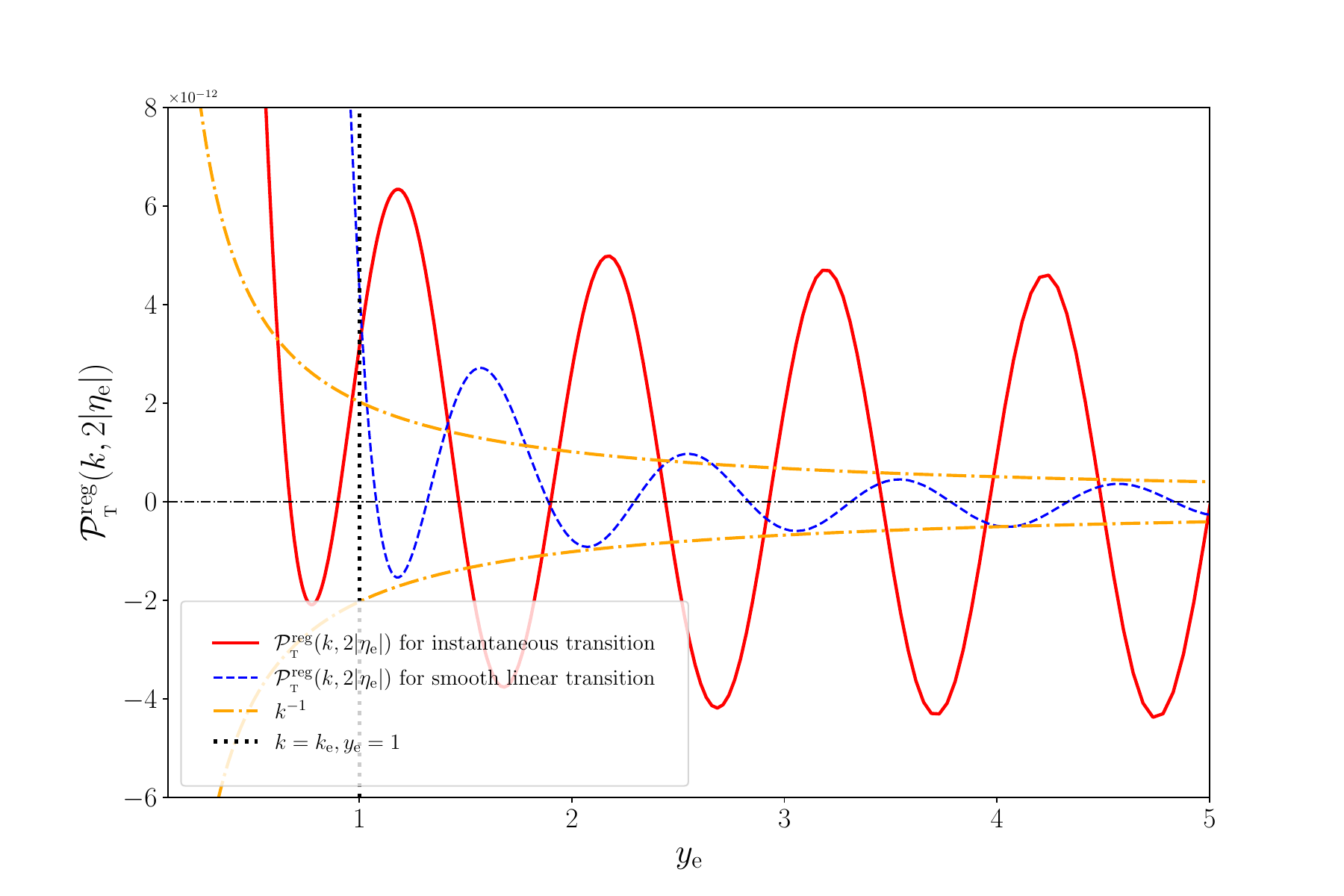}
\caption{The regularized PS of PGWs~$\ptr(k,\eta)$, evaluated during the 
early stages of the epoch of radiation domination at the conformal time 
$\eta=2\vert\ee \vert$, has been plotted in the cases of instantaneous 
and smoother transitions from inflation to radiation domination (in red 
and blue). 
The nature of the transition from inflation to radiation domination affects 
the PS only over wave numbers such that~$y_{\e}\gtrsim 1$.
Hence, as in the previous figure, we have illustrated the PS only over this
domain.
Note that, while the regularized PS has a constant amplitude over small scales 
(such that over $y_{\e}\gg 1$) in the case of the instantaneous transition,
clearly, the amplitude of the oscillations decreases as~$k^{-1}$ in the case 
of the smoother linear transition.} 
\label{fig:neg-PS_smooth_instant}
\end{figure}

Let us now discuss the effects of the smoother transition from 
inflation to radiation domination on the PS of PGWs.
The actual and regularized PS of PGWs during the epoch of radiation 
domination, after the smoother transition from inflation, can be 
obtained by substituting the results~\eqref{eq:akbk-slt} for 
$(\alpha_k^{\mathrm{r}},\beta_k^{\mathrm{r}})$ in 
Eqs.~\eqref{eq:generalpt} and~\eqref{eq:generalptregulated},
respectively.
In Fig.~\ref{fig:neg-PS_smooth}, we have plotted the actual and the 
regularized PS evaluated during the early stages of the epoch of
radiation domination, after the smoother transition from inflation.
In Fig.~\ref{fig:neg-PS_smooth_instant}, we have compared the regularized
PS in the cases of instantaneous and smoother transitions from inflation,
when they are evaluated during the early stages of radiation domination.
Two points should be clear from these figures.
Firstly, in the case of the smoother transition, over wave numbers 
$k\gtrsim \ke$, the regularized PS oscillates about zero as in the 
case of the instantaneous transition.
Secondly, while the amplitude of the oscillations remain a constant in the
case of the instantaneous transition, the amplitude decreases as $k^{-1}$
in the case of the smooth transition.

\subsection{PS during matter domination, after the smoother transition}

\begin{figure}[!t]
\centering
\includegraphics[width=1.0\textwidth]{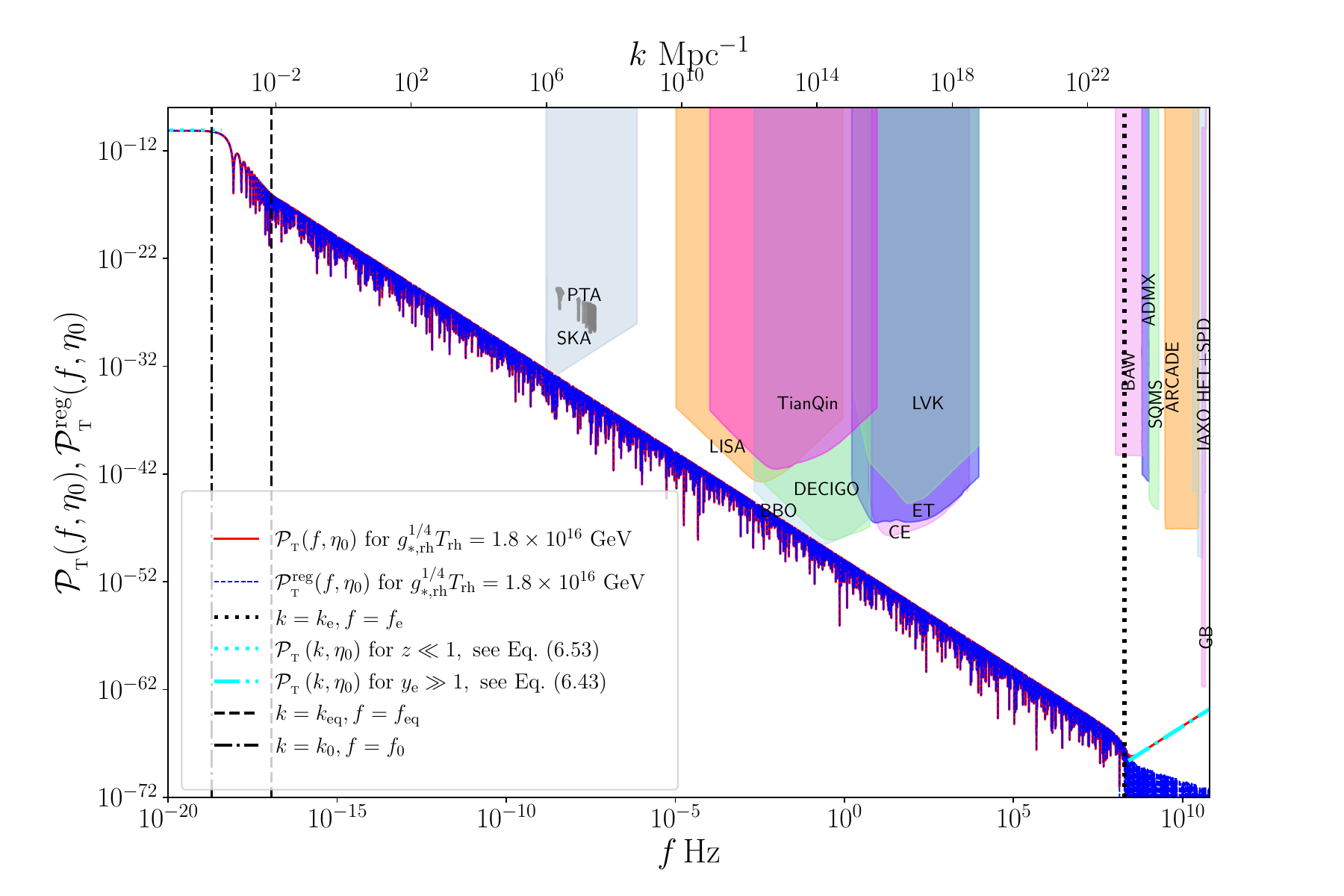}
\caption{The actual and the regularized PS of PGWs, viz. $\pt(k,\eta)$
and $\ptr(k,\eta)$, evaluated today, i.e. during the late stages of matter 
domination at the conformal time $\eta_0$, have been plotted~(in red 
and blue), assuming a smoother linear transition from de Sitter inflation 
to the epoch of radiation domination.
In plotting the figure, we have assumed that the transition from radiation 
to matter domination is instantaneous.
As in Figs.~\ref{fig:PT-hf} and~\ref{fig:ps-rps-ds-rd-it}, the vertical
lines indicate the frequencies associated with the wave numbers~$(k_0/a_0, 
\keq/a_0, \ke/a_0)$. 
We have illustrated the PS only until the frequency of $f \simeq 10^{11}\,
\mathrm{Hz}$, since, at higher frequencies, the numerical errors in plotting 
the regularized PS become significant.
Note that the process of regularization does not alter the PS over wave numbers
$k \lesssim \ke$.
However, as in Fig.~\ref{fig:ps-rps-ds-rd-it}, it is clear that regularization
truncates the $k^2$ rise in the PS over $k \gtrsim \ke$. 
In addition, the smoother linear transition from inflation to radiation domination
suppresses the amplitude of the oscillations in the PS over $k \gtrsim \ke $ 
as~$k^{-1}$.}\label{fig:PS-ds-sd-rd-it-f}
\end{figure}

In this section, we shall discuss the behavior of PS of PGWs during the 
epoch of matter domination, after the smoother linear transition from 
inflation to the epoch of radiation domination.
Earlier, we had expressed the actual and the regularized PS of PGWS during 
matter domination in terms of the Bogoliubov coefficients $(\alpha_k^{\mathrm{m}},
\beta_k^{\mathrm{m}})$ [see Eqs.~\eqref{eq:generalptmatunregulated}
and~\eqref{eq:generalptmatregulated}].
Recall that, these coefficients were, in turn, dependent on the Bogoliubov
coefficients~$(\alpha_k^{\mathrm{r}},\beta_k^{\mathrm{r}})$ during the 
epoch of radiation domination [see Eqs.~\eqref{eq:alpham_betam}].
In the case of the smoother linear transition from inflation to radiation
domination, we can substitute~$(\alpha_k^{\mathrm{r}},\beta_k^{\mathrm{r}})$
from Eqs.~\eqref{eq:akbk-slt} in Eqs.~\eqref{eq:alpham_betam} to arrive at
the coefficients~$(\alpha_k^{\mathrm{m}},\beta_k^{\mathrm{m}})$.
The corresponding actual and regularized PS of PGWs during the epoch of 
matter domination can be arrived at by substituting these 
expressions for $(\alpha_k^{\mathrm{m}},\beta_k^{\mathrm{m}})$
in Eqs.~\eqref{eq:generalptmatunregulated}
and~\eqref{eq:generalptmatregulated}.
In Fig.~\ref{fig:PS-ds-sd-rd-it-f}, we have plotted the actual and regularized 
PS of PGWs evaluated today.
Evidently, the process regularization as well as the smoother transition from 
inflation to radiation domination affects the PS only over wave numbers such 
that $k \gtrsim \ke$.
Note that the process of regularization truncates the rise in the PS over 
$k \gtrsim \ke$, as we have already discussed.
The regularized PS oscillates about zero over this domain.
The smoother linear transition leads to a suppression in the amplitude of the 
oscillations in the regularized PS (as $k^{-1}$) at such high wave numbers.

In the remainder of this section, we shall analytically arrive at the 
behavior of the PS during the epoch of matter domination on different 
scales.
For simplicity, we shall assume de Sitter inflation corresponding to~$q=-1$.
Let us first focus on wave numbers such that $y_\e\gg1$.
In such a limit, since $z \gg 1$ as well, the PS in 
Eq.~\eqref{eq:generalptmatunregulated} simplifies to be 
\begin{align}
\pt(k,\eta)
& \simeq \frac{2}{\pi^2}\left(\frac{H_\mathrm{e}}{\Mpl}\right)^2
\left(\frac{a_\mathrm{e}}{a}\right)^2 y_\mathrm{e}^2
\Biggl[\left(1+2\,\left \vert\f{\beta_k^\mathrm{m}}{A_{k}}\right\vert ^2\right)+2 
\Re\left(\frac{\alpha_k^\mathrm{m}\beta_k^\mathrm{m}{}^*}{\vert 
A_k\vert^2}\right)\cos(2z)\nn\\ 
&\quad +2 \Im\left(\frac{\alpha_k^\mathrm{m}\beta_k^\mathrm{m}{}^*}{\vert 
A_k\vert^2}\right)\sin(2z)\Biggr].
\end{align}
Also, in the limit $\xeq \gg 1$, we obtain that [cf.
Eqs.~\eqref{eq:alphamod2_matter_exact}]
\begin{subequations}
\begin{align}
\Re\left(\frac{\alpha_k^\mathrm{m}\beta_k^\mathrm{m}{}^*}{\vert A_k\vert^2}\right)
\simeq \frac{-2}{16 x_\mathrm{eq}^2}
\left \vert \frac{\alpha_k^\mathrm{r}}{ A_k}\right \vert ^2 \cos(4 x_\mathrm{eq})
-\Re\left(\frac{\alpha_k^\mathrm{r}\beta_k^\mathrm{r}{}^*}{\vert 
A_k\vert^2}\right)\cos(2x_\mathrm{eq})
+\Im\left(\frac{\alpha_k^\mathrm{r}\beta_k^\mathrm{r}{}^*}{\vert 
A_k\vert^2}\right)\sin(2x_\mathrm{eq}),\\
\Im\left(\frac{\alpha_k^\mathrm{m}\beta_k^\mathrm{m}{}^*}{\vert A_k\vert^2}\right)
\simeq \frac{-2}{16 x_\mathrm{eq}^2}
\left \vert \frac{\alpha_k^\mathrm{r}}{ A_k}\right \vert ^2
\sin(4 x_\mathrm{eq})
-\Re\left(\frac{\alpha_k^\mathrm{r}\beta_k^\mathrm{r}{}^*}{\vert 
A_k\vert^2}\right)\sin(2x_\mathrm{eq})
-\Im\left(\frac{\alpha_k^\mathrm{r}\beta_k^\mathrm{r}{}^*}{\vert 
A_k\vert^2}\right)\cos(2x_\mathrm{eq}).
\end{align}    
\end{subequations}
In arriving at these expressions, we have also made use of the fact that 
$1/x_\mathrm{eq}\gg 1/z_0$ and $1/y_\mathrm{e}^3\gg 1/(y_\mathrm{e}^2z_0)$ 
since, $1/y_\mathrm{e}\gg 1/z_0$.
Moreover, in such a limit, we have $\left \vert \alpha_k^\mathrm{m}/A_k\right
\vert\simeq \left \vert \alpha_k^\mathrm{r}/A_k\right \vert \simeq 1$.
Using the above results, we can write the PS during matter domination as
\begin{align}
\pt(k,\eta) 
& \simeq \frac{2}{\pi^2}\left(\frac{H_\mathrm{e}}{\Mpl}\right)^2
\left(\frac{a_\mathrm{e}}{a}\right)^2 y_\mathrm{e}^2
\Biggl[1-\frac{1}{4x_\mathrm{eq}^2}\cos(4 x_\mathrm{eq}-2z)
-2\Re\left(\frac{\alpha_k^\mathrm{r}\beta_k^\mathrm{r}{}^*}{\vert 
A_k\vert^2}\right)\cos(2x_\mathrm{eq}-2z)\nn\\ 
&\quad+2\Im\left(\frac{\alpha_k^\mathrm{r}\beta_k^\mathrm{r}{}^*}{\vert 
A_k\vert^2}\right)\sin(2x_\mathrm{eq}-2z)\Biggr].
\end{align}
Further, from Eqs.~\eqref{eq:cwdwsmallscales}, in the limit $y_{\e}\gg 1$,
we have
\begin{subequations}
\begin{align}
\Re\left(\frac{\alpha_k^\mathrm{r}\beta_k^{\mathrm{r}*}}{\vert A_k\vert^2}\right)
&\simeq \frac{1}{2y_\mathrm{e}^3} 
\Biggl[\sin\left(2\zeta_\mathrm{e}-2\zeta_\mathrm{tr}-2Yy_\mathrm{e}\right)\nn\\ 
&\quad+\frac{X}{2}\cos\left(\zeta_\mathrm{e}-\zeta_\mathrm{tr}
-2Yy_\mathrm{e}\right)\sin\left(\zeta_\mathrm{e}-\zeta_\mathrm{tr}\right)\Biggr],\\
\Im\left(\frac{\alpha_k^\mathrm{r}\beta_k^{\mathrm{r}*}}{\vert A_k\vert^2}\right)
&\simeq \frac{1}{2y_\mathrm{e}^3} 
\Biggl[-\cos\left(2\zeta_\mathrm{e}-2\zeta_\mathrm{tr}-2Yy_\mathrm{e}\right)\nn\\ 
&\quad+\frac{X}{2}\sin\left(\zeta_\mathrm{e}-\zeta_\mathrm{tr}
-2Yy_\mathrm{e}\right)\sin\left(\zeta_\mathrm{e}-\zeta_\mathrm{tr}\right)\Biggr].
\end{align}
\end{subequations}
As a result, the PS in the limit $y_{\e}\gg 1$ can be written as
\begin{align}
\pt (k,\eta)
&=\frac{2}{\pi^2}\left(\frac{H_\mathrm{e}}{\Mpl}\right)^2
\left(\frac{a_\mathrm{e}}{a}\right)^2 y_\mathrm{e}^2
\Biggl\{1-\frac{1}{4 x_\mathrm{eq}^2}\cos(4 x_\mathrm{eq}-2z)\nn\\ 
&\quad-\frac{1}{y_\mathrm{e}^3}
\sin \left[2\l(x_\mathrm{eq}-Yy_\mathrm{e}-z+\zeta_\mathrm{e}
-\zeta_\mathrm{tr}\r)\right]\nn\\ 
&\quad-\frac{X}{2y_\mathrm{e}^3}
\cos\left(2x_\mathrm{eq}-2Yy_\mathrm{e}-2z+\zeta_\mathrm{e}
-\zeta_\mathrm{tr}\right)
\sin\left(\zeta_\mathrm{e}-\zeta_\mathrm{tr}\right)\biggr\}.
\end{align}
Note that, if we now consider the regularized PS, the first term will 
not be present.
On comparing the last two terms with the second term, we find that, 
around $k\gtrsim\ke$, the last two terms will dominate.
These terms oscillate about zero as in the regularized 
PS over this domain in the case of the instantaneous transition
from inflation to radiation domination [see Eq.~\eqref{eq:ps-md-lwn}].
But, because of the smoother transition from inflation to the epoch
of radiation domination, the amplitude of the oscillations is 
suppressed as $k^{-1}$.
However, interestingly, we find that the second term eventually becomes 
dominant in the regularized PS over $k\gtrsim\ke^3/\keq^2$. 
Therefore, at ultra-high wave numbers, we find that the regularized PS 
behaves as
\begin{align}
\ptr(k,\eta) 
& = -\frac{2}{\pi^2}\left(\frac{H_\mathrm{e}}{\Mpl}\right)^2
\left(\frac{a_\mathrm{e}}{a}\right)^2 
\frac{y_\mathrm{e}^2}{4 x_\mathrm{eq}^2}
\cos(4 x_\mathrm{eq}-2z).
\end{align}
In other words, at ultra-high wave numbers, the amplitude of the oscillations 
in the regularized PS ceases to decrease as $k^{-1}$ and settles down to a 
constant value.

Let us now turn to the limit $y_{\e} \ll 1$.
These correspond to wave numbers which are outside the Hubble radius 
at the end of inflation.
In this regime, we find that the Bogoliubov coefficients 
$(\alpha_k^\mathrm{r},\beta_k^\mathrm{r})$ can be expanded as
\begin{subequations}
\label{eq:radsmallscales}
\begin{align}
\alpha_k^\mathrm{r}&\simeq \mathcal{F}\f{i A_k}{y_\e^2}
+i \mathcal{G}\f{i A_k}{y_\e}+\mathcal{O}(y_\e^0),\\
\beta_k^\mathrm{r}&\simeq \mathcal{F}\f{i A_k}{y_\e^2}
-i \mathcal{G}\f{i A_k}{y_\e}+\mathcal{O}(y_\e^0),
\end{align}
\end{subequations}
where $\mathcal{F}$ and $\mathcal{G}$ are dimensionless real functions 
which depend on the parameters of the smooth transition.
They are given by
\begin{subequations}
\begin{align}
\mathcal{F}
&= -\f{\pi}{6\Gamma(1/3)}\bigg[3^{7/6}\Ai(\tau_\e)
+3^{2/3}\Bi(\tau_\e)+3^{7/6}X^{1/3}\Ai'(\tau_\e)
+3^{2/3}X^{1/3}\Bi'(\tau_\e)\bigg],\\
\mathcal{G}
&= -\f{1}{4\times 3^{1/3}}\bigg\{\left[-3^{2/3}\Ai'(\tau_\e)+3^{1/6}\Bi'(\tau_\e)
-3^{2/3}X^{-1/3}\Ai(\tau_\e)+3^{1/6}X^{-1/3}\Bi(\tau_\e)\right]
\Gamma\left(\frac{1}{3}\right)\nn\\ &\quad+\left[3\Ai(\tau_\e)+3^{1/2}\Bi(\tau_\e)
+3X^{1/3}\Ai'(\tau_\e)+3^{1/2}X^{1/3}\Bi'(\tau_\e)\right]Y\, 
\Gamma\left(\frac{2}{3}\right)\bigg\},
\end{align}   
\end{subequations}
where $\tau_\e=\tau(\ee)\simeq2/X^{2/3}$ [cf. Eq.~\eqref{eq:tau-e}]. 

We can now distinguish three different cases. 
The first corresponds to $y_\mathrm{e}\ll 1$, $x_\mathrm{eq}\gg 1$ and 
$z_0\gg 1$. 
These limits correspond to wave numbers $k_\mathrm{eq}\lesssim k \lesssim 
k_\mathrm{e}$, i.e. wave numbers which have re-entered the Hubble radius 
before the time of radiation-matter equality. 
In this case, the relations between the coefficients $(\alpha _k^\mathrm{m}, 
\beta_k^\mathrm{m})$ and $(\alpha _k^\mathrm{r}, \beta_k^\mathrm{r})$ are 
given by Eqs.~\eqref{eq:alphamod2_matter_approx_abrupt}. 
On using the fact that $z\gg 1$, we find that the PS during matter domination, 
as given by Eq.~\eqref{eq:generalptmatunregulated}, reduces to the 
following form:
\begin{align}
\pt(k,\eta) 
& \simeq \frac{2}{\pi^2} \left(\frac{H_\mathrm{e}}{\Mpl}\right)^2
\left(\frac{a_\mathrm{e}}{a}\right)^2 y_\mathrm{e}^2
\biggl\{1+2 \left\vert \frac{\beta_k^\mathrm{r}}{A_k}\right\vert^2
+2\biggl[-\Re \left(\frac{\alpha_k^\mathrm{r}\beta_k^\mathrm{r}{}^*}
{\vert A_k\vert^2}\right) \cos\left(2x_\mathrm{eq}\right)\nn\\ 
&\quad+\Im\left(\frac{\alpha_k^\mathrm{r}\beta_k^\mathrm{r}{}^*}
{\vert  A_k\vert^2}\right)
\sin\left(2x_\mathrm{eq}\right)\biggr]\cos\left(2 z\right)
+2\biggl[-\Re \left(\frac{\alpha_k^\mathrm{r}\beta_k^\mathrm{r}{}^*}
{\vert A_k\vert^2}\right)
\sin\left(2x_\mathrm{eq}\right)\nn\\ 
&\quad-\Im\left(\frac{\alpha_k^\mathrm{r}\beta_k^\mathrm{r}{}^*}
{\vert A_k\vert^2}\right)
\cos\left(2x_\mathrm{eq}\right)\biggr]\sin\left(2 z\right)
\biggr\}.
\end{align}
The expressions~\eqref{eq:radsmallscales} for
$(\alpha _k^\mathrm{r}, \beta_k^\mathrm{r})$ imply that, at leading 
order in the limit $y_{\e} \ll 1$, 
\begin{align}
\l\vert \f{\alpha_k^\mathrm{r}}{A_k}\r\vert ^2
=\l\vert \f{\beta_k^\mathrm{r}}{A_k}\r\vert ^2
=\f{\alpha_k^\mathrm{r}
\beta_k^{\mathrm{r}*}}{\vert A_k\vert^2}
\simeq \f{{\cal F}^2}{y_\mathrm{e}^4}.
\end{align}
It then follows that the above PS simplifies to
\begin{align}
\pt(k,\eta)\simeq\f{8 \mathcal{F}^2}{\pi^2}
\left(\frac{H_\mathrm{e}}{\Mpl}\right)^2
\left(\frac{a_\mathrm{e}}{a}\right)^2\f{1}{y_\e^2}\sin^2(x_{\rm eq}-z).
\label{eq:one}
\end{align}

The second case corresponds to $y_\mathrm{e}\ll 1$, $x_\mathrm{eq}\ll 1$ 
and $z_0\gg 1$. 
These limits correspond to the wave numbers $k_0\lesssim k \lesssim
k_\mathrm{eq}$, which re-enter the Hubble radius during the epoch 
of matter domination. 
Recall that, in the domain $x_\mathrm{eq}\ll 1$, the relation between the 
coefficients $(\alpha _k^\mathrm{m}, \beta_k^\mathrm{m})$ 
and $(\alpha _k^\mathrm{r}, \beta_k^\mathrm{r})$ are given by 
Eqs.~\eqref{eq:matxeqsmall}. 
On making use of the expressions~\eqref{eq:radsmallscales} for 
$(\alpha _k^\mathrm{r}, \beta_k^\mathrm{r})$, we obtain that
\begin{align}
\label{eq:alphabetasecondcase}
\alpha_k^\mathrm{m}\simeq \beta_k^\mathrm{m}
\simeq \frac{2iA_k}{8x_\mathrm{eq}^2y_\mathrm{e}^2}
\left(-{\cal G}y_\mathrm{e}+3x_\mathrm{eq}{\cal F}\right),
\end{align}
where we have retained the leading and next-to-leading orders in 
Eqs.~\eqref{eq:matxeqsmall}. 
If we now make use of these results 
in Eq.~\eqref{eq:generalptmatunregulated} and consider the limit
$z\gg 1$, we obtain the PS during the epoch of matter domination 
to be 
\begin{align}
\pt(k,\eta) 
&\simeq \frac{2}{\pi^2}\left(\frac{H_\mathrm{e}}{\Mpl}\right)^2
\left(\frac{a_\mathrm{e}}{a}\right)^2y_\mathrm{e}^2
\biggl[1+2 \left\vert \frac{\beta_k^\mathrm{m}}{A_k}\right\vert^2
+2 \Re \left(\frac{\alpha_k^\mathrm{m}\beta_k^\mathrm{m}{}^*}
{\vert A_k\vert^2}\right)
\cos\left(2z\right)\nn\\ 
&\quad+2 \Im \left(\frac{\alpha_k^\mathrm{m}\beta_k^\mathrm{m}{}^*}
{\vert A_k\vert^2}\right)\sin\left(2z\right)\biggr]\nn\\
&\simeq \frac{2}{\pi^2}\left(\frac{H_\mathrm{e}}{\Mpl}\right)^2
\left(\frac{a_\mathrm{e}}{a}\right)^2y_\mathrm{e}^2
2\, \left\vert \frac{\beta_k^\mathrm{m}}{A_k}\right\vert^2
\left[1+\cos(2z)\right]\nn\\
&\simeq \frac{2}{\pi^2}\left(\frac{H_\mathrm{e}}{\Mpl}\right)^2
\left(\frac{a_\mathrm{e}}{a}\right)^2y_\mathrm{e}^2
\frac{1}{4x_\mathrm{eq}^4y_\mathrm{e}^4}\left(-{\cal G}y_\mathrm{e}
+3x_\mathrm{eq}{\cal F}\right)^2\cos^2 z\nn\\
&\simeq \frac{9{\cal F}^2}{2\pi^2}
\left(\frac{H_\mathrm{e}}{\Mpl}\right)^2
\left(\frac{a_\mathrm{e}}{a}\right)^2
\frac{1}{x_\mathrm{eq}^2y_\mathrm{e}^2} \cos^2 z,
\label{eq:two}
\end{align}
where in the last step we have used the fact that $x_\mathrm{eq}\gg y_\mathrm{e}$.

The third and last case corresponds to $y_\mathrm{e}\ll 1$, 
$x_\mathrm{eq}\ll 1$ and $z_0\ll 1$.
These limit correspond to wave numbers $k \lesssim k_0$, which are 
yet to re-enter the Hubble radius. 
In this domain, the PS during the epoch of matter domination can be
written as
\begin{align}
\pt (k,\eta) 
&\simeq \frac{2}{\pi^2}\left(\frac{H_\mathrm{e}}{\Mpl}\right)^2
\left(\frac{a_\mathrm{e}}{a}\right)^2y_\mathrm{e}^2
\left \vert \frac{\alpha_k^\mathrm{m}}{A_k}\right\vert^2
\biggl[2\left(1+\frac{1}{z^2}\right)+2\left(1-\frac{1}{z^2}\right)
\cos(2z)-\frac{4}{z}\sin (2z)\biggr]\nn\\
&\simeq \frac{1}{18\pi^2}\left(\frac{H_\mathrm{e}}{\Mpl}\right)^2
\left(\frac{a_\mathrm{e}}{a}\right)^2
\frac{z^4}{x_\mathrm{eq}^4y_\mathrm{e}^2}
 \left(-{\cal G}y_\mathrm{e}+3x_\mathrm{eq}{\cal F}\right)^2,
\end{align}
where, in the second expression, we have made use of the fact 
that $z\ll 1$. 
It follows that
\begin{align}
\pt(k,\eta) \simeq \frac{{\cal F}^2}{2\pi^2}
\left(\frac{H_\mathrm{e}}{\Mpl}\right)^2
\left(\frac{a_\mathrm{e}}{a}\right)^2
\frac{z^4}{x_\mathrm{eq}^2y_\mathrm{e}^2},\label{eq:three}
\end{align}
which is scale-invariant.

Finally, let us point out that, in the limit of instantaneous transition
from de Sitter inflation to the epoch of radiation domination, we have 
$X\rightarrow\infty,~Y\rightarrow-1,~(\zeta_\mathrm{e}
-\zeta_{\mathrm{tr}})\rightarrow2y_\e/X$.
In such a limit, we find that $\mathcal{F}\simeq -1/2$ and $\mathcal{G}
\simeq 2/(3X^2)$ and the above expressions~\eqref{eq:one},  \eqref{eq:two} 
and~\eqref{eq:three} for the PS reduce to the expressions~\eqref{eq:ptradenter},
\eqref{eq:pteqtodaycale} and~\eqref{eq:PTlarge} we had obtained earlier.


\section{Summary and outlook}\label{sec:so}

In this final section, after a quick summary of the results we have 
obtained, we shall provide a brief outlook.


\subsection{Summary}

The direct or indirect detection of PGWs remains the holy grail in 
cosmology.
The detection of PGWs can help us determine the energy scale at which
inflation occurred.
As we discussed in the introduction, on the largest scales, the CMB 
carries imprints of PGWs.
However, currently, there exists only an upper bound on the contributions
of the primordial GWs to the anisotropies in the CMB~\cite{Planck:2015sxf,
Planck:2018jri,Paoletti:2022anb}.
The latest CMB observations constrain the tensor-to-scalar ratio to be
$r<0.036$~\cite{BICEP:2021xfz}, which corresponds to the upper bound on
the Hubble scale during inflation to be $\HI/\Mpl\simeq 10^{-5}$.
The recent detection of a SGWB by the PTAs has provided hope that direct
observations of PGWs at small scales or high frequencies may be possible
in the near future.
In fact, as we outlined in the introduction, a variety of GW observatories 
have been proposed, which are expected to operate at frequencies even as 
high as a few hundred GHz.
In such a situation, it becomes imperative to understand the behavior of 
the PS of GWs at very high frequencies.

In this work, we have examined the effects of regularization and smoothing
of the transition from inflation to the epoch of radiation domination on
the PS of GWs.
We have argued that regularization is essential if we are to avoid the~$k^2$ 
rise in the PS on small scales.
We have adopted the method of adiabatic subtraction to regularize the PS.
We have evolved the PGWs until the time of radiation-matter equality and 
today, and regularized the PS at these instances.
Importantly, we find that regularization does not affect the standard 
predictions over the wave numbers $k \lesssim \ke$. 
Moreover, we find that, in the case of instantaneous transitions from de
Sitter inflation to the epochs of radiation and matter domination, the 
regularized PS of PGWs does not exhibit a rise over wave numbers such 
that $k\gtrsim \ke$, i.e. over wave numbers that always remain inside the
Hubble radius.
Instead, the PS oscillates about zero with a constant amplitude in this 
domain.

We also showed that, when transitions are involved in the cosmological 
scenarios of interest, it becomes important to take into account the 
effects due to the smoothness of the transitions.
We should clarify that the process of regularization and smoothing of
the transitions play distinct roles in the determining structure of 
the primordial correlation functions.
It seems reasonable to expect that, across transitions, the combination 
of regularization and smooth transitions can ensure that generic 
correlation functions in real space are well-behaved.
This is of course natural since realistic transitions are expected to 
be smooth and, in fact, they are expected to be infinitely continuous.
We worked with a linear function to smooth the `effective potential' $U(\eta)$ 
during the transition from inflation to the epoch of radiation domination.
We obtained exact solutions to the mode functions describing the PGWs 
during the transition and evaluated the PS of GWs at the time of
radiation-matter equality as well as today.
We showed that, in contrast to the case of instantaneous transition, 
the smoother transition leads to a suppression in the regularized PS 
of PGWs over wave numbers~$k\gtrsim \ke$. 
Over such high wave numbers, while the regularized PS continues to oscillate 
about zero as in the case of the instantaneous transition, it does so with a 
decreasing amplitude.
Further, we encountered an interesting behavior in the regularized PS 
evaluated today, after a smoother transition from inflation to radiation
domination and an instantaneous transition from radiation to matter 
domination.
In this scenario, we found that the suppression in the regularized PS 
ceases at ultra-high wave numbers such that $k\gtrsim \ke^3/\keq^2$ and 
the amplitude of the oscillations in the PS becomes constant thereafter.
This behavior can be attributed to the assumption of an instantaneous 
transition from radiation to matter domination.
The result clearly suggests that, unless all the intermediate transitions 
are completely smoothed, the PS of PGWs evaluated in a given epoch will 
carry imprints of earlier, unrealistic, instantaneous transitions.


\subsection{Outlook}

There are a few directions in which the calculations we have carried out 
need to be extended.
To begin with, we need to extend our analysis for transitions from inflation
to radiation domination that are even more smooth.
In fact, we have to work with $U(\eta)$ that are infinitely continuous.
However, it is not easy to obtain solutions for the scale factor~$a(\eta)$ 
and the rescaled mode functions~$\mu_k(\eta)$ in such situations.
We may have to consider some approximations to determine the Bogoliubov
coefficients~$(\alpha_k^{\mathrm{r}},\beta_k^{\mathrm{r}})$ in these 
situations.
Secondly, we need to introduce a phase of reheating (between inflation
and the epoch of radiation domination) and evolve the PGWs across the 
period of reheating as well as the epochs of radiation and matter 
domination, and evaluate the regularized PS of PGWs today.
These are relatively straightforward exercises and can be implemented 
easily. 
Thirdly, we need to smooth the different transitions from the end of
inflation until today and determine the corresponding imprints on the
PS of PGWs.
It seems clear that the suppression in the PS of PGWs over $k\gtrsim \ke$ 
arises due to the smoothing of the transition from inflation to the 
epoch that succeeds it.
The smoothing of the later transitions can be expected to lead to some 
features around wave numbers that re-enter the Hubble radius at the time 
of these transitions, say, $\kre$, corresponding to the transition from
the phase of reheating to the epoch of radiation domination, and~$\keq$
corresponding to the time of transition from the epoch of radiation
domination to that of matter domination~\cite{Ng:1993pv}.
Fourthly, we have seen that the smoother transition from inflation to the
epoch that follows will lead to a suppression in the regularized PS of PGWs 
over the wave numbers $k \gtrsim \ke$.
In the simplest of scenarios, a smooth transition from inflation to 
reheating and eventually to the epoch of radiation domination is 
achieved by coupling the inflaton to radiation with the help of a 
constant coupling parameter.
The manner in which the regularized PS of PGWs will be suppressed at high 
wave numbers can be expected to depend on the details of the transition.
As is well known, the nature of the epoch of reheating will depend on 
the behavior of the inflationary potential near the minimum. 
It seems worthwhile to systematically examine the relation between the
form of the suppression in the regularized PS and the behavior of the 
inflationary potential near its minimum and the coupling of the inflaton 
to radiation.
We are currently working on these problems.


\section*{Acknowledgements}

The authors wish to thank Debika Chowdhury, Fabio Finelli, John Giblin, Vincent 
Vennin and Masahide Yamaguchi for discussions.
AH wishes to thank Indian Institute of Technology (IIT) Madras, Chennai, India,
for supporting a visit to Institut d'Astrophysique de Paris~(IAP), France, 
through the International Immersion Experience Program.
AP, JM and LS would like to thank the Indo-French Centre for the Promotion of 
Advanced Research (IFCPAR/CEFIPRA), New Delhi, India, for support of the proposal
6704-4 titled `Testing flavors of the early universe beyond vanilla models with 
cosmological observations’ under the Collaborative Scientific Research Programme.
LS would also like to thank IAP and F{\' e}d{\' e}ration de Recherche 
Interactions Fondamentales for supporting a visit to IAP, where part 
of this work was completed.
The authors also thank IAP and IIT Madras for hospitality. 


\appendix

\section{Equation of motion of PGWs}\label{app:eom-gws}

In this appendix, we shall derive the equation of motion governing PGWs.
Though this is a standard result, we shall present the derivation for 
the sake of completeness.
Also, the derivation helps us to set the notation.

Let us expand the metric tensor as 
\begin{equation}
g_{\mu \nu}=\bar{g}_{\mu\nu}+\th_{\mu \nu},\label{eq:dm}
\end{equation}
where $\bar{g}_{\mu \nu}$ is the background metric and $\th_{\mu \nu}$ is 
the perturbation in the metric at the first order.
Using this perturbative expansion, we can express the Ricci tensor as
follows:
\begin{equation}
R_{\mu \nu}=\bar{R}_{\mu \nu}+R_{\mu \nu}^{(1)}(\bar{g},\tilde{h}),\label{eq:rt}
\end{equation}
where the first term is the Ricci tensor associated with the background,
while the second term is linear in $\th_{\mu\nu}$.

To arrive at the equation of motion of GWs, we begin with the following 
expression for the Ricci tensor $R_{\mu \nu}^{(1)}$ at the first order
in the perturbations~(in this context, see Ref.~\cite{Misner:1973prb},
Sec.~35.13):
\begin{equation}
R_{\mu \nu}^{(1)} 
= \f{1}{2} \l(-\nabla_{\nu}\nabla_{\mu} \tilde{h}
-\nabla^{\alpha}\nabla_{\alpha} \tilde{h}_{\mu \nu}
+\nabla^{\alpha}\nabla_{\nu} \tilde{h}_{\alpha \mu}
+\nabla^{\alpha}\nabla_{\mu} \tilde{h}_{\alpha \nu}\r),\label{eq:Rmunu1}
\end{equation}
where the covariant derivatives are to be taken with respect to the 
background metric~$\bar{g}_{\mu\nu}$.
Let us now assume that~$\bar{g}_{\mu\nu}$ denotes the 
line-element~\eqref{eq:flrw-le} expressed in terms of the cosmic time~$t$.
The time-time-component of $R_{\mu \nu}^{(1)}$ is given by
\begin{equation}
R_{tt}^{(1)} 
= \f{1}{2} \l(-\nabla_{t}\nabla_{t} \tilde{h}
-\nabla^{\alpha}\nabla_{\alpha} \tilde{h}_{tt}
+\nabla^{\alpha}\nabla_{t} \tilde{h}_{\alpha t}
+\nabla^{\alpha}\nabla_{t} \tilde{h}_{\alpha t}\r).
\end{equation}
Note that, in the case of GWs, $\th_{t\mu}=\th^{t\mu}=0$ 
[cf. Eq.~\eqref{eq:flrw-wgws}].
Also, the transverse and traceless conditions~\eqref{eq:ttc} correspond 
to $\nabla_{\mu}\tilde{h}^{\mu\nu}=0$ and $\tilde{h}=\tilde{h}^{\mu}_{\mu}=0$.
On using these conditions, the different terms of $R_{tt}^{(1)}$ can be
calculated to be
\begin{subequations}
\begin{align}
\nabla_{t}\nabla_{t} \tilde{h}&=0,\\
\nabla^{\alpha}\nabla_{\alpha} \tilde{h}_{tt}&=0,\\
\nabla^{\alpha}\nabla_{t} \tilde{h}_{\alpha t}&=0,
\end{align}
\end{subequations}
which lead to 
\begin{equation}
R_{tt}^{(1)} = 0.
\end{equation}
The time-space~components of $R_{\mu \nu}^{(1)}$ are given by
\begin{equation}
R_{ti}^{(1)} 
= \f{1}{2} \l(-\nabla_{i}\nabla_{t} \tilde{h}
-\nabla^{\alpha}\nabla_{\alpha} \tilde{h}_{t i}
+\nabla^{\alpha}\nabla_{i} \tilde{h}_{\alpha t}
+\nabla^{\alpha}\nabla_{t} \tilde{h}_{\alpha i}\r).
\end{equation}
The different terms of $R_{ti}^{(1)}$ can be evaluated to be
\begin{subequations}
\begin{align}
\nabla_{i}\nabla_{t} \tilde{h}&=0,\\
\nabla^{\alpha}\nabla_{\alpha} \tilde{h}_{t i}&=0,\\
\nabla^{\alpha}\nabla_{i} \tilde{h}_{\alpha t}&=0,\\
\nabla^{\alpha}\nabla_{t} \tilde{h}_{\alpha i}&=0,
\end{align}
\end{subequations}
leading to
\begin{equation}
R_{ti}^{(1)} = 0.
\end{equation}
Lastly, the space-space~components of $R_{\mu \nu}^{(1)}$ are given by
\begin{equation}
R_{ij}^{(1)} 
= \f{1}{2} \l(-\nabla_{j}\nabla_{i} \tilde{h}
-\nabla^{\alpha}\nabla_{\alpha} 
\tilde{h}_{i j}+\nabla^{\alpha}\nabla_{j} \tilde{h}_{\alpha i}
+\nabla^{\alpha}\nabla_{i}\tilde{h}_{\alpha j}\r).\label{eq:rt-fo}
\end{equation}
The different terms of $R_{ij}^{(1)}$ can be calculated to be
\begin{subequations}
\begin{align}
\nabla_{j}\nabla_{i} \tilde{h}&=0,\\
\nabla^{\alpha}\nabla_{\alpha} \tilde{h}_{ij}
&=-\ddot{\th}_{ij}+H \dot{\th}_{ij}+2 \dot{H} \th_{ij}
+4 H^2 \th_{ij}+\f{1}{a^2}\pa^2\th_{ij},\\
\nabla^{\alpha}\nabla_{j} \tilde{h}_{\alpha i}
&=\dot{H} \th_{ij}+4 H^2 \th_{ij},\\
\nabla^{\alpha}\nabla_{i} \tilde{h}_{\alpha j}
&=\dot{H} \th_{ij}+4 H^2 \th_{ij},
\end{align}
\end{subequations}
where, recall that, $H=\dot{a}/a$ is the Hubble parameter
and $\pa^2=\delta^{ij}\pa_{i}\pa_j$.
These results lead to
\begin{equation}
R_{ij}^{(1)} = \f{1}{2} \l(\ddot{\th}_{ij}
-H \dot{\th}_{ij}+4 H^2 \th_{ij}-\f{1}{a^2} \pa^2\th_{ij}\r).
\end{equation}
and, since $\tilde{h}_{ij}=a^2 h_{ij}$ [cf. Eqs.~\eqref{eq:dm} 
and~\eqref{eq:flrw-wgws}], this expression translates to
\begin{equation}
R_{ij}^{(1)} = \f{a^2}{2}\,\l(\ddot{h}_{ij} +3 H \dot{h}_{ij}
+4 H^2 h_{ij}+2 \f{\ddot{a}}{a} h_{ij}-\f{1}{a^2}\pa^2h_{ij}\r).
\end{equation}

At the first order in the perturbations, the Einstein tensor is
given by 
\begin{equation}
G_{\mu\nu}^{(1)}
=R_{\mu\nu}^{(1)}-\f{1}{2} g_{\mu\nu}^{(1)} R^{(0)}
=R_{\mu\nu}^{(1)}-3 a^2 \l(\f{\ddot{a}}{a}+H^2\r) h_{\mu\nu}.
\end{equation}
Since $h_{t \mu}=0$ and $R_{t\mu}^{(1)}=0$, we obtain that $G_{t\mu}^{(1)}=0$.
At the same order, the spatial components of the Einstein tensor 
can be evaluated to be
\begin{equation}
G_{ij}^{(1)}=\f{a^2}{2} \l(\ddot{h}_{ij}
+3 H \dot{h}_{ij}-\f{1}{a^2} \pa^2 h_{ij}\r)
-a^2 \l(2 \f{\ddot{a}}{a}+H^2\r) h_{ij}.
\end{equation}
Recall that the stress-energy tensor associated with a perfect fluid is 
given by 
\begin{equation}
T_{\mu\nu}=(\rho+p) u_\mu  u_\nu+p g_{\mu\nu},
\end{equation}
where $\rho$ and $p$ denote the energy density and pressure of the fluid
and $u^{\mu}$ represents its four velocity, which satisfies the condition
$g_{\mu\nu}u^\mu u^\nu=-1$.
We should mention that $(\rho,p,u^{\mu})$ include the perturbations as well.
According to the decomposition theorem, which is valid at the first order
in perturbation theory, the GWs will be sourced {\it only}\/ by the tensor 
component of the above stress-energy tensor (for a discussion in this 
context, see the textbook~\cite{Weinberg:2008zzc}).
At the first order in the perturbations, the tensor component of the 
stress-energy tensor in the FLRW universe is given by 
\begin{subequations}
\begin{align}
T_{t\mu}^{(1)} & =0,\\
T_{ij}^{(1)} &=\bar{p} a^2 h_{ij},
\end{align}
\end{subequations}
where $\bar{p}$ is the background pressure that is driving the expansion.
Note that, according to the zeroth order Friedmann equation 
\begin{equation}
2 \f{\ddot{a}}{a}+H^2=-8 \pi G \bar{p}.
\end{equation}
On taking into account the above results, the first order Einstein equation
$G_{ij}^{(1)}=8\pi G T_{ij}^{(1)}$ leads to the following equation of 
motion governing the PGWs:
\begin{equation}
\ddot{h}_{ij}+3 H \dot{h}_{ij}-\f{1}{a^2} \pa^2 h_{ij}=0.
\end{equation}
In terms of the conformal time $\eta$, this equation reduces to
\begin{equation}
h_{ij}'' + 2 {\cal H} h_{ij}'-\pa^2h_{ij}=0,
\end{equation}
where ${\cal H}=a'/a=a H$ is the conformal Hubble parameter.


\section{Characteristic strain, sensitivity curves and the 
PS today}\label{app:strain}

The PGWs will appear as a stochastic and isotropic background of GWs today.
The characteristic strain~$h_{\mathrm{c}}(f)$, the spectral density~$S_h(f)$ 
and the PS~$\pt(f)$ are equivalent theoretical measures to describe a 
stochastic and isotropic GW background~\cite{Maggiore:1999vm,
Romano:2016dpx,Maggiore:2007ulw}.
In this appendix, we shall discuss the relation between these quantities.

\begin{figure}[!t]
\centering
\includegraphics[width=1.0\textwidth]{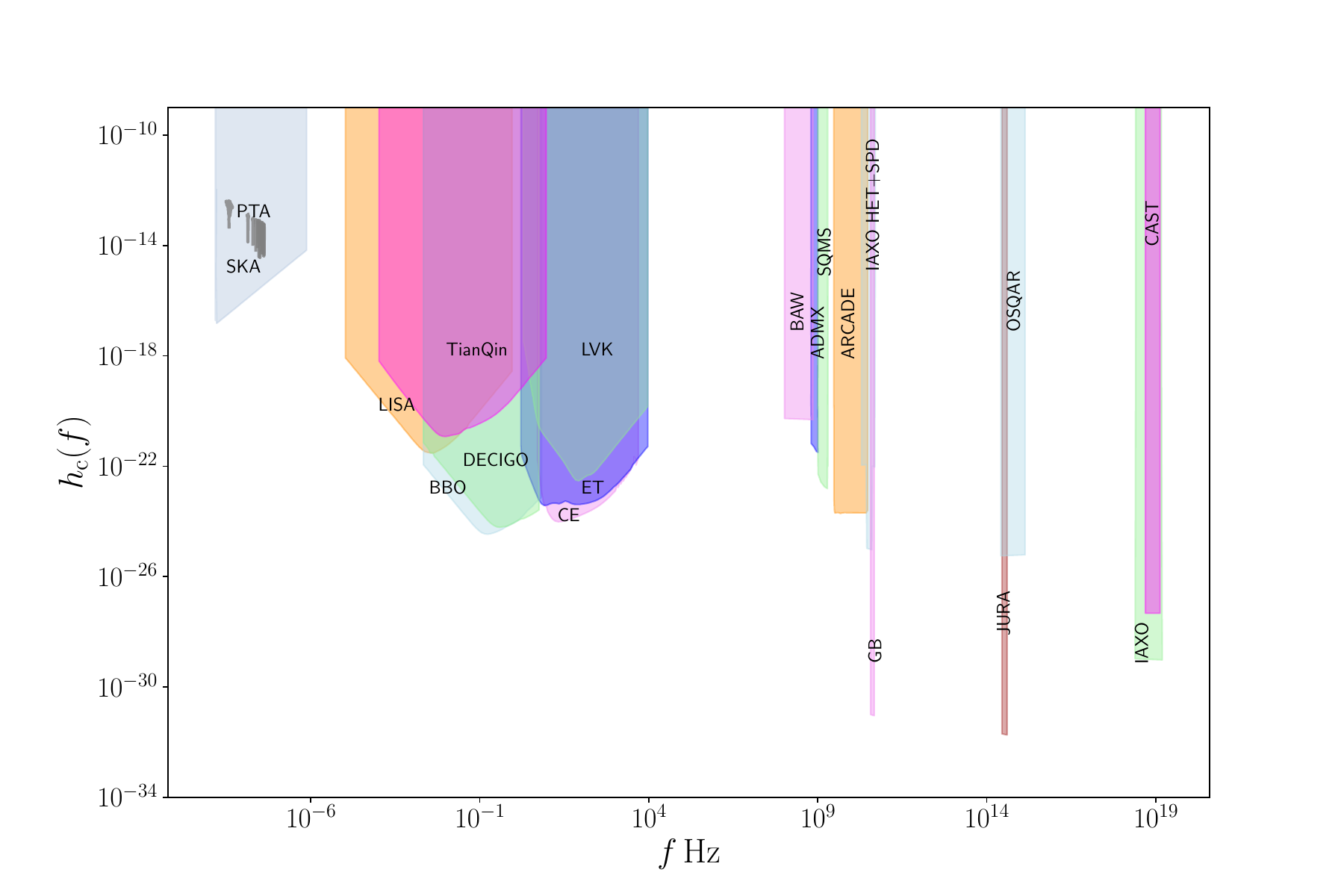}
\caption{The sensitivity curves for the dimensionless characteristic 
strain $h_{\mathrm{c}}(f)$ have been illustrated for the different GW
observatories~\cite{Moore:2014lga,Franciolini:2022htd}.}
\label{fig:strain}
\end{figure}

Let us decompose the stochastic GWs at a given spatial point, say,
${\bm x}=0$, as follows (in this regard, see Refs.~\cite{Maggiore:1999vm,
Romano:2016dpx,Maggiore:2007ulw}):
\begin{align}\label{eq:hij-d}
h_{ij}(t)=\sum_{\lambda={(+,\times)}}
\int_{-\infty}^{\infty}\d f\int \d \hat{\bm n}\,
\varepsilon^{\lambda}_{ij}(\hat{\bm n})
\tilde{h}^{\lambda}_f(\hat{\bm n}) \e^{-2\pi i f t},
\end{align}
where $\varepsilon^{\lambda}_{ij}(\hat{\bm n})$ is the polarization 
tensor we had introduced in Eqs.~\eqref{eq:ptd}.
Recall that the polarization tensor satisfies the transverse and 
traceless conditions~\eqref{eq:ttc}.
Also, it is subject to the normalization condition~\eqref{eq:n}.
The fact that $h_{ij}(t)$ is real implies that
$\tilde{h}_f^{\lambda\ast}(\hat{\bm n})
=\tilde{h}_{-f}^{\lambda}(\hat{\bm n})$.
If we assume that the GW background is stochastic, isotropic, unpolarized
and stationary, then the quantity $\tilde{h}_f^{\lambda}(\hat{\bm n})$
satisfies the condition~\cite{Maggiore:1999vm,Romano:2016dpx,
Maggiore:2007ulw}
\begin{equation}\label{eq:hlhlAvg}
\l\langle\tilde{h}_f^{\lambda}(\hat{\bm n})
\tilde{h}_{f'}^{\lambda'\ast}(\hat{\bm n}')\r\rangle
=\delta^{(1)}(f-f') \f{1}{4\pi}\delta^{(2)}(\hat{\bm n},\hat{\bm n}')
\delta_{\lambda\lambda'}\f{1}{2}S_h(f),
\end{equation}
where the angular brackets on the left-hand side denote averaging
over the random variables, and the quantity $S_h(f)$ is referred 
to as the spectral density of GWs.
We should point out that the delta function~$\delta^{(2)}(\hat{\bm n},
\hat{\bm n}')$ in the above expression implies that the stochastic
background is isotropic.
Note that the delta function satisfies the condition
\begin{equation}
\int \d^{2}\hat{\bm n}\, \delta^{(2)}(\hat{\bm n},\hat{\bm n}')
=\int_{0}^{2\pi} \d\phi\, \delta^{(1)}(\phi-\phi')
\int_{0}^{\pi} \d\theta \sin\theta\,\delta^{(1)}(\cos\theta-\cos\theta')=1.
\end{equation}
If we now substitute the decomposition~\eqref{eq:hij-d} in
Eq.~\eqref{eq:hlhlAvg}, we obtain that
\begin{equation}\label{eq:hijhijavg1}
\delta^{im}\delta^{jn} \langle h_{mn}(t)h_{ij}(t)\rangle
=2\int_{-\infty}^{\infty}\d f S_h(f)=4 \int_{0}^{\infty}
\d (\ln f) f S_h(f),
\end{equation}
where we have made use of the fact that $S_h(f)=S_h(-f)$.
It should be clear from the above relation that $S_h(f)$
has the same dimensions as~$f^{-1}$.
The so-called dimensionless characteristic strain~$h_{\mathrm{c}}(f)$,
which is a measure of the amplitude of GWs as a function of frequency,
is defined through the relation~\cite{Maggiore:1999vm,Maggiore:2007ulw} 
\begin{equation}
\delta^{im}\delta^{jn} 
\langle h_{mn}(t)h_{ij}(t)\rangle
=2 \int_{0}^{\infty}\d (\ln f) h_{\mathrm{c}}^2(f)\label{eq:sh}
\end{equation}
and, from the above two equations, we obtain that
\begin{equation}
h_{\mathrm{c}}^2(f)=2f S_h(f).
\end{equation}
On comparing Eqs.~\eqref{eq:twopoint} and~\eqref{eq:hijhijavg1}, 
we arrive at the following relation between the PS of PGWs today, 
viz. $\pt(f,\eta_0)$, the spectral density~$S_h(f)$ and the
characteristic strain~$h_{\mathrm{c}}(f)$:
\begin{equation}\label{eq:pt-hc}
\pt(f,\eta_0)=4 f S_h(f)=2 h_{\mathrm{c}}^2(f).
\end{equation}
Often, the sensitivity curves of the GW observatories are indicated
in terms of the characteristic strain~$h_{\mathrm{c}}(f)$ (for 
related discussions, see Ref.~\cite{Moore:2014lga} and the associated
web-page; in this regard, also see Refs.~\cite{Romano:2016dpx,
Franciolini:2022htd}).
In Fig.~\ref{fig:strain}, we have illustrated these sensitivity curves 
for the different GW observatories.
We have made use of the sensitivity curves for $h_{\mathrm{c}}(f)$ and 
the relation~\eqref{eq:pt-hc} above to arrive at the corresponding 
sensitivity curves for the PS of GWs.
It is these sensitivity curves that we have included in 
Figs.~\ref{fig:PT-hf},  \ref{fig:ps-rps-ds-rd-it}
and~\ref{fig:PS-ds-sd-rd-it-f}. 


\bibliographystyle{JHEP}
\bibliography{references}
\end{document}